\documentclass[12pt]{article}

\usepackage{amsmath}
\usepackage[hidelinks]{hyperref}
\usepackage{graphicx,psfrag,epsf}
\usepackage{enumerate}
\usepackage{natbib}
\usepackage{url} 
\usepackage{amsmath} 
\usepackage{amssymb}
\usepackage[section]{placeins}
\usepackage{booktabs}
\usepackage[toc,page]{appendix}
\usepackage{multirow}
\usepackage{hyperref}
\usepackage{xurl}
\usepackage{natbib}
\usepackage{float}
\usepackage{comment}
\usepackage{graphicx}
\graphicspath{{figures/}}
\usepackage{dsfont}
\usepackage{rotating}
\usepackage{lscape}
\usepackage{amsthm}
\usepackage{caption}
\usepackage{subcaption}
\usepackage{dsfont}
\usepackage[showframe=false]{geometry}
\usepackage{changepage}

\usepackage{tikz}
\usetikzlibrary{bayesnet}
\usepackage[frame,graph,arrow]{xypic}
\usepackage{pifont}

\usepackage{makecell} 

\usepackage{setspace}
\usepackage{authblk}

\usepackage[english]{babel}
\newcommand{\sym}[1]{\ifmmode^{#1}\else\(^{#1}\)\fi}
\usepackage{geometry}
\usepackage{rotating}
\usepackage{setspace} 
\usepackage{footmisc}
\usepackage{subcaption} 
\usepackage{longtable}
\usepackage{pdflscape}
\usepackage{array}

\newif\ifmain
\newif\ifonlineappendix

\maintrue
\onlineappendixtrue

\title{Sanctions, Evasion, and Deterrence: Evidence from Speeding Enforcement\thanks{This research was supported by the Australian Research Council under the Discovery Projects funding scheme (project number DP240100895). We thank the Queensland Department of Transport and Main Roads for providing access to data derived from the TRAILS database. We are grateful to Jeffrey Alcorn, Craig Mitchell, and Toby Scholl for their assistance in answering our questions and helping us better understand the data. We also thank Andrew Ireland, Andrea La Nauze, Johannes Kunz, and participants at the 17th Workshop on the Economics of Health and Wellbeing, the 2026 Australian Conference of Economists, and the Third Freiburg PhD-Workshop on the Economics of Criminal Behavior for valuable comments and suggestions.}
}
\author[]{Romain Gauriot\thanks{romain.gauriot@deakin.edu.au}\ }
\author[]{Maria Wisniewska\thanks{maria.wisniewska@deakin.edu.au}}
\affil{Department of Economics, Deakin University, Melbourne, Australia}

\date{\today}

\begin{document}

\maketitle

\begin{abstract}
			\noindent
           The design of sanctions plays a central role in shaping reoffending behavior. We study the effects of fixed monetary fines and accumulating penalty points on speeding recidivism and liability transfer. Exploiting a large increase in sanctions in Queensland, Australia, we estimate the causal effects of these sanctions on subsequent offending and nomination of other drivers. Higher demerit points increase fine renomination, consistent with strategic blame shifting. Estimated reductions in reoffending are small among offenders facing higher fines alone. Reoffending falls by a substantial 4.6 percentage points when higher fines are accompanied by additional penalty points.
\end{abstract}
{\small \noindent\textbf{Keywords}:  Deterrence, Evasion, Recidivism, Regression Discontinuity in Time.}
 
					\bigbreak

\newpage
\section{Introduction}

A central question in the economics of crime is how the design of sanctions affects reoffending behavior. In Becker’s deterrence framework, penalties deter offending when their expected cost, determined by both sanction severity and the probability of detection, exceeds the expected benefit from the offense \citep{becker1968}. In many enforcement settings, sanctions escalate with repeated violations, potentially strengthening incentives for compliance as offenders accumulate a record \citep{polinsky1991}. This escalation can be structured explicitly through administrative demerit point systems, such as those used in road traffic enforcement and occupational licensing where repeated infractions accumulate toward license suspensions. Related dynamics arise in financial settings, where repeated defaults progressively restrict future borrowing capacity through deteriorating credit scores. Escalation may emerge less formally within judicial systems, where judges exercise discretion to impose harsher sentences on repeat offenders.

Cumulative sanction systems can only operate as intended when violations are attributed to the individuals who committed them. When attribution is imperfect, offenders may avoid escalating penalties by transferring liability to others. This is particularly relevant in camera-based traffic enforcement, where registered vehicle owners can transfer fines and associated demerit points to another person by nominating them as the driver. Similar incentives arise in taxation, where households reallocate asset ownership between spouses to reduce their total tax burden \citep{stephens2004impact}. In cumulative sanction systems, these opportunities for reassignment create an additional behavioral margin: harsher sanctions may deter future violations, but they may also induce strategic liability transfer.

We examine these two margins of behavioral adjustment, deterrence and strategic liability transfer,  in the context of a sharp increase in speeding fines and demerit points introduced in Queensland, Australia. In July 2022, fine amounts rose substantially ranging from 123\% to 225\% of pre-reform levels depending on the severity of the offense. Drivers travelling 11–12 km/h above the speed limit also became subject to higher demerit points. We exploit differential changes in penalties across speed bands to estimate the behavioral response to alternative sanction regimes, distinguishing between categories affected by fine increases only and those subject to simultaneous increases in fines and demerit points.

Our study yields two main findings. First, increases in demerit points led to a substantial rise in driver nomination, consistent with strategic efforts to avoid license suspension. Second, reoffending fell sharply in speed bands with increases in both fines and demerit points, but remained unchanged in bands where penalties increased through fines alone.

Our empirical analysis draws on administrative data covering the universe of speeding infringements issued in Queensland. The dataset comprises violations detected by speed cameras, which provide objective and highly precise measurements of vehicle speed. We consider both fixed and mobile cameras, which differ in an important respect: because mobile cameras may photograph vehicles from the rear, the driver is not always identifiable, creating greater scope to avoid attribution by nominating another driver. Fixed cameras, by contrast, capture a frontal image that more readily identifies the driver, limiting opportunities for such reassignment.

We estimate the effect of the penalty increase on nomination and reoffending using a Regression Discontinuity in Time (RDiT) design. This approach exploits the fact that the policy change was introduced sharply and applied uniformly to all drivers in the state. Because the reform took effect on a precise date, individuals who committed an offense just before the cutoff were exposed to the old penalty schedule, whereas otherwise similar individuals who offended just after were subject to substantially higher fines. By comparing drivers who offend just before and just after this temporal threshold, the RDiT framework allows us to estimate the causal effect of the reform on both driver nomination and subsequent reoffending.

For the RDiT design to yield credible causal estimates, offenses occurring just before and just after the reform must be comparable, and all other determinants of reoffending risk should evolve smoothly at the cutoff. While driving behavior and enforcement outcomes fluctuate substantially on a day-to-day basis, these high-frequency changes do not threaten identification so long as they are unrelated to the policy change and do not induce systematic manipulation of offending around the reform date \citep{hausman2018}.

We assess the credibility of the design through a series of balance checks on predetermined characteristics and prior offending behavior around the cutoff. For fixed cameras, we find no discontinuities in covariates around the reform, supporting a causal interpretation of our reoffending estimates. 

For mobile cameras, however, we detect differences in driver characteristics on either side of the cutoff. 
These differences are driven in part by an increase in fine transfers following the introduction of higher demerit points (see Section \ref{sec:blame-shifting}). The remaining differences may reflect changes in mobile-camera enforcement or greater avoidance of mobile cameras by drivers after the reform (see Section \ref{sec:gendetterenceandenforcement_mobile}). For fixed cameras, by contrast, we find no evidence of a contemporaneous change in enforcement intensity or detections around the reform (see Section \ref{sec:gendetterenceandenforcement_fixed}). We therefore use the fixed-camera sample to estimate the effect of sanctions on recidivism and use the mobile-camera sample, where photographic identification of the driver is less certain, to study nomination behavior.

Using the fixed-camera sample, we find that increases in penalties lead to a 1.5 percentage-point decrease in reoffending, corresponding to a 5\% reduction relative to the pre-reform mean. The effect is concentrated among offenses attracting the largest increases in both financial penalties and demerit points, namely those involving speed excesses of 11 or 12~km/h.\footnote{The fine for exceeding the speed limit by 11 or 12~km/h increased from \$183 to \$431, while the demerit point allocation rose from 1 to 3. See Table~\ref{tab:fee_change}.} For this subsample, the reform reduced reoffending by 4.6 percentage points, equivalent to a 13\% decline relative to the pre-reform mean. We find no comparable effect for offenses involving speed excesses below 11~km/h or above 12~km/h, where monetary fine increased but demerit point allocations remained unchanged. This pattern suggests that the behavioral response is driven by the combined increase in financial penalties and demerit points, rather than by higher fines alone.

To study nomination behavior, we exploit the mobile-camera sample, where the driver is not always unequivocally identifiable from photographic evidence, creating scope for liability reassignment. Renomination rates increase for mobile-camera fines involving 11–12 km/h speed excesses, consistent with stronger incentives to reassign liability in response to higher demerit points. We detect no corresponding effect for other speed categories, where demerit points do not increase, or for fixed-camera detections, where frontal imaging substantially limits opportunities for such reassignment.  

Understanding these responses is of particular importance. Transport-related injuries are estimated to cost approximately AUD~1.5 billion annually in Australia.\footnote{\url{https://www.aihw.gov.au/reports/injury/transport-accidents}, accessed on 02.12.2025} These costs may increase with the growing adoption of electric vehicles, which tend to be quieter and heavier than comparable petrol cars, potentially creating additional risks for pedestrians and cyclists.

Traffic enforcement relies primarily on monetary fines and escalating consequences for repeat offenders, making behavioral responses to sanction severity a key determinant of road safety outcomes. Because speeding is both pervasive and closely linked to accident risk \citep{deangelo2014}, understanding whether harsher penalties reduce subsequent violations, or instead induce strategic liability transfers, is essential for evaluating the effectiveness of road safety policy. More broadly, our setting provides a real-world context in which to assess the empirical relevance of specific deterrence and the conditions under which  sanctions achieve their intended behavioral effects.

\section{Related Literature}

We contribute to four strands of literature.

First, we contribute to the literature on evasion and strategic responses to enforcement. A large literature, primarily focused on taxation, studies how individuals adapt to audits, enforcement intensity, and monitoring technologies  \citep{andreoni1998,slemrod2019tax}. Influential work shows that compliance depends critically on the ability to verify attribution: tax evasion is close to zero for income subject to third-party reporting but substantial for self-reported income \citep{kleven2011}; transactions with a verifiable third-party paper trail respond less to announcements of increased monitoring than transactions without one \citep{pomeranz2015}. Related theoretical and empirical work outside taxation examines how monitoring and contract design shape fraudulent reporting in unemployment insurance \citep{fuller2015} and automobile insurance \citep{dionne2001}. Comparatively little attention, however, has been paid to how cumulative sanction systems shape incentives for attribution avoidance. In the context of traffic enforcement, we are not aware of prior work examining whether demerit point regimes induce strategic reassignment of liability. We show that escalating non-monetary sanctions can create incentives to avoid attribution by nominating another driver.

Second, we contribute to the literature on deterrence and sanctions' effectiveness. While economic theory predicts that higher expected penalties should reduce offending, the empirical evidence on sanction severity remains mixed. Some studies find that stronger sanctions reduce offending \citep{helland2007does,abrams2012estimating,bhuller2020incarceration,dusek2021,gonccalves2023police,Kaila2024}, whereas others document limited or unintended behavioral responses to monetary penalties \citep{gneezy2000fine,ALEXEEV2022555,finlay2024}.

The papers closest to ours are \citet{abay2018}, \citet{dusek2021}, \citet{gonccalves2023police}, and \citet{Kaila2024}. \citet{dusek2021} exploit discontinuities in speeding sanctions around legal thresholds to study deterrence, focusing primarily on the extensive margin of punishment (receiving a ticket versus none). They show large effects of receiving a ticket on subsequent speeding, while the additional effect of higher fines is small and imprecisely estimated. \citet{Kaila2024} studies the intensive margin of fines using similar discontinuities and finds strong deterrent effects of income-based fines for extreme offenders. In contrast, our setting allows us to study sanction severity among more common offenders with fixed, rather than income-based, fines. \citet{gonccalves2023police} exploit variation in police officers’ discretionary use of discounted fines and show that harsher sanctions reduce future traffic offending and crash involvement. \citet{abay2018} studies the effect of demerit points on reoffending using a difference-in-differences framework and finds that demerit points lead to a 9-34\% reduction in frequency of offending, depending on the accumulated amount. Online Appendix \ref{app:littreview} provides a detailed overview of the literature on speeding recidivism. Relative to this literature, our paper makes two contributions. First, whereas the existing literature tends to focus on either monetary fines or demerit points, we study a reform affecting both sanctions. Second, we document strategic evasion through liability transfer, an adjustment margin that, to the best of our knowledge, has not been studied previously in the speeding enforcement literature.

Third, we contribute to the literature exploiting traffic enforcement settings as a powerful empirical environment for studying enforcement, risky behavior, and public policy. \citet{ashenfelter2004using} exploit changes in highway speed limits to estimate the value of a statistical life. Other work uses traffic enforcement data to study deterrence and recidivism \citep{hansen2015punishment}, discrimination in policing \citep{Goncalves2021,grosjean2023inflammatory}, behavioral responses to non-linear penalty schedules \citep{traxler2018bunching}, and the effects of targeted regulation on risky behavior \citep{moore2024shaping}. Our study adds to this literature by showing how cumulative sanction regimes shape both deterrence and strategic responses.

Fourth, we contribute to the literature on manipulation in regression discontinuity designs (RDD). A central identifying assumption of RDD is that individuals on either side of the cutoff are comparable \citep{Cattaneo2020a}. In many settings, however, agents can manipulate their positions relative to the threshold, leading to bunching in the distribution of the running variable. For example, teachers inflate grades to move students above performance thresholds \citep{Diamond2016,Dee2019}, police officers disproportionately report speed excesses below sanction cutoffs \citep{Goncalves2021,gonccalves2023police}, and medical screening outcomes are adjusted around eligibility thresholds to permit blood donations \citep{rosenman2027donor}. We document a different form of manipulation: individuals manipulate not the running variable itself, but the assignment of outcomes across individuals through liability transfer. We discuss this mechanism and its implications for RDD in Section \ref{sec:manipulationRD}.

\section{Background}

\subsection{Regulatory setting}

Any monetary penalty in Queensland is defined in terms of a base unit, the Penalty Unit (PU), which is indexed annually at the start of each financial year. The fine for a given offense is determined by applying offense-specific weights to the prevailing value of the PU, and policymakers may adjust penalties either through routine indexation or by revising these weights. For speeding offenses, weights vary discretely with the degree of speed excess above the posted limit. Over our study period (1 July 2014-31 December 2024), penalty amounts grew through routine indexation between July 2014 and June 2022, and again from July 2023 onwards. On 1 July 2022, however, the government implemented a discrete increase in offense weights, resulting in a substantial rise in fines across speeding and other traffic offenses (see Table~\ref{tab:pu_change} in the Appendix).\footnote{\url{https://www.parliament.qld.gov.au/Work-of-the-Assembly/Tabled-Papers/docs/5722t571/5722t571-1155.pdf}, accessed on 10.09.2025} The reform was accompanied by a reduction in the lowest penalty threshold from 13~km/h to 11~km/h above the speed limit, with the stated objective of advancing the government's long-term road safety strategy of reducing serious injuries and fatalities to zero by 2050.\footnote{\url{https://statements.qld.gov.au/statements/94934}, accessed on 11.09.2025}

The reform was publicly announced several months before its implementation, with media coverage detailing the new penalties as early as April 2022.\footnote{See \url{https://www.carexpert.com.au/car-news/queensland-increases-fines-for-driving-offences}, accessed 17 August 2026.} However, knowledge of the precise penalties appears to have been limited: a survey done a year after the reform found that 66.7\% of drivers could not correctly identify the new lowest speeding bracket \citep{qldtmr2023speeding}.

\begin{figure}[h!]
  \centering
  \includegraphics[width=0.9\textwidth]{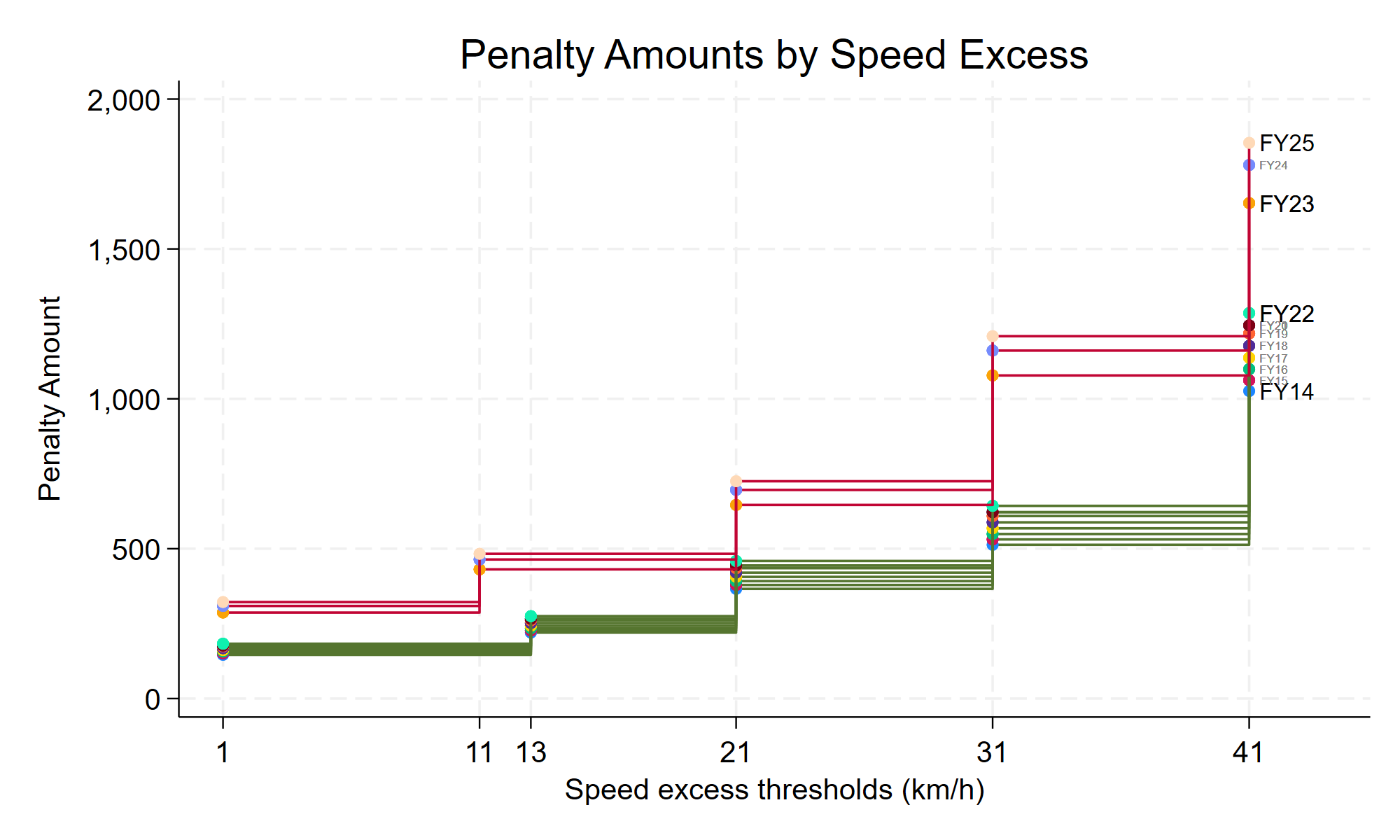}
  \caption{Penalty amount by speed excess over time. Before the rule change in green and after it in red.}
  \label{fig:penalty_amounts}
\end{figure}

The resulting changes in fines and demerit points are summarised in Table~\ref{tab:fee_change} and illustrated in Figure~\ref{fig:penalty_amounts}. Moving from the lowest to the highest speed excess categories, the 1-10~km/h range saw a fine increase of \$104 (from \$183 to \$287), with no change in demerit points. The 11--12~km/h range experienced a fine increase of \$248 (from \$183 to \$431) and, crucially, a demerit point increase of 2 points (from 1 to 3)---the only category where non-monetary sanctions were tightened. The 13-20~km/h range saw a fine increase of \$156 (from \$275 to \$431) with no change in demerit points. For higher speed excess categories, the 21-30~km/h range increased by \$187 and the 31-40~km/h range by \$435, both with unchanged demerit point allocations. The fact that demerit points changed only for the 11--12~km/h range, while all categories faced higher monetary fines, is the key source of variation in our empirical design---it allows us to partially disentangle the effects of monetary and non-monetary sanctions on subsequent behavior.

\begin{table}[h!]
\centering
\begin{tabular}{c|ccc|ccc}
\hline
 & \multicolumn{3}{c}{Fine amount (AUD)} & \multicolumn{3}{c}{Demerit points} \\
Speed excess & Before & After & $\Delta$ & Before & After & $\Delta pt$ \\
\hline
1-10 & \multirow{2}{*}{183 (6\%)} & 287 (8\%) & +104 & \multirow{2}{*}{1} & 1 & 0 \\
\cline{1-1}\cline{3-4}\cline{6-7}
11--12 & & \multirow{2}{*}{431 (13\%)} & +248  & & \multirow{2}{*}{3} & +2 \\
\cline{1-2}\cline{4-5}\cline{7-7}
13--20 & 275 (8\%) & & +156 & 3 & & 0 \\
\hline
21--30 & 459 (14\%) & 646 (19\%) & +187 & \multicolumn{2}{c}{4} &  0 \\
\hline
31--40 & 643 (20\%) & 1,078 (38\%) & +435 & \multicolumn{2}{c}{6}  &  0 \\
\hline
$>$40 & 1,286 (40\%) & 1,653 (48\%) & +367 & \multicolumn{2}{c}{8}  &  0\\
\hline
\multicolumn{7}{p{14cm}}{\footnotesize Notes: Fine amounts are in AUD. Percentages are relative to the statutory minimum monthly wage in effect on 1 July 2021 (AUD 20.33 per hour) and 1 July 2022 (AUD 21.38 per hour).} \\
\end{tabular}
\caption{Fine amount and demerit points for speeding offenses in Queensland before and after the change of 1 July 2022.}
\label{tab:fee_change}
\end{table}

Demerit points are a particularly consequential sanction: a driver who accumulates 12 or more points within a three-year period becomes liable for license suspension. The increase in demerit points for the 11--12~km/h range therefore accelerated progression toward suspension for drivers in this category, raising the non-monetary cost of repeated offending.

When an offense is detected by a fixed or mobile camera, the infringement notice is automatically issued to the registered owner of the vehicle, who is taken to have committed the offense even if they were not the actual driver at the time.\footnote{\url{https://everydaylawinqld.org.au/resource/camera-detected-infringement-notices/}, accessed on 11.09.2025} If the registered owner was not the driver, they may transfer liability to the actual driver by submitting a nomination within 28 days of the issue date.\footnote{\url{https://www.qld.gov.au/transport/safety/fines/how-to-transfer-a-fine}, accessed on 11.09.2025} The nominated person must hold a valid Australian driver license; if no nomination is submitted within this window, the registered owner is deemed to have been the driver and is liable for both the monetary fine and the associated demerit points.\footnote{\url{https://qro.qld.gov.au/fines-and-infringements/about/}, accessed on 11.09.2025}

Making a false declaration in a nomination constitutes a criminal offense under Queensland law, punishable by a fine or imprisonment.\footnote{\url{https://www.qld.gov.au/transport/safety/fines/how-to-transfer-a-fine}, accessed on 11.09.2025} Nonetheless, the transfer system creates a margin for strategic behavior: registered owners with a higher demerit point count have an incentive to nominate another driver regardless of who was actually behind the wheel. We refer to such behavior as an \textit{untruthful transfer}, distinguishing it from a \textit{truthful transfer} in which the nomination correctly identifies the actual driver.

\section{Data}

We use administrative data from the Queensland Department of Transport and Main Roads covering the universe of detected speeding offenses in the state \citep{qld_trails}. The dataset records the method of detection, distinguishing between on-road police enforcement, fixed speed cameras, mobile speed cameras, road safety camera trailers, point-to-point speed cameras, and combined red-light/speed cameras.\footnote{\url{https://www.qld.gov.au/transport/safety/fines/enforcement}
, accessed on 11.09.2025.} We focus on offenses detected by fixed and mobile speed cameras, distinguishing between the two detection methods throughout.

In 2022, speed cameras detected a total of 743,539 speeding offenses in Queensland, of which 203,319 were recorded by fixed speed cameras and 540,220 by mobile speed cameras. We exclude offenses involving speeds of 40 km/h or more above the limit (which trigger automatic license suspension), observations where the offender’s recorded age is below 16, and drivers who had accumulated more than 11 demerit points in the three years preceding the offense.\footnote{Accumulated demerit points are limited to speeding offenses in our analysis. Drivers may have accumulated demerit points from other offenses---such as red light offenses.} Applying a 150-day window around the July 2022 penalty reform yields final estimation samples of 149,755 offenses for fixed cameras and 438,590 for mobile cameras.

Each offense is linked to a unique driver identifier, allowing us to track offenders over time. The dataset contains information on the offense date and time, recorded speed, offense category, penalty amount, location, detection method, and the offender’s age. We derive the applicable local speed limit from the offense description. For instance, if a driver is recorded at 70 km/h and the offense category indicates that the speed exceeded the limit by “at least 13 km/h and not more than 20 km/h,” we infer a local speed limit of 50 km/h.

Fixed cameras provide clearer photographic evidence of driver identity than mobile cameras, making untruthful transfers more difficult. We therefore use fixed cameras to study recidivism and mobile cameras to study blame-shifting.

\section{Empirical strategy}
\label{sec:empirical_strategy}

We exploit a sharp, state-wide increase in speeding penalties that took effect on 1 July 2022. We first present event study estimates that trace the evolution of reoffending and transfer rates across 30-day periods relative to the reform date. These estimates provide a descriptive picture of the dynamics around the reform and complement the causal RDiT analysis that follows. We use the following equation:

\begin{equation}
y_{it} = \sum_{k \neq 0} \beta_k \cdot \mathbf{1}[P_{t} = k] + \gamma X_{it} + \alpha_i + \delta_{d} + \eta_{n} + \varepsilon_{it}
\label{eq:dynamic}
\end{equation}
where $y_{it}$ is an indicator equal to one if offense $i$ committed on day $t$ was followed by a subsequent offense within one year, in the recidivism specifications,\footnote{Using the unique identifier of each offender, we track whether an individual commits a subsequent speeding offense following an initial offense. To account for the possibility that drivers only learn about the penalty amount upon receipt of the fine (via e-mail or post), our main reoffending measure excludes offenses occurring within the first 14 days after the initial offense.} or was transferred, in the transfer specifications. $\mathbf{1}[P_{t} = k]$ is an indicator for the k-th 30-day period relative to the reform date, with k$<$0 denoting pre-reform periods and k$>$0 denoting post-reform periods. The omitted category is k=0, corresponding to the 30-day window immediately before the reform, which serves as the baseline. $X_{it}$ is a vector of controls including age at offense, school holidays, and public holidays; $\alpha_i$ denotes camera fixed effects; $\delta_{d}$ denotes day-of-week fixed effects; and $\eta_{n}$ denotes day/night fixed effects. Standard errors are clustered at the driver level. The pre-reform coefficients allow us to assess whether outcomes exhibit systematic trends prior to the reform; in the absence of such trends, these coefficients should be statistically indistinguishable from zero.

To move from descriptive patterns to causal estimates, we exploit the discrete timing of the reform, which creates a sharp discontinuity in penalty severity. Following \citet{hausman2018}, we implement a Regression Discontinuity in Time (RDiT) design, using time as the running variable to identify the causal effect of higher penalties on subsequent behavior. This approach has been widely used in economics, including in studies of transport and traffic-related policies. For example, \citet{anderson2014} uses a transit strike in Los Angeles to estimate the effect of public transport suspension on highway congestion, while \citet{bento2014} examine the welfare effects of a policy encouraging adoption of low-emission vehicles that indirectly penalised carpoolers through increased congestion. More closely related to our setting, \citet{finlay2024} study the impact of discrete increases in financial penalties for criminal traffic offenses on subsequent convictions, exploiting sharp policy introductions across US states.

Our empirical setting raises several identification considerations. Because the reform was implemented simultaneously across the entire state, we cannot exploit cross-sectional or geographic variation, and identification relies entirely on temporal variation around the reform date. The key identifying assumption is that the composition of offenders and their underlying propensity to reoffend evolve smoothly through the cutoff, so that any discontinuity in outcomes at the reform date can be attributed to the policy change rather than to coincident shifts in driving conditions, enforcement intensity, or offender characteristics. We assess this assumption through a series of balance checks on predetermined characteristics, placebo tests, and bandwidth sensitivity analyses, all discussed below.

We estimate the following equation to derive a causal effect of the penalty increase on reoffending:
\begin{equation}
y_{i} = \alpha + \beta \, policy\_change_i + \gamma_1 \, day_i + \gamma_2 \, day_i \times policy\_change_i + \delta X_i + \varepsilon_i,
\label{eq:rdit}
\end{equation}
where $y_{i}$ denotes the outcome of interest: reoffending, the placebo outcome, or the transfer indicator. $policy\_change_i$ is an indicator for whether the infringement occurred after the reform, and $day_i$ measures the number of days between the infringement date and the reform date. The coefficient $\beta$ is our parameter of interest and captures the effect of the penalty increase on the outcome.

As in Eq. (\ref{eq:dynamic}), the vector of controls $X_i$ accounts for systematic variation in traffic conditions, enforcement, and driver characteristics.\footnote{Controls include day-of-week fixed effects and day indicator, speed camera indicator, school and public holidays indicators, and driver's age.} 

\section{Results}

\FloatBarrier
\subsection{Descriptive statistics}

Table \ref{tab:desc_table} reports descriptive statistics and balance checks for observations within a 150-day window on either side of the policy change, separately for fixed and mobile speed cameras. The table compares mean characteristics of offenses occurring before and after the reform and presents RDiT estimates to assess the continuity of covariates at the cutoff.

\begin{table}[ht]
\centering

\begin{tabular}{p{3cm}ccp{2cm}ccp{2cm}}
\hline
& \multicolumn{3}{c}{Fixed} & \multicolumn{3}{c}{Mobile} \\
\cmidrule(lr){2-4}\cmidrule(lr){5-7}
Variable & Mean & Mean & RDiT & Mean & Mean & RDiT \\ 
& 150 days & 150 days & & 150 days & 150 days & \\
& before & after & & before & after & \\
\hline
Speeding before & 0.216 & 0.194 & -0.001 \newline (0.004) & 0.266 & 0.248 & -0.009** \newline (0.003) \\
Age of offenders & 43.68 & 44.11 & -0.040 \newline (0.168) & 43.34 & 43.84 & 0.222* \newline (0.101) \\
Demerit points acc. & 2.301 & 2.187 & -0.038 \newline (0.030) & 2.019 & 1.904 & -0.063*** \newline (0.016) \\
Speed & 83.49 & 80.49 & 0.023 \newline (0.037) & 73.57 & 72.94 & 0.028 \newline (0.028) \\
Days from last off.$^\ddagger$ & 631 & 680 & 8.066 \newline (8.723) & 709 & 758 & 13.00* \newline (5.364) \\
Transfer & 0.020 & 0.020 & 0.000 \newline (0.002) & 0.016 & 0.017 & 0.001 \newline (0.001) \\
\bottomrule
\multicolumn{7}{p{15cm}}{\footnotesize \emph{Notes:} $^\dagger$ p$<$0.10, * p$<$0.05, ** p$<$0.01, *** p$<$0.001. $^\ddagger$ Days from last offense is measured on a sub-sample of drivers who offended before in our dataset; total observations are 60,174 and 55,549.} \\
\end{tabular}
\caption{Descriptive statistics and balance check.}
\label{tab:desc_table}
\end{table}

The pre-reform means indicate that fixed camera offenders travel at higher recorded speeds, and have accumulated more demerit points than mobile camera offenders. Transfer rates are low and stable across both subsamples, at around 1.6--2.0\% of offenses in the pre-period.

The RDiT estimates show that for fixed speed cameras, none of the estimated discontinuities are statistically significant, and point estimates are close to zero across all characteristics, providing clean evidence of covariate balance. For mobile cameras, several covariates show statistically significant discontinuities at the cutoff: the share of offenders with a prior speeding offense falls by 0.9 percentage points, accumulated demerit points decline, and average age increases slightly. The discontinuity in days since the last offense is also significant for mobile cameras, suggesting some differences in the composition of offenders around the threshold in this subsample.

For mobile-camera offenses, the significant discontinuities in predetermined covariates indicate that the RDiT design is not valid for studying recidivism. By contrast, for fixed speed cameras, the absence of significant discontinuities in predetermined characteristics supports the validity of the design for identifying the causal effect of the penalty change on recidivism.

\FloatBarrier
\subsection{Blame-shifting}
\label{sec:blame-shifting}

We begin by examining the effect of the policy on the probability that a fine is transferred. We conduct this analysis separately for two sub-groups of offenses: infringements caught by mobile speed cameras and infringements caught by fixed speed cameras. We hypothesise that there are heterogeneous effects depending on the type of camera used to detect offenses. Mobile cameras offer less precision in identifying the driver with photographic evidence than fixed cameras. 

\begin{figure}[htbp]
    \centering
     \begin{subfigure}[b]{0.48\textwidth}
        \centering
        \caption{Mobile: 11 and 12 km/h speed excess}
        \includegraphics[width=\textwidth]{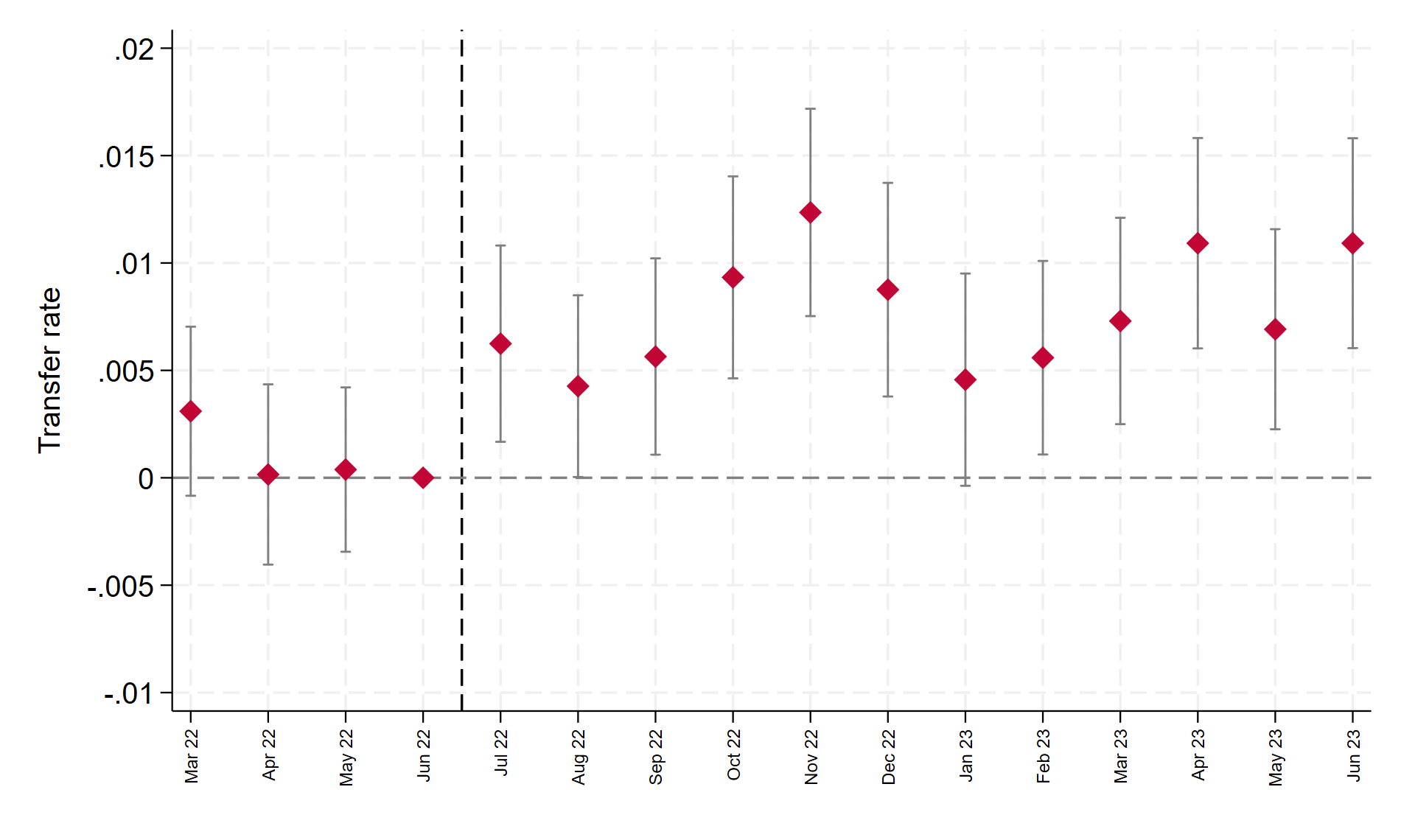}
        \label{fig:dynamic_transfer_mobile_11_12}
    \end{subfigure}
    \hfill
    \begin{subfigure}[b]{0.48\textwidth}
        \centering
        \caption{Mobile: Other speed excess}
        \includegraphics[width=\textwidth]{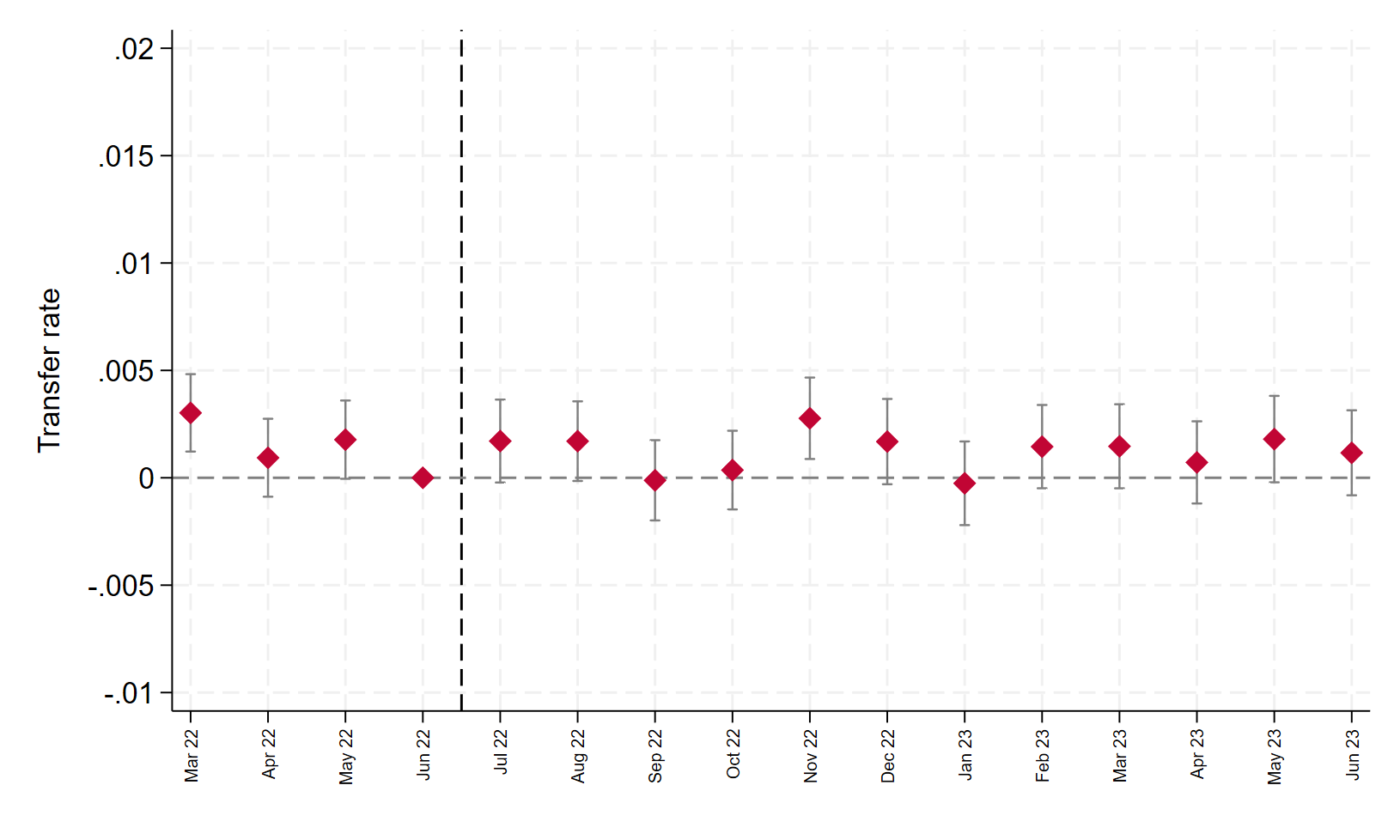}
        \label{fig:dynamic_transfer_mobile_not_11_12}
    \end{subfigure}
    \begin{subfigure}[b]{0.48\textwidth}
        \centering
        \caption{Fixed: 11 and 12 km/h speed excess}
        \includegraphics[width=\textwidth]{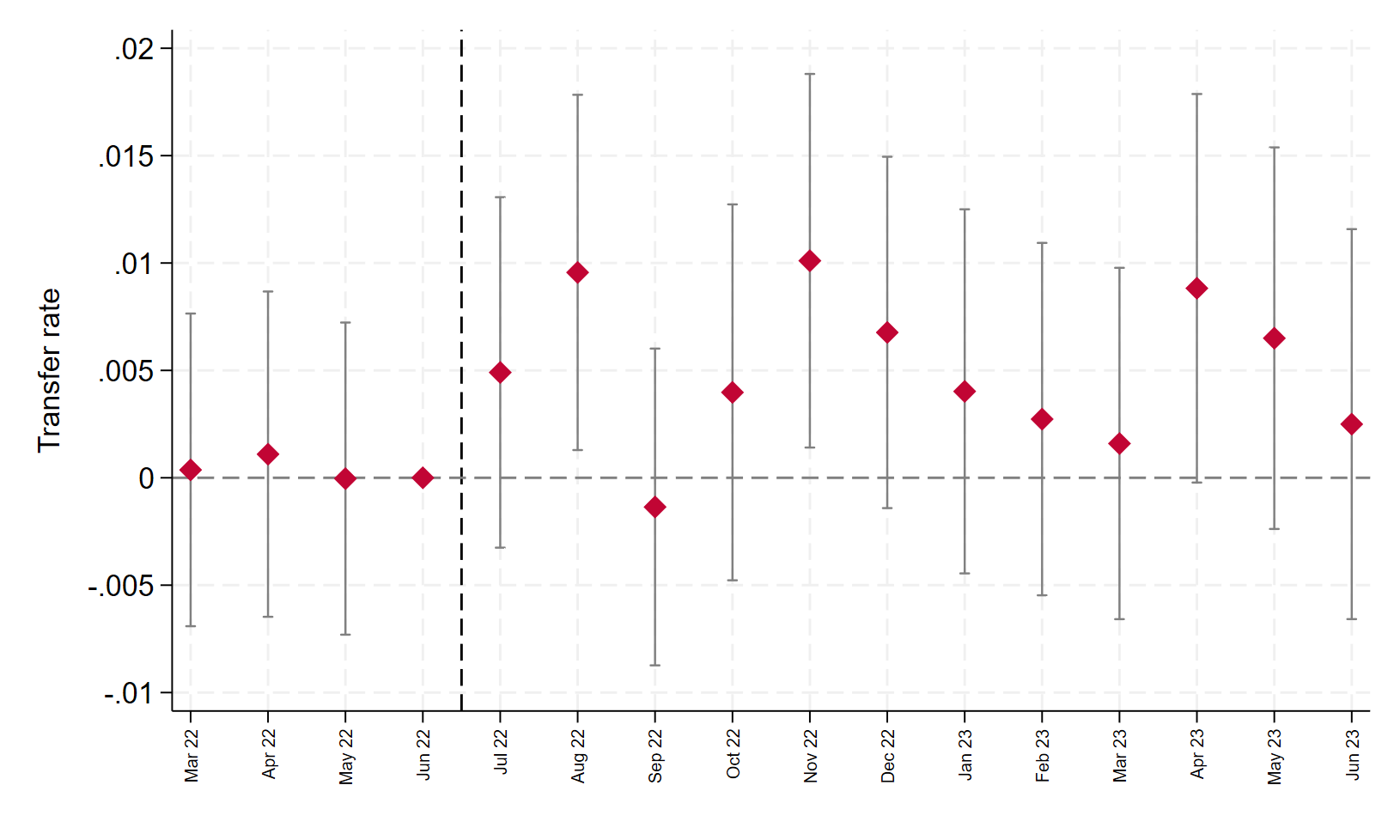}
        \label{fig:dynamic_transfer_fixed_11_12}
    \end{subfigure}
    \hfill
    \begin{subfigure}[b]{0.48\textwidth}
        \centering
        \caption{Fixed: Other speed excess}
        \includegraphics[width=\textwidth]{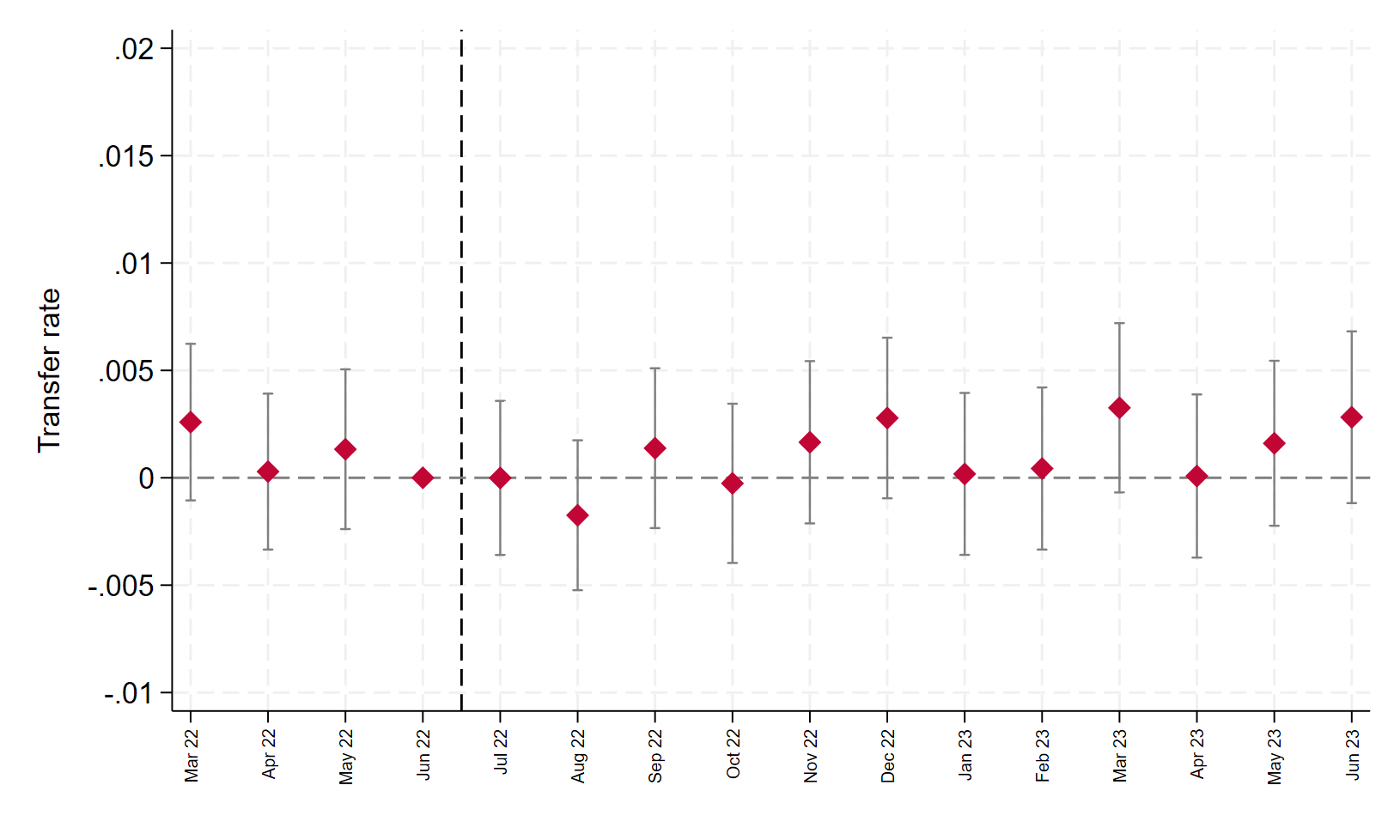}
        \label{fig:dynamic_transfer_fixed_not_11_12}
    \end{subfigure}
   
    \caption{Dynamic effects of the July 2022 penalty reform on transfer rate for mobile speed cameras (top) and fixed cameras (bottom) for speed excess at 11--12 km/h (left) and for other speeds (right).}
    \label{fig:dynamic_transfer}
\end{figure}

Figure \ref{fig:dynamic_transfer} shows event studies of transfer rates before and after the change of rule. The figure presents results for four subsamples: transfer rates, separately for mobile and fixed speed cameras and for the 11--12 km/h speed excess tier and all other tiers. Each point represents the average transfer rate in a 30-day period relative to the baseline period (the 30 days immediately preceding the reform), estimated from Eq. (\ref{eq:dynamic}).

Panel (a) shows a clear and persistent increase in transfer rates following the July 2022 reform for mobile speed cameras at the 11--12 km/h speed excess tier, where both the fine amount and demerit point allocation increased. The pre-reform coefficients are small and statistically insignificant, providing no evidence of pre-existing trends in transfer behavior prior to the reform and supporting the validity of the identification strategy. The post-reform coefficients are consistently positive and range between around 0.005 and 0.01, implying a relative effect of 30-60\%. In Panel (b), the post-reform coefficients are smaller and not statistically significant, suggesting that the increase in monetary fines alone was insufficient to generate a comparable nomination response.

Panel (c) and (d) show the corresponding results for fixed cameras. In contrast to the mobile camera results, there is no evidence of a systematic increase in transfer rates following the reform for either speed excess tier. While some individual post-reform coefficients are positive in panel (c),  the estimates are noisy and mostly statistically indistinguishable from zero throughout the post-reform window. The absence of a transfer response for fixed cameras is an important feature of our setting---it confirms that the reoffending estimates based on fixed camera data are unbiased by strategic nomination behavior. 

\begin{figure}[h!]
  \centering
   \begin{subfigure}[b]{0.48\textwidth}
        \centering
        \caption{Fixed}
        \includegraphics[width=\textwidth]{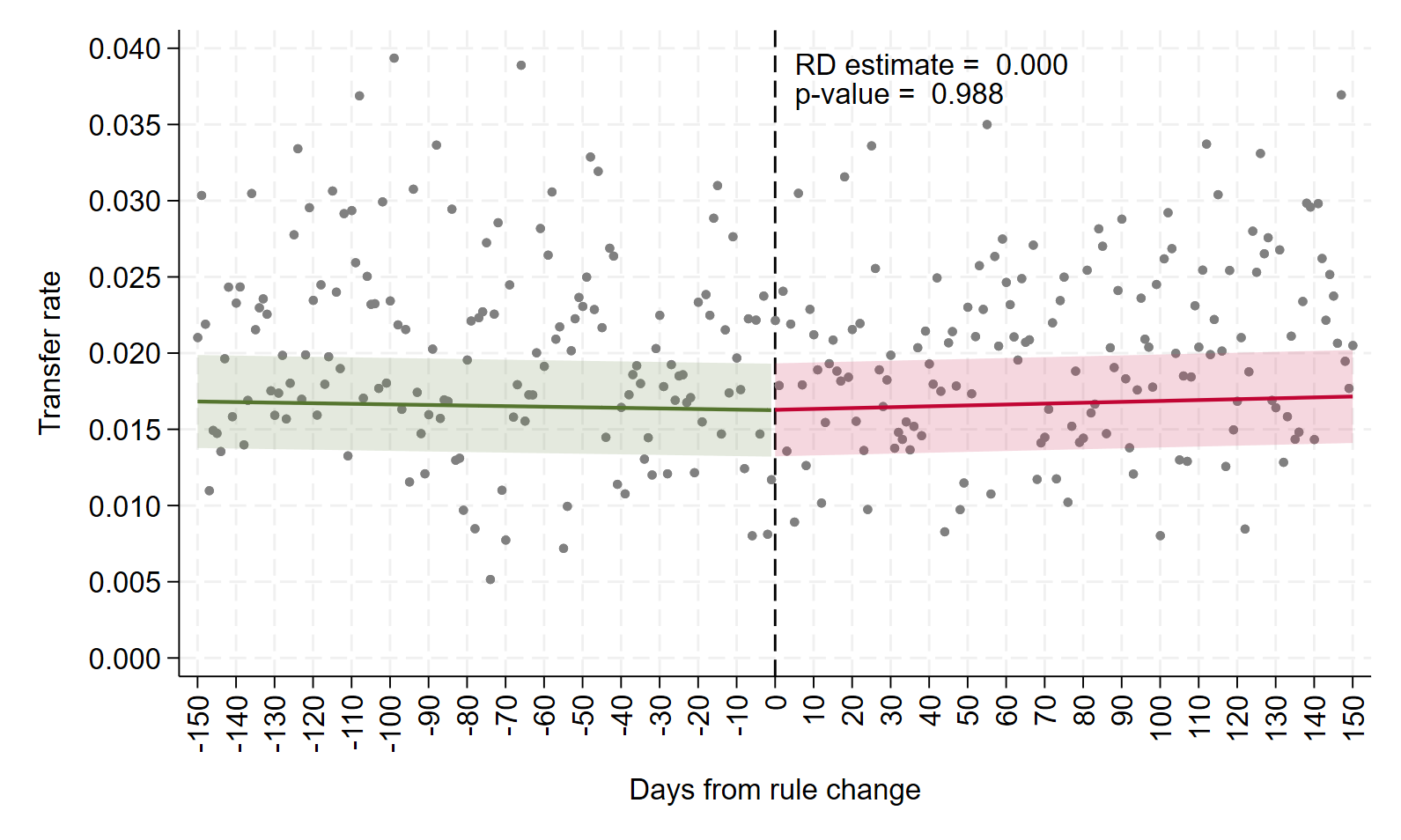}
        \label{fig:rdit_transfer_fixed_2022}
    \end{subfigure}
    \begin{subfigure}[b]{0.48\textwidth}
        \centering
        \caption{Mobile}
        \includegraphics[width=\textwidth]{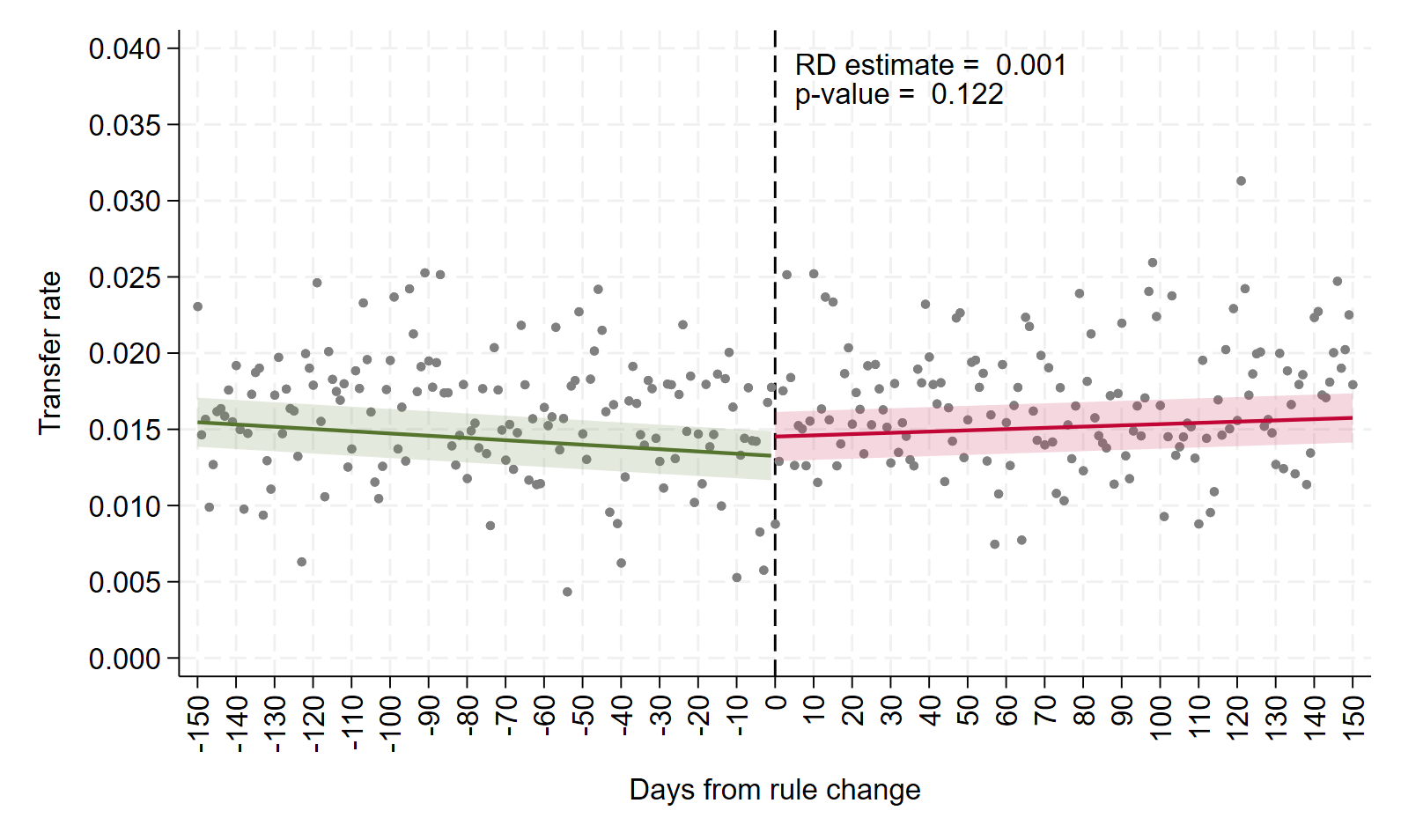}
        \label{fig:rdit_transfer_mobile_2022}
    \end{subfigure}
  \caption{Mean transfer rate around the July 2022 reform for fines detected by fixed speed cameras (a) and mobile speed cameras (b).}
  \label{fig:transfer_mobile_fixed}
\end{figure}

Subsequently, we apply our main methodological approach, described in Section \ref{sec:empirical_strategy}, to transfer rates. Figure \ref{fig:transfer_mobile_fixed} plots the mean daily transfer rate around the July 2022 reform, separately for fixed cameras  (panel a) and mobile cameras (panel b). A positive discontinuity is visible at the cutoff for mobile cameras, with the transfer rate rising after the penalty change, while no such effect is present for fixed cameras.

\begin{figure}[h!]
  \centering
   \begin{subfigure}[b]{0.48\textwidth}
        \centering
        \caption{Mobile: 11 and 12 km/h speed excess}
        \includegraphics[width=\textwidth]{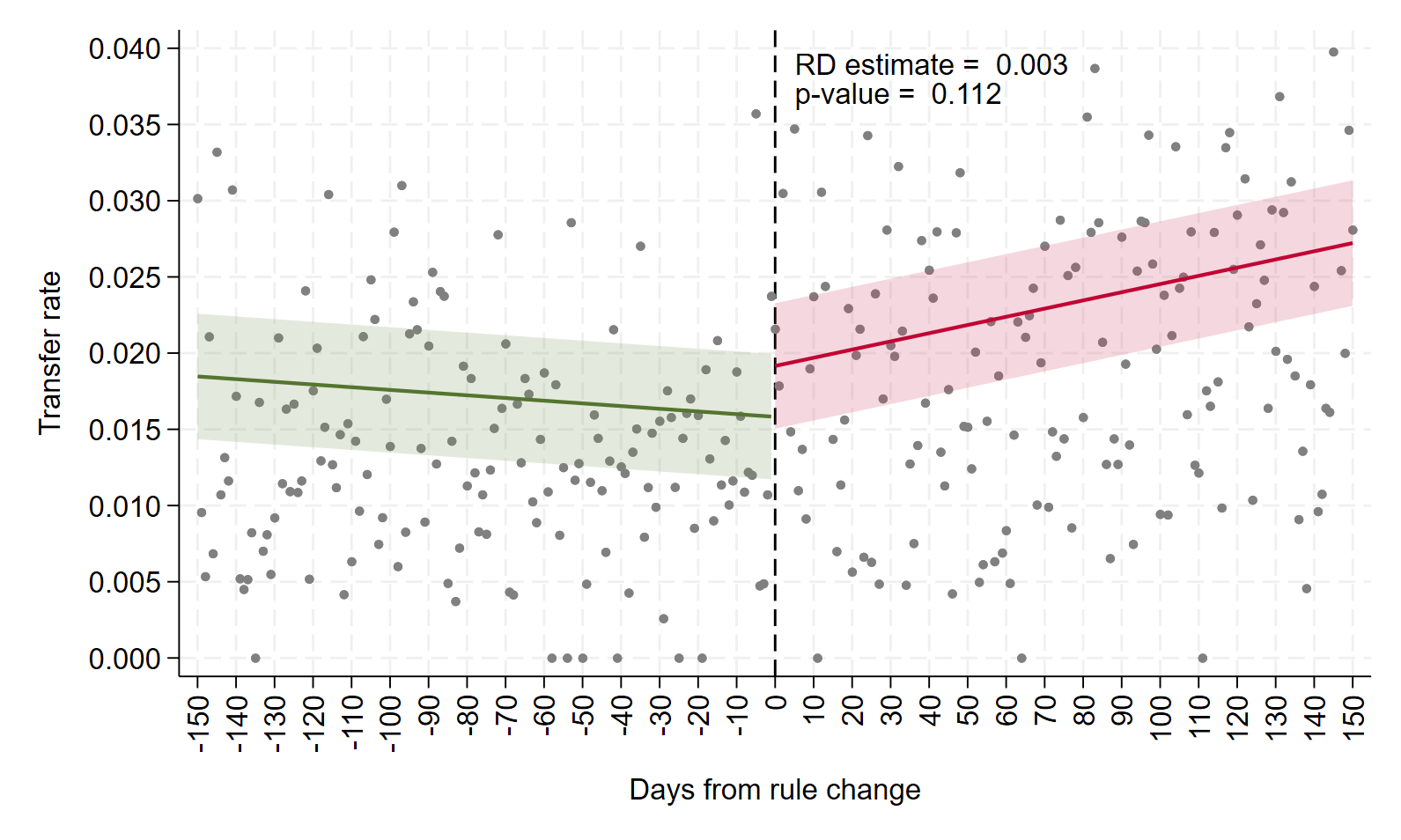}
        \label{fig:rdit_transfer_mobile_11_12_2022}
    \end{subfigure}
    \begin{subfigure}[b]{0.48\textwidth}
        \centering
        \caption{Mobile: Other speed excess}
        \includegraphics[width=\textwidth]{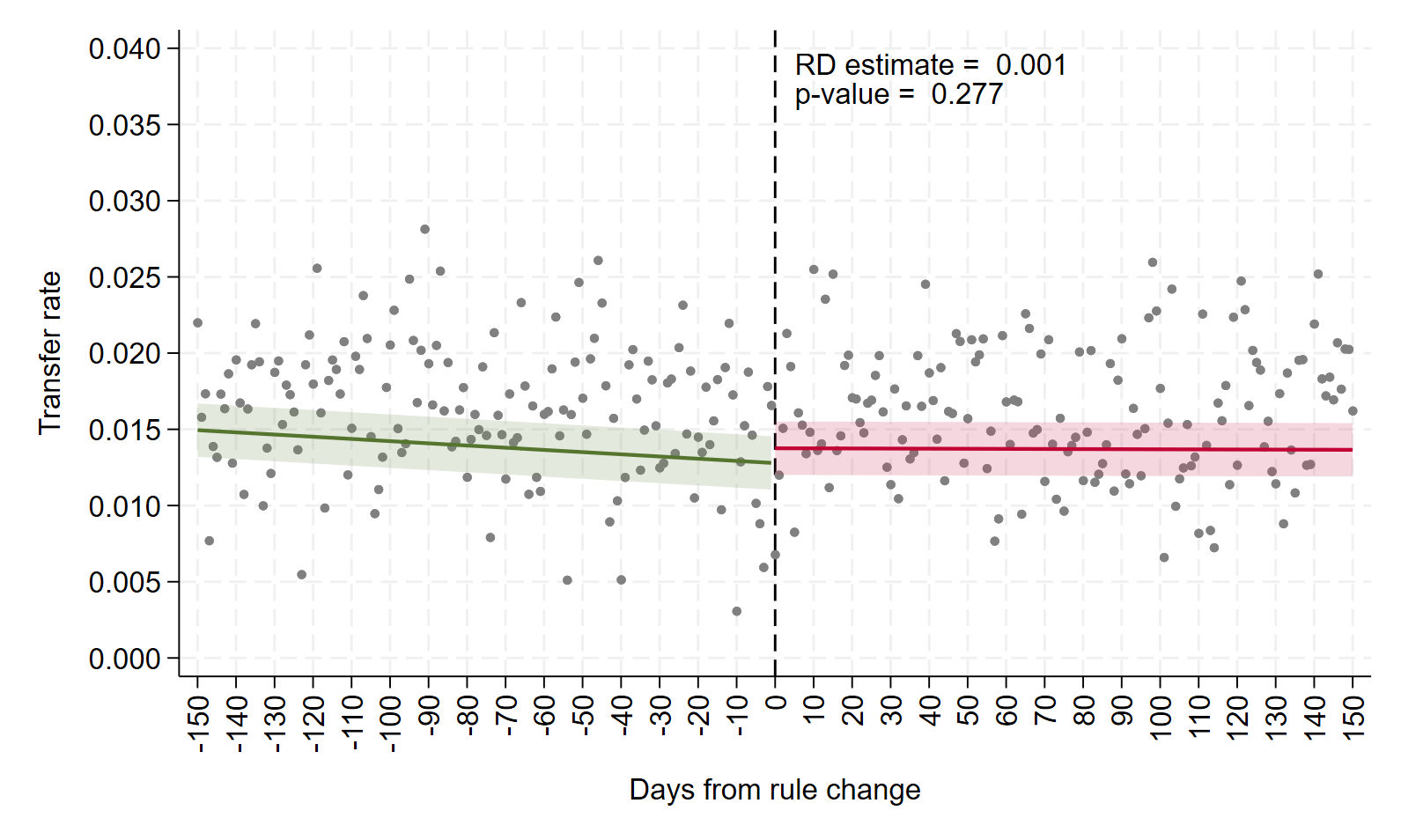}
        \label{fig:rdit_transfer_mobile_not_11_12_2022}
    \end{subfigure}
\caption{Mean transfer rate around the July 2022 reform for mobile-camera offenses involving speed excesses of 11--12 km/h (a) and all other speed excesses (b).}
  \label{fig:transfer_mobile_11_12}
\end{figure}

Figure \ref{fig:transfer_mobile_11_12} further decomposes the mobile camera results by speed tier, restricting to 11--12 km/h offenders (panel a) and all other speeds (panel b). The effect is concentrated among 11--12 km/h offenders, where the reform simultaneously increased the monetary fine and added demerit points, producing the largest combined increase in sanctions. 

This is consistent with blame-shifting behavior: faced with substantially higher penalties, registered owners became more likely to reassign liability to another driver rather than absorb the penalty themselves. Together, these patterns suggest that a meaningful share of the behavioral response to the reform operated through the transfer margin rather than through genuine adjustment of driving behavior.

While the RD estimate in Figure \ref{fig:rdit_transfer_mobile_11_12_2022} does not reach conventional levels of statistical significance at our chosen bandwidth (p = 0.112), Figure \ref{fig:bandwidth_check_speeds} (a) demonstrates that the result is sensitive to bandwidth selection. Across nearly all bandwidths above approximately 50 days, the estimated effect is positive and stable, with the confidence interval lying at or above zero and statistical significance achieved over a wide range of bandwidths. We therefore interpret the evidence as consistent with a positive effect of the reform on transfer rates among mobile camera offenders at 11--12 km/h. The lack of significance at our preferred bandwidth reflects estimation noise arising from the local linear approximation rather than an absence of a true effect.

One possible explanation for these relatively modest discontinuity estimates is that transfer behavior continued to increase after the reform, as suggested by the upward trend after the reform in Figure \ref{fig:rdit_transfer_mobile_11_12_2022}. If drivers gradually learned about the new penalties and the possibility of avoiding demerit points through nomination, the immediate discontinuity at the cutoff would understate the full behavioral response. Consistent with this interpretation, the event-study estimates in Figure \ref{fig:dynamic_transfer_mobile_11_12} reveal a larger and more persistent increase in transfer rates over time than is captured by the local RDiT estimates.

\subsubsection{Who are the fines transferred to?}

Our data consist of records of final fine payers---that is, the individual ultimately responsible for paying the fine after any transfer has taken place---along with an indicator for whether each fine was transferred. A driver identifier allows us to follow individuals over time. However, the data does not include a vehicle identifier, nor does it allow to distinguish between truthful transfers---where the fine is reassigned to the actual offender---and untruthful transfers, where it is reassigned to a third party in order to avoid demerit points.

We compare the individuals receiving the transfer to the broader population of drivers before and after the rule change. Table \ref{tab:char_transfers} shows the results. Both before and after the reform, drivers who received a transfer are slightly younger than those who did not, with a mean age of approximately 41 years compared to 43-44 years, and the difference is statistically significant in both periods. The sum of demerit points is marginally lower for transfer recipients, though the difference is small in magnitude and only weakly significant in the post-period. Nominated drivers also tend to travel at slightly higher speeds. For the initial offenses, there is a weakly significant difference in the post-period, showing a higher share among nominated drivers. 

\begin{table}[htbp]\centering
\def\sym#1{\ifmmode^{#1}\else\(^{#1}\)\fi}

\begin{tabular}{p{3cm}p{1.75cm}p{1.75cm}p{1.75cm}p{1.75cm}p{1.75cm}p{1.75cm}}
\toprule
 & \multicolumn{3}{c}{Pre Reform} & \multicolumn{3}{c}{Post Reform} \\
 \cmidrule(lr){2-4} \cmidrule(lr){5-7}
 & Mean (No Transfer) & Mean (Transfer) & Diff
 & Mean (No Transfer) & Mean (Transfer) & Diff \\
\midrule
Age at offense & 42.85 & 40.79 & -2.059\sym{**} & 43.55 & 41.37 & -2.178\sym{***} \\
 & & & (0.688) & & & (0.638) \\
Sum of   & 2.15 & 2.08 & -0.07 & 1.92 & 1.75 & -0.168$^\dagger$ \\
demerit points & & & (0.114) & & & (0.099) \\
Speed & 79.02 & 79.17 & 0.154 & 78.16 & 78.73 & 0.565 \\
 & & & (0.769) & & & (0.692) \\
Initial offense & 0.72 & 0.71 & -0.014 & 0.75 & 0.78 & 0.028$^\dagger$ \\
 & & & (0.020) & & & (0.017) \\
\midrule
N & 37,164 & 517 & & 30,047 & 621 & \\
\bottomrule
\multicolumn{7}{l}{\footnotesize Standard errors from two-sample t-test in parentheses.}\\
\multicolumn{7}{l}{\footnotesize $^\dagger$ p$<$0.10, * p$<$0.05, ** p$<$0.01, *** p$<$0.001}
\end{tabular}
\caption{Driver characteristics by transfer status of the fines caught by
mobile speed cameras 150 days before and after the rule change.}
\label{tab:char_transfers}
\end{table}

\FloatBarrier
\subsubsection{Threat to identification\label{sec:manipulationRD}}

The policy increases the prevalence of fine transfers, which threatens our identification strategy of reoffending. We estimate recidivism using the regression discontinuity design in equation (\ref{eq:rdit}), where $i$ denotes the recorded offender and $day_i$ is the running variable.

The RD literature has focused on the manipulation of the running variable. Examples in which the running variable is manipulated abound. For example, teachers inflate grades to push students above cutoffs \citep{Diamond2016,Dee2019}, police officers report speed excess below fines' threshold \citep{Goncalves2021,gonccalves2023police}, close union certification elections are manipulated \citep{Frandsen2017},  medical test results are adjusted to allow blood donations \citep{rosenman2027donor} and school administrators  strategically adjust reported enrollment around class size rule to increase their budget \citep{Angrist2019}. Ignoring such behavior can bias standard regression discontinuity estimates \citep{McCrary2008,Frandsen2017}. Recent work addresses this concern using partial identification approaches that remain valid under manipulation \citep{Gerard2020,rosenman2027donor}.

However, the concern here is not manipulation of the running variable but misclassifications of the observational unit. When a fine is transferred, an offender $j$ shifts the sanction to offender $i$, so that we observe $y_i$ while the relevant outcome is $y_j$. This generates measurement error in the dependent variable that is systematically affected by the policy.

The problem arises for both untruthful and truthful transfers. Even when transfers are legitimate, we observe a mixture of correctly and incorrectly assigned offenders, depending on whether transfers occur. Because the policy changes the likelihood of transfer, the composition of observed offenders varies discontinuously at the cutoff, introducing selection bias.

Importantly, this issue persists whether offenders are indexed by the fine recipient $i$ or by the registered vehicle $j$, as both are affected by endogenous transfer behavior. Indexing observations at the vehicle level does not resolve the problem when the outcome of interest is person-level recidivism: a vehicle may be driven by multiple individuals, so subsequent violations need not be committed by the original offender. Because the policy affects transfer behavior, it alters the mapping between the observed unit and the underlying offender, generating selection bias in recidivism estimates.

To the best of our knowledge, this form of unit misclassification induced by policy-driven reassignment has not been previously highlighted in the RD literature. Related studies using data from the Czech Republic and Finland \citep{dusek2021,Kaila2024} are unlikely to be affected, as these settings do not feature demerit point systems. However, in Finland, where fines are income-based, higher-income offenders may have incentives to transfer fines to lower-income individuals.

\subsubsection{Discontinuity in Covariates}

Table~\ref{tab:desc_table} reports descriptive statistics and balance checks for predetermined characteristics around the reform date, separately for fixed and mobile camera detections. For fixed cameras, none of the covariates exhibit a statistically significant discontinuity at the cutoff, supporting a causal interpretation of our main estimates. For mobile cameras, several predetermined characteristics show significant discontinuities at the threshold. These discontinuities are driven in part by the increase in fine transfers following the reform: as more registered owners nominate another driver, the composition of recorded offenders changes mechanically. However, as discussed in Section~\ref{sec:gendetterenceandenforcement_mobile}, the remaining differences may also reflect changes in mobile-camera enforcement or greater avoidance of mobile cameras by drivers after the reform. Given these compositional changes, we do not use the mobile-camera sample to estimate specific deterrence. Instead, we rely on fixed-camera detections for our main recidivism estimates and use mobile-camera detections to study nomination behavior.

\FloatBarrier
\subsection{Recidivism}
\label{sec:results}

Given that the policy raised the number of transfers for mobile cameras and not fixed cameras, we focus on fixed cameras for our estimates of recidivism.

\begin{figure}[h!]
  \centering
   \begin{subfigure}[b]{0.48\textwidth}
        \centering
        \caption{Fixed: 11 and 12 km/h speed excess}
  \includegraphics[width=\textwidth]{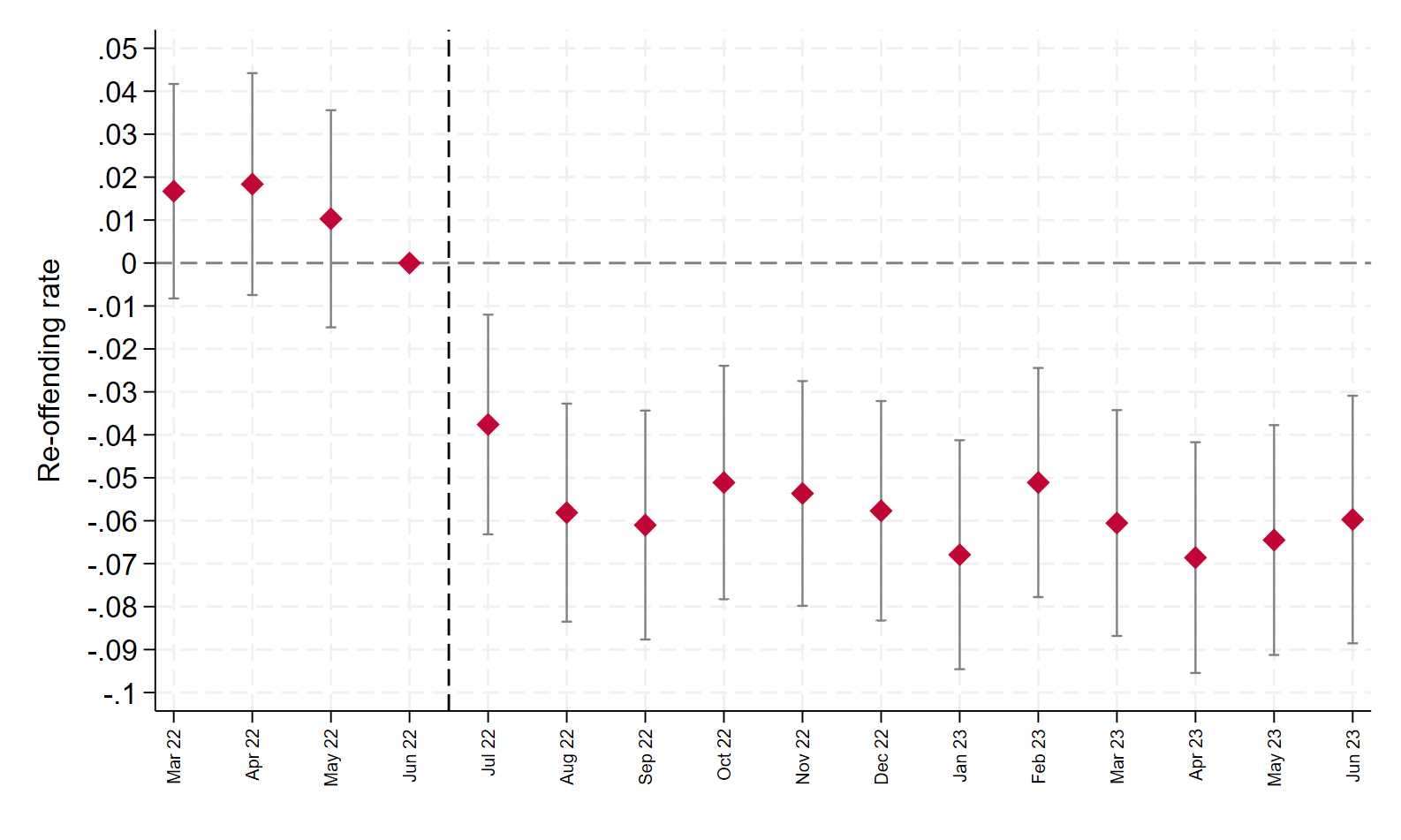}
      \end{subfigure}
    \begin{subfigure}[b]{0.48\textwidth}
        \centering
        \caption{Fixed: Other speed excess}
  \includegraphics[width=\textwidth]{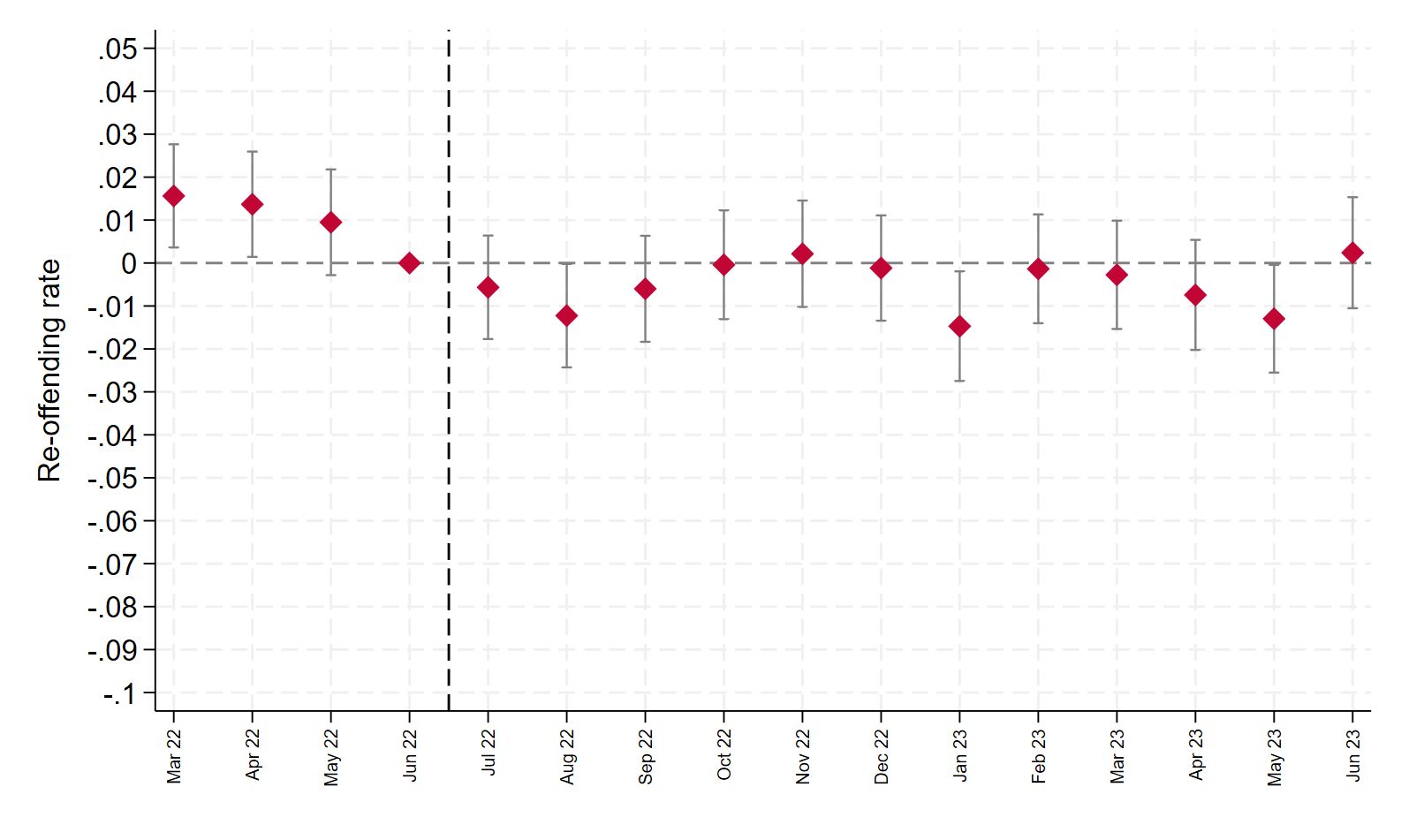}
  \end{subfigure}
 \caption{Dynamic effects of the July 2022 penalty reform on reoffending within one year of the initial offense, for fixed-camera offenses involving speed excesses of 11--12 km/h (a) and all other speed excesses (b).}
  \label{fig:dynamic_fixed}
\end{figure}

Figure \ref{fig:dynamic_fixed} presents event study estimates for reoffending separately by speed excess groups.\footnote{We use a specification for reoffending as described in Eq. (\ref{eq:dynamic}).} For drivers caught at 11--12 km/h above the speed limit (panel a), the coefficients are close to zero and statistically indistinguishable from the baseline in the pre-reform periods, consistent with the absence of pre-existing trends. Following the penalty increase, reoffending falls persistently and significantly, with the effect remaining stable across subsequent months. For other speed excess tiers (panel b), where only the monetary fine increased and demerit point allocations were unchanged, the estimated coefficients are small and largely insignificant throughout, with no discernible pattern. 

Table \ref{tab:main_results} presents the results from the RDiT analysis. Each of the five models reports estimates for our main outcome of interest---reoffending within one year following the offense (first row)---as well as for a placebo outcome measuring speeding offenses in the year prior to the focal offense (“offending before”, second row). The first column reports results without controls, while the second includes controls. The third column excludes transferred fines, and the fourth and fifth columns report estimates separately for offenses involving speed excesses of 11–12 km/h and all other speed excesses, respectively. All specifications except the first include controls.

For reoffending, the overall results for models with and without controls (first two columns of the first row)  are similar---showing an effect of 1.4 percentage point decrease and 1.6 percentage point decrease, respectively. The effects differ by speed excess and are substantially larger for offenses involving an excess speed of 11 or 12 km/h (fourth column). For these offenses, reoffending decreases by 4.6 percentage points, corresponding to a 13\% reduction relative to the pre-reform reoffending rate of 34.8\%. We find no effects of the rule change on reoffending at speed excess different than 11 or 12 km/h (last column).\footnote{The estimation sample is restricted to offenses detected by fixed cameras, ensuring that the initial offense generating variation around the cutoff cannot be strategically transferred. The recidivism outcome, however, is constructed using all subsequent offenses, including those detected by mobile cameras. Restricting subsequent offenses to fixed-camera detections yields similar results, ruling out bias from strategic liability transfer; see Table \ref{tab:main_results_fixed} in the Appendix.}

The second row reports placebo estimates using an indicator for whether the driver committed an offense in the year preceding the current offense. The coefficients are close to zero in all specifications, providing reassurance that the discontinuity is not driven by pre-existing differences across drivers.

The last row, show the effect of the reform on transfer. Recall, that in this table the estimation is done with the fixed camera sample. In contrast to the mobile cameras sample (see Section \ref{sec:blame-shifting}) all estimates are precisely estimated at 0.

\begin{table}[htbp]
\centering
\resizebox{\textwidth}{!}{%
\begin{tabular}{>{\raggedright\arraybackslash}p{3cm} *{5}{>{\centering\arraybackslash}p{1.8cm}}}
\toprule
 & \multicolumn{3}{c}{Overall} 
 & \multicolumn{2}{c}{Speed excess: 11--12 km/h} \\
\cmidrule(lr){2-4}\cmidrule(lr){5-6}
 &  &  & Exclude transfers & Only & Excluded \\
\midrule
Reoffending  & \mbox{-0.014**} & \mbox{-0.016**} & \mbox{-0.016**} & \mbox{-0.046***}  & -0.009  \\
after & (0.005) & (0.005) & (0.005) & (0.012) & (0.006) \\
\midrule
Offending before  & -0.006 & -0.001 & -0.001 & -0.010  & 0.001  \\ 
(placebo) & (0.004) & (0.004) & (0.004) & (0.010) & (0.005) \\
\midrule
Transfer  & -0.001 & 0.000 & - & 0.004  & -0.001  \\ 
 & (0.001) & (0.002) & - & (0.004) & (0.002) \\
\midrule
N & 149,755 & 149,755 & 146,752 & 27,077 & 122,678 \\
\midrule 
Controls &  & \checkmark & \checkmark & \checkmark & \checkmark \\
\bottomrule
\end{tabular}
}
\caption{Effect of policy change on reoffending in the next year, placebo, and transfer rates. Standard errors clustered at driver level in parentheses. $^\dagger$ p$<$0.10, * p$<$0.05, ** p$<$0.01, *** p$<$0.001
\label{tab:main_results}}
\end{table}

\begin{figure}[h!]
  \centering
  \includegraphics[width=0.7\linewidth]{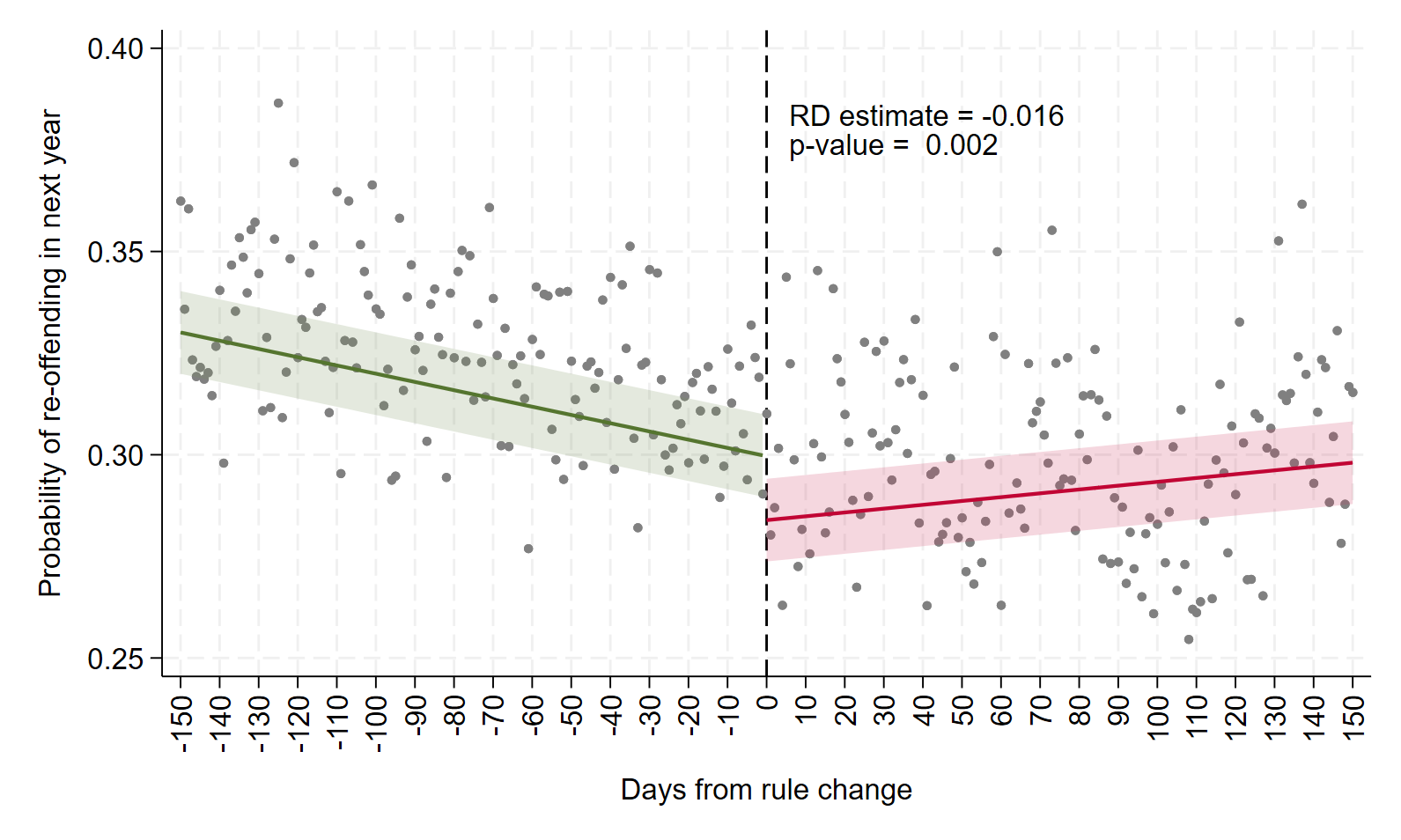}
  \caption{Effects of the increase in penalty on reoffending (RDiT).}
  \label{fig:pre_post_trends}
\end{figure}

Figure \ref{fig:pre_post_trends} shows a visual representation of the effect, with plotted daily mean reoffending and predicted fit lines for pre- and post-trends. The difference between both lines at 0 represents the overall effect of the rule change. 

\FloatBarrier
\subsection{Effect evolution over time}

\begin{figure}[h!]
  \centering
   \begin{subfigure}[b]{0.48\textwidth}
  \centering
  \caption{Fixed: 11 and 12 km/h speed excess}
  \includegraphics[width=\textwidth]{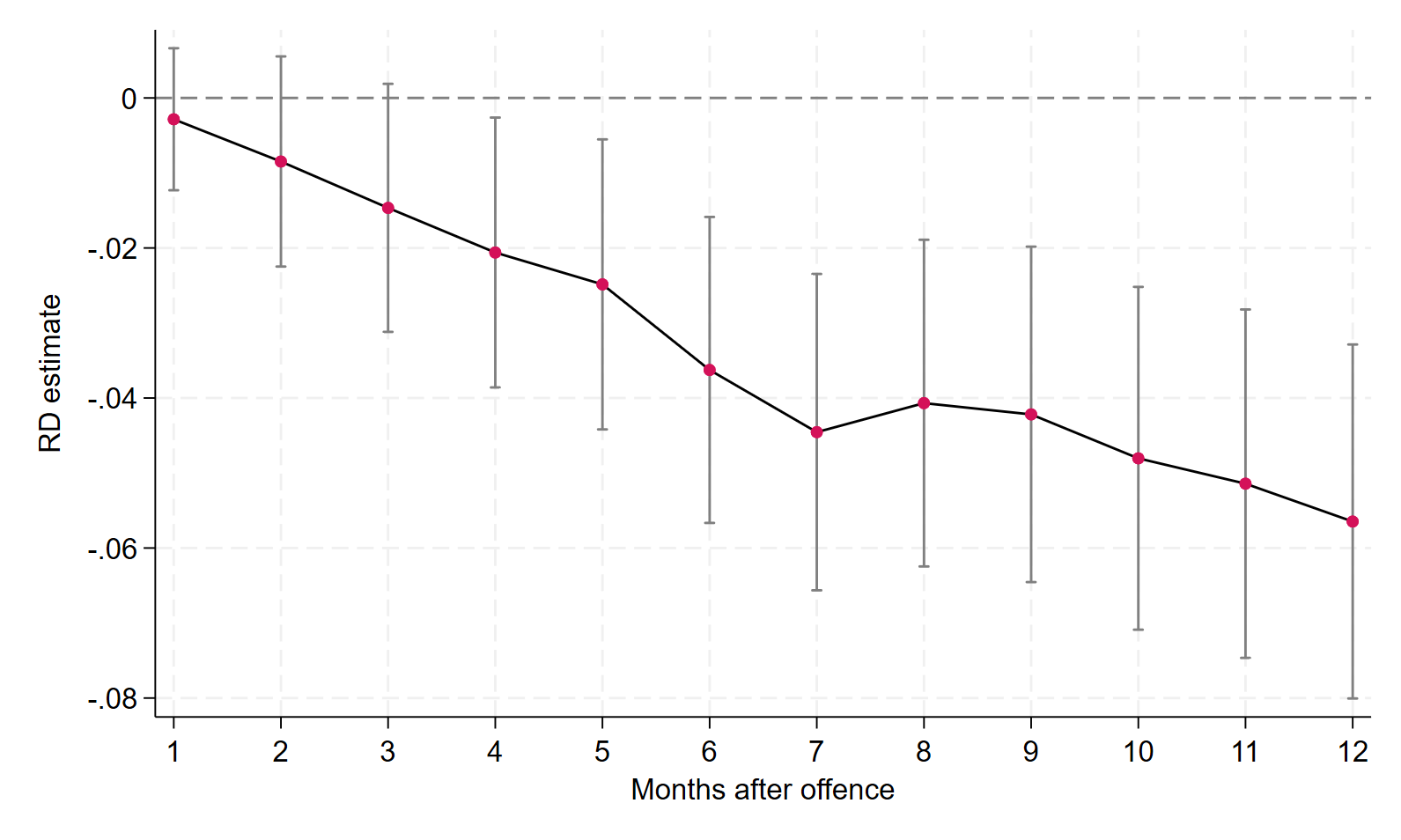}
  \end{subfigure}
    \begin{subfigure}[b]{0.48\textwidth}
        \centering
        \caption{Fixed: Other speed excess}
  \includegraphics[width=\textwidth]{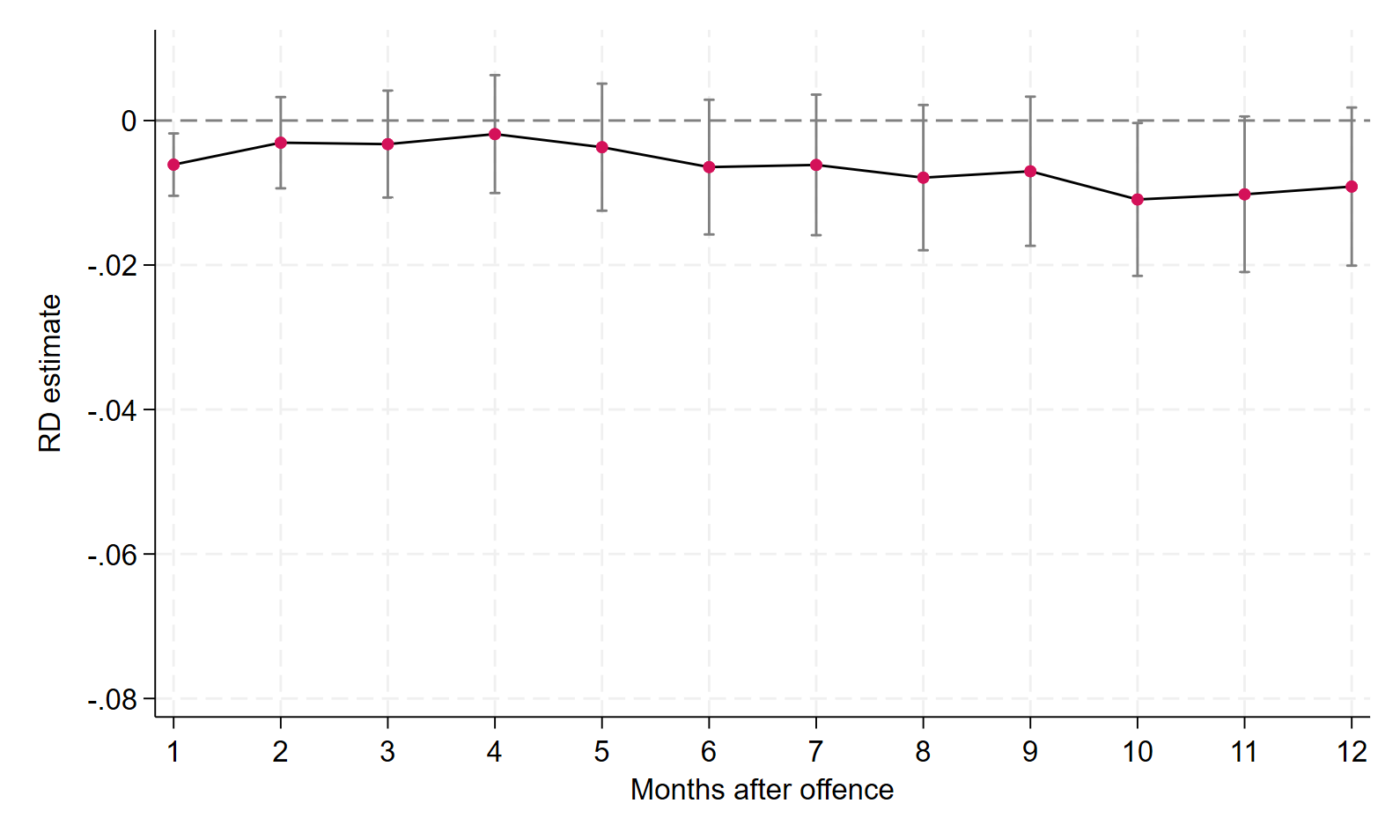}
  \end{subfigure}
  \caption{Effects of the penalty increase on reoffending at horizons ranging from 1 to 12 months after the initial offense. Panel (a) shows drivers who exceeded the speed limit by 11 or 12 km/h, and panel (b) shows all other drivers.}
  \label{fig:reoffending_by_outcome_range}
\end{figure}

Figure \ref{fig:reoffending_by_outcome_range} presents the results for our main outcome of interest, reoffending, measured over varying periods following the initial offense, from 1 to 12 months.  Panel (a) shows reoffending among drivers who exceeded the speed limit by 11 or 12 km/h, while panel (b) shows the results for all other drivers.

Consistent with our main results, for offenses involving speed excesses of 11–12 km/h (panel a), we observe a negative effect on reoffending. The effect becomes larger as the time horizon used to define the outcome variable increases. In contrast, for drivers detected at other speed excess categories (panel b), the effect is substantially smaller and remains statistically insignificant regardless of the time horizon considered. 

\FloatBarrier
\subsection{Heterogeneity}

\begin{figure}[h!]
\centering
\begin{subfigure}{0.49\linewidth}
    \centering
    \caption{Transfer rates (11 and 12 km/h speed excess)}
    \includegraphics[width=\linewidth]{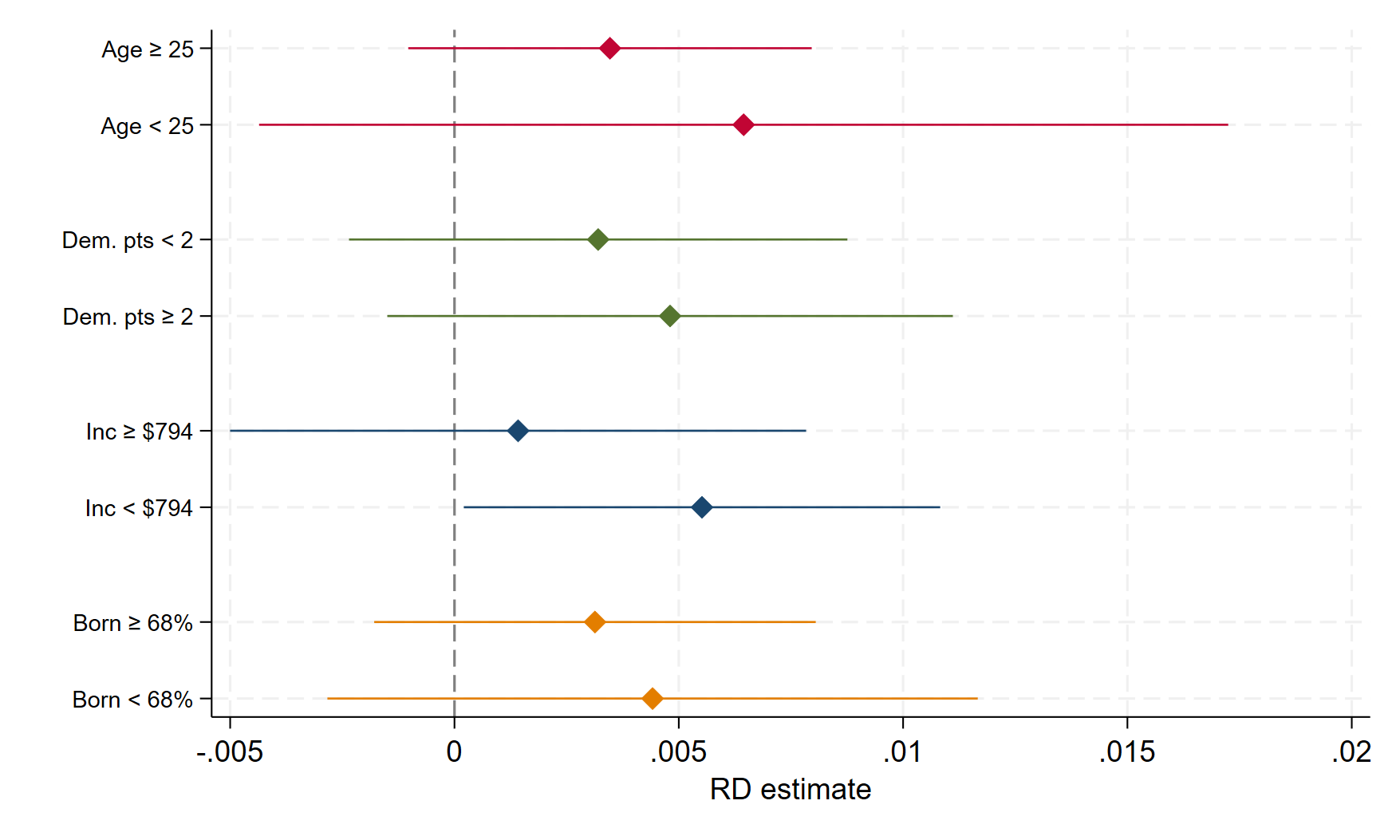}
\end{subfigure}
\hfill
\begin{subfigure}{0.49\linewidth}
    \centering
    \caption{Transfer rates (other speeds)}
    \includegraphics[width=\linewidth]{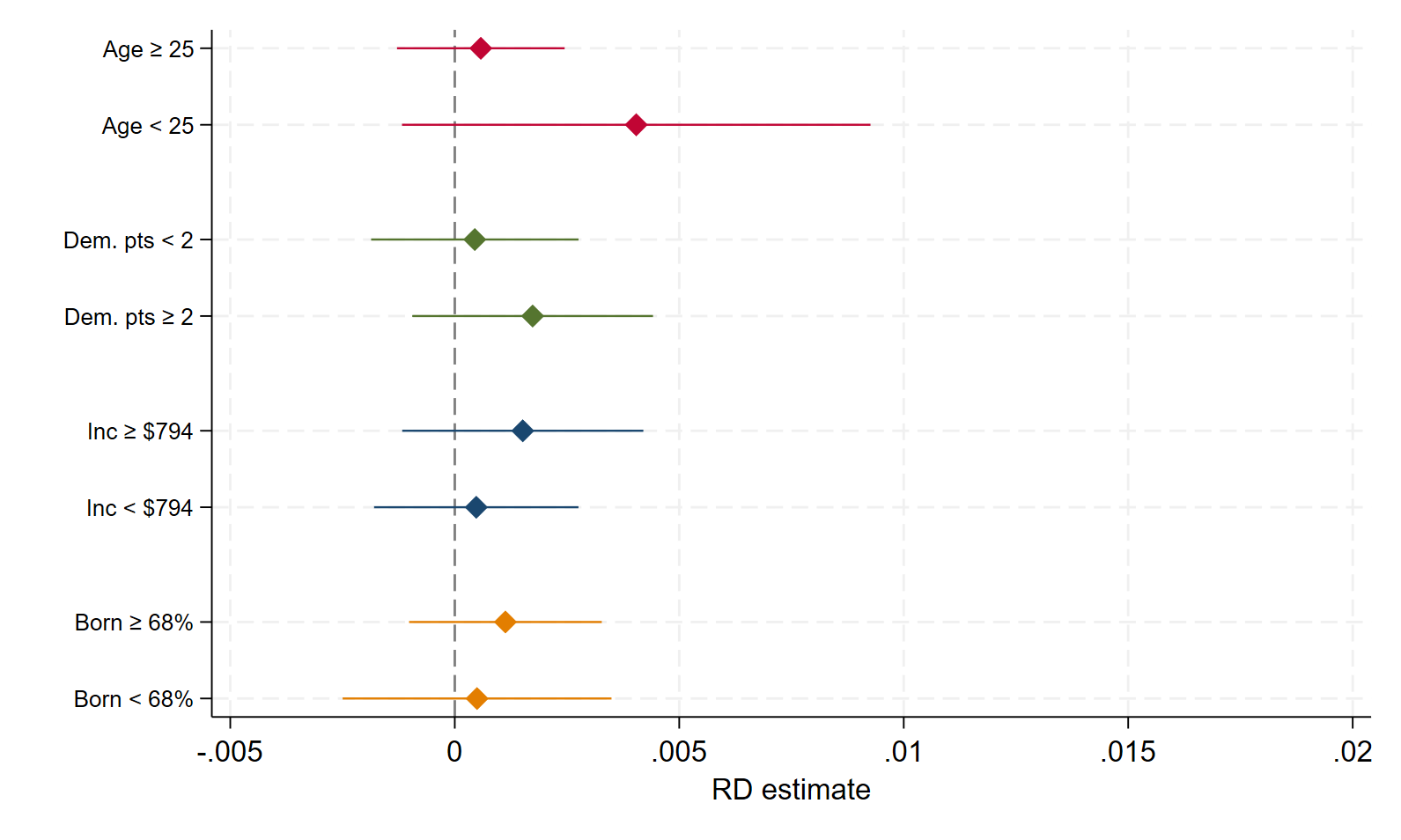}
\end{subfigure}
\caption{Estimates for liability transfer using the subsample of offenses detected by mobile speed cameras. Panel (a) reports estimates for subsample fines issued at 11 and 12 km/h of speed excess. Panel (b) reports estimates for fines issued at other speed excess.}
\label{fig:heterogeneity_transfer}
\end{figure}

\begin{figure}[h!]
\centering
\begin{subfigure}{0.49\linewidth}
    \centering
    \caption{Reoffending rates (11 and 12 km/h speed excess).}
    \includegraphics[width=\linewidth]{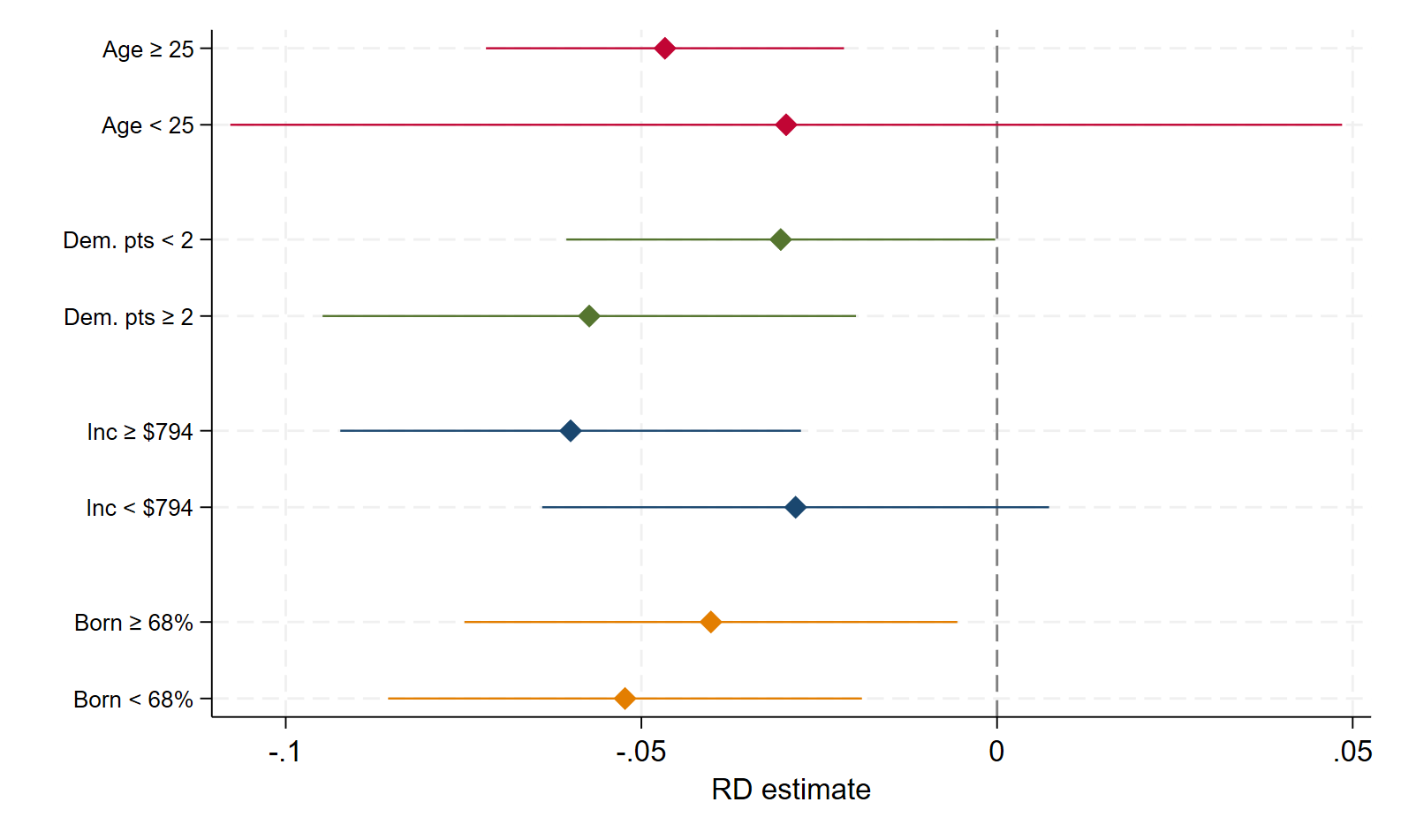}
\end{subfigure}
\hfill
\begin{subfigure}{0.49\linewidth}
    \centering
    \caption{Reoffending rates (other speeds).}
    \includegraphics[width=\linewidth]{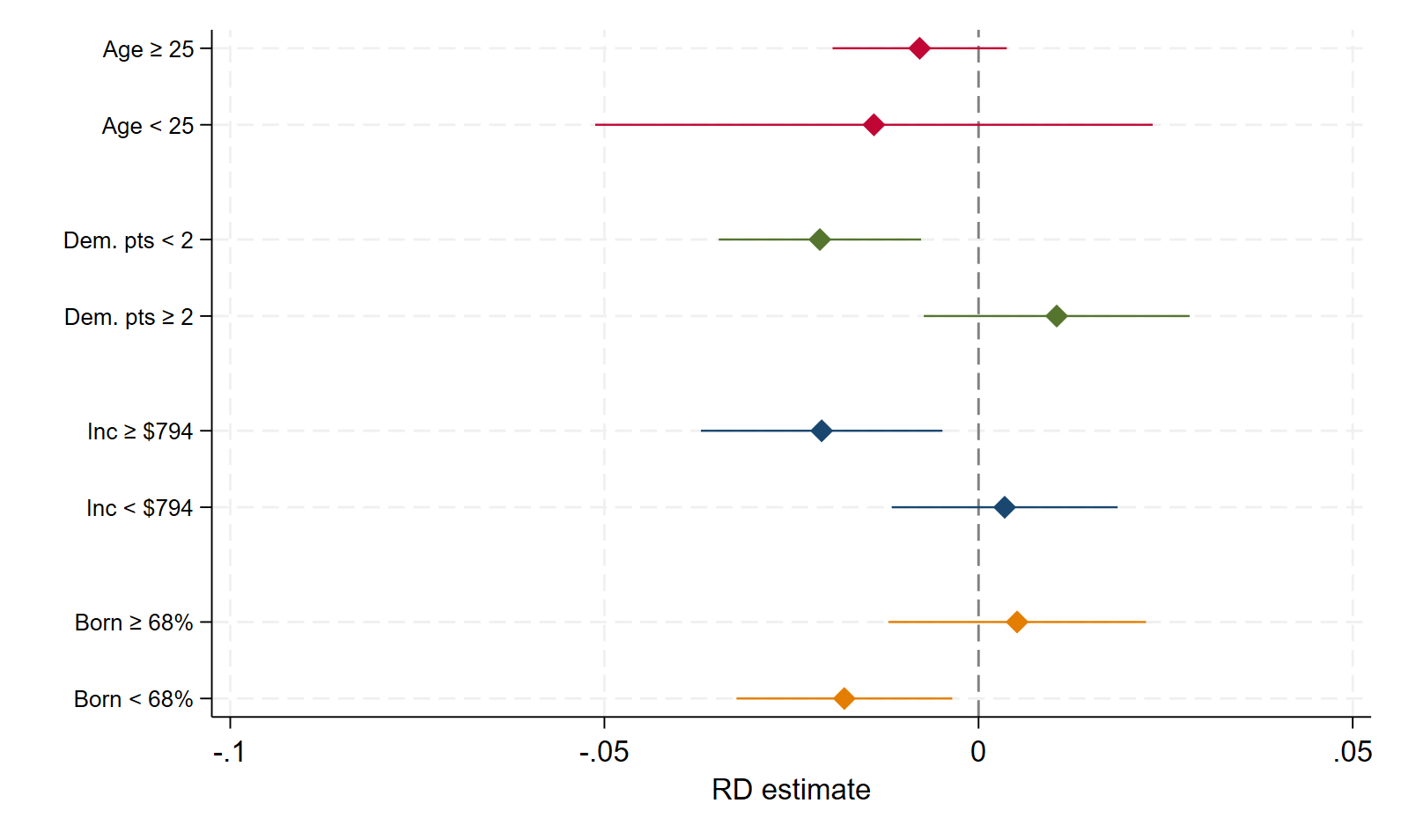}
\end{subfigure}
\caption{Estimates for reoffending using the subsample of offenses detected by fixed speed cameras. Panel (a) reports estimates for subsample fines issued at 11 and 12 km/h of speed excess. Panel (b) reports estimates for fines issued at other speed excess.}
\label{fig:heterogeneity_reoffending}
\end{figure}

Figures \ref{fig:heterogeneity_transfer} and \ref{fig:heterogeneity_reoffending} present heterogeneity analysis for the RDiT estimates of transfer rates (mobile camera subsample) and reoffending (fixed camera subsample) respectively. We examine how the effect of the July 2022 penalty reform varies across subgroups defined by individual and offense location characteristics. 

At the individual level, we use the age of the offender and their accumulated demerit points at the time of the offense, both of which are recorded directly in the infringement dataset. As proxies for offenders’ socioeconomic characteristics, we use the location of the offense, under the assumption that drivers are more likely to be detected in areas close to where they reside. Specifically, we link infringement records to 2021 Australian Census data at the Statistical Area Level 1 (SA1) using the offense location, and use median weekly personal income and the share of Australian-born residents as local area characteristics.

Figure \ref{fig:heterogeneity_transfer} presents heterogeneity estimates for transfer rates on the mobile camera sample, separately for offenses at 11--12 km/h speed excess (panel a) and other speed excess tiers (panel b). For the 11--12 km/h tier (panel a), the estimated effects are positive across all subgroups, though confidence intervals are wide and most estimates are statistically insignificant. The exception is offenders in lower-income areas (Inc $<$ \$794), where the estimated effect is positive and statistically significant. The point estimates are otherwise of similar magnitude across age, demerit points, and nativity subgroups, suggesting no strong differential response along these dimensions. For other speed excess tiers (panel b), the estimated effects are smaller and statistically indistinguishable from zero across all subgroups.

Figure \ref{fig:heterogeneity_reoffending} presents the corresponding heterogeneity estimates for reoffending on the fixed camera sample. For the 11--12 km/h tier (panel a), the estimated effects are negative across all subgroups, with larger estimates for older offenders (Age $>=$ 25), those with higher accumulated demerit points (Dem. pts $\geq$ 2), offenders in higher-income areas (Inc $>=$ \$794), and those in areas with a lower share of Australian-born residents (Born $<$ 68\%). The estimate for younger offenders (Age $<$ 25) is negative but imprecise. For other speed excess tiers (panel b), the estimated effects are smaller, with statistically significant effects for drivers with fewer accumulated demerit points, offenders in higher-income areas and those in areas with a lower share of Australian-born residents.

\FloatBarrier
\subsection{General deterrence and enforcement\label{sec:gendetterenceandenforcement}}

Our analysis so far relies on two assumptions. First, drivers who have not been fined do not change their behavior around the cutoff, implying no general deterrence effect; and second, enforcement remains stable around the cutoff. 

In this section, we assess these two assumptions separately for fixed and mobile cameras.

\FloatBarrier
\subsubsection{Fixed Cameras\label{sec:gendetterenceandenforcement_fixed}}

The balance tests reported in Table \ref{tab:desc_table} are reassuring. For fines issued by fixed cameras, we find no discontinuity in observable driver or offense characteristics around the rule change.

\begin{figure}[ht]
    \centering
    \includegraphics[width=0.6\linewidth]{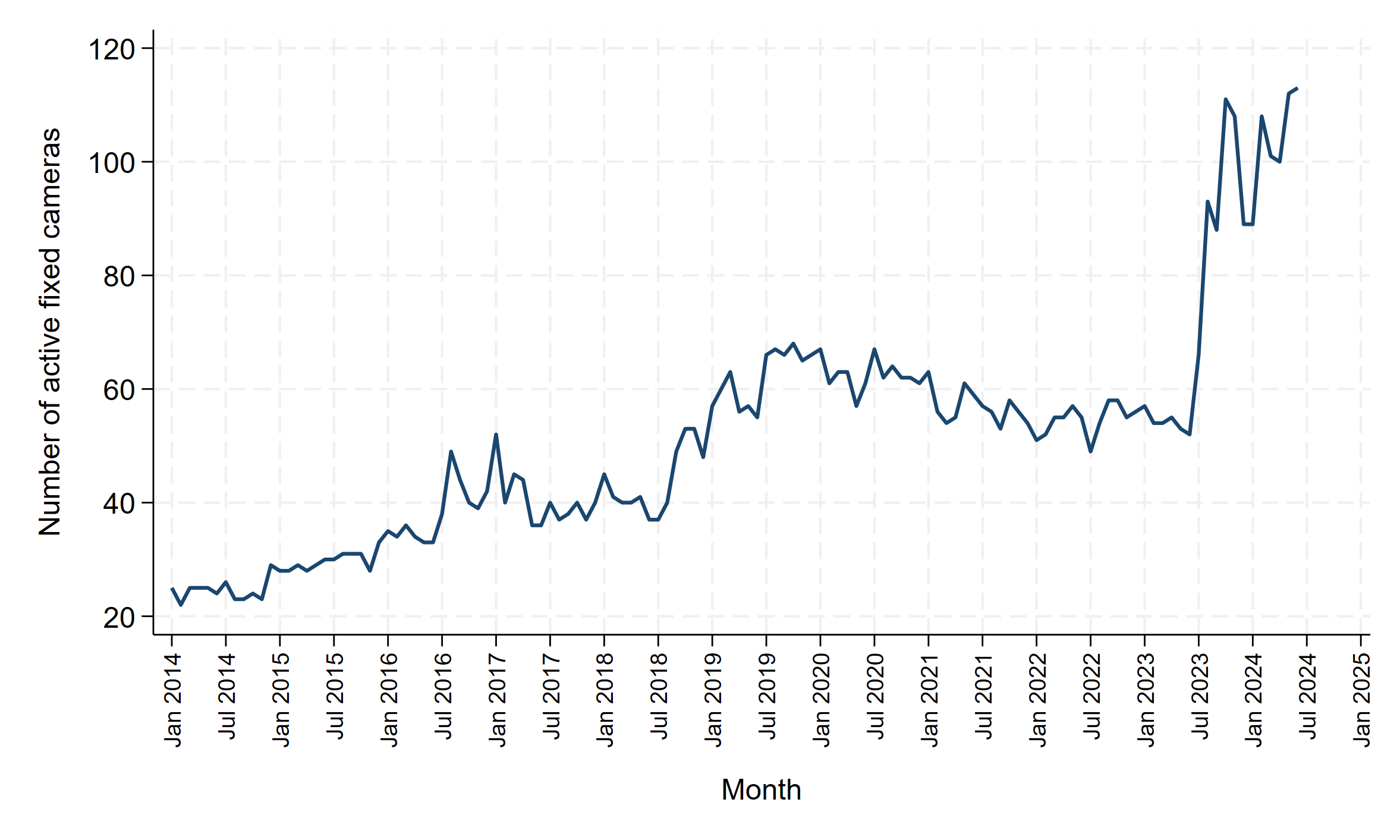}
    \caption{Number of active fixed cameras.\label{fig:active_fixed_cameras}}
\end{figure}

We first examine overall enforcement intensity over time. Figure \ref{fig:active_fixed_cameras} presents the number of active fixed cameras between January 2014 and June 2024.\footnote{We define active cameras as those issuing at least one fine during a given month.} The number of active fixed cameras increased steadily from around 20 in 2014 to approximately 50--60 by 2021, before rising sharply from July 2023 and reaching more than 100 by mid-2024. Importantly, there is no visible change around the July 2022 policy reform.

The sharp increase in camera coverage occurred 12 months after the reform, and therefore, does not affect the composition of drivers entering our RDiT estimation sample. It may, however, affect our measure of subsequent recidivism. For example, drivers fined two months before the reform reach the ten-month follow-up horizon in March 2023, whereas those fined two months after reach it in July 2023, when fixed-camera coverage began expanding. The two cohorts may therefore face different detection risks over their follow-up periods. This would bias our estimated reduction in recidivism toward zero. However, the left panel of Figure \ref{fig:reoffending_by_outcome_range} shows that our results are robust when restricting the recidivism outcome to offenses occurring within seven months of the initial offense, before the expansion in fixed-camera coverage.

\begin{figure}[ht]
  \centering
  \begin{subfigure}{0.49\textwidth}
    \centering
    \caption{Fines per camera per day}
    \label{fig:fines_per_camera_fixed}
    \includegraphics[width=\textwidth]{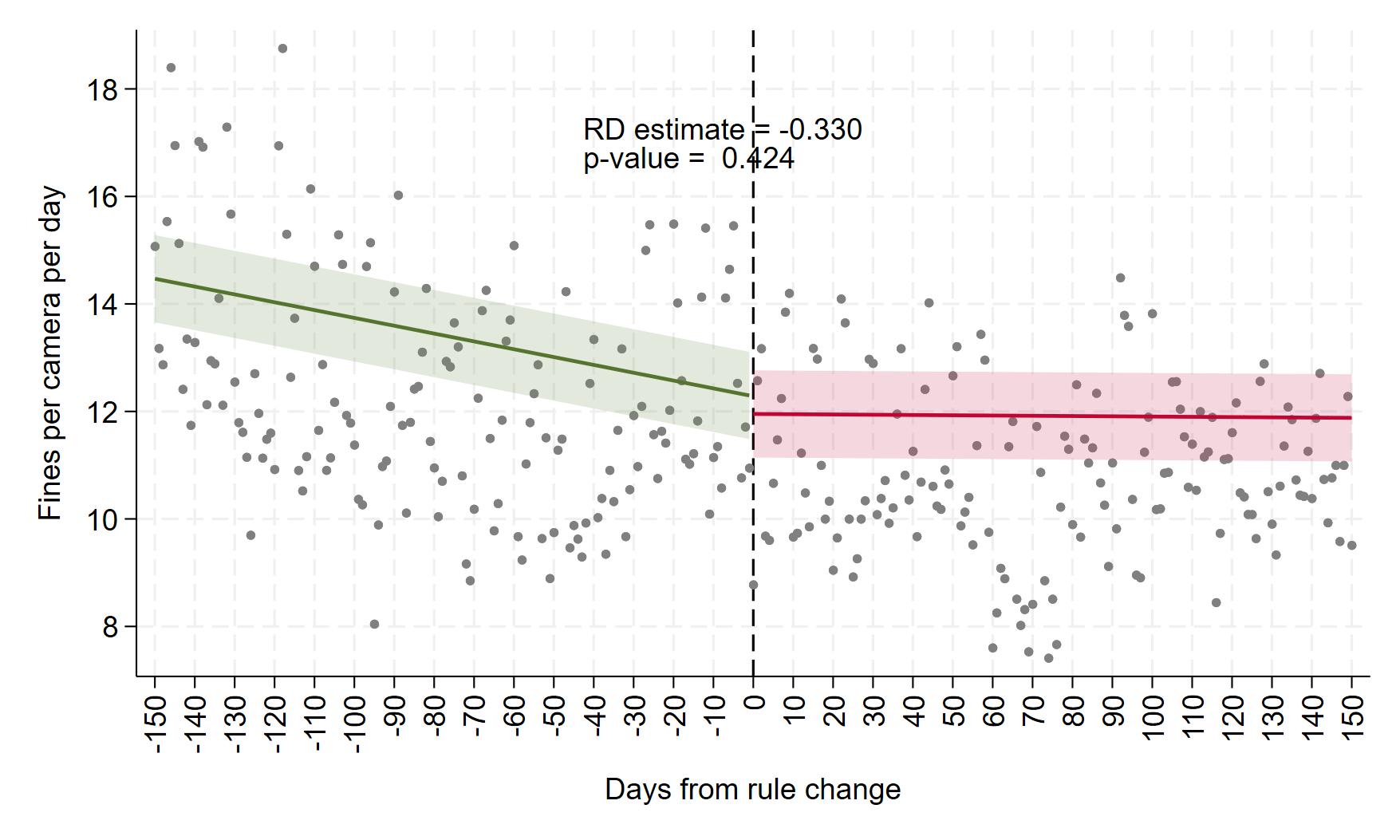}
    
  \end{subfigure}
  \hfill
  \begin{subfigure}{0.49\textwidth}
    \centering
    \caption{Event study: fines per camera}
    \label{fig:event_study_fines_fixed}
    \includegraphics[width=\textwidth]{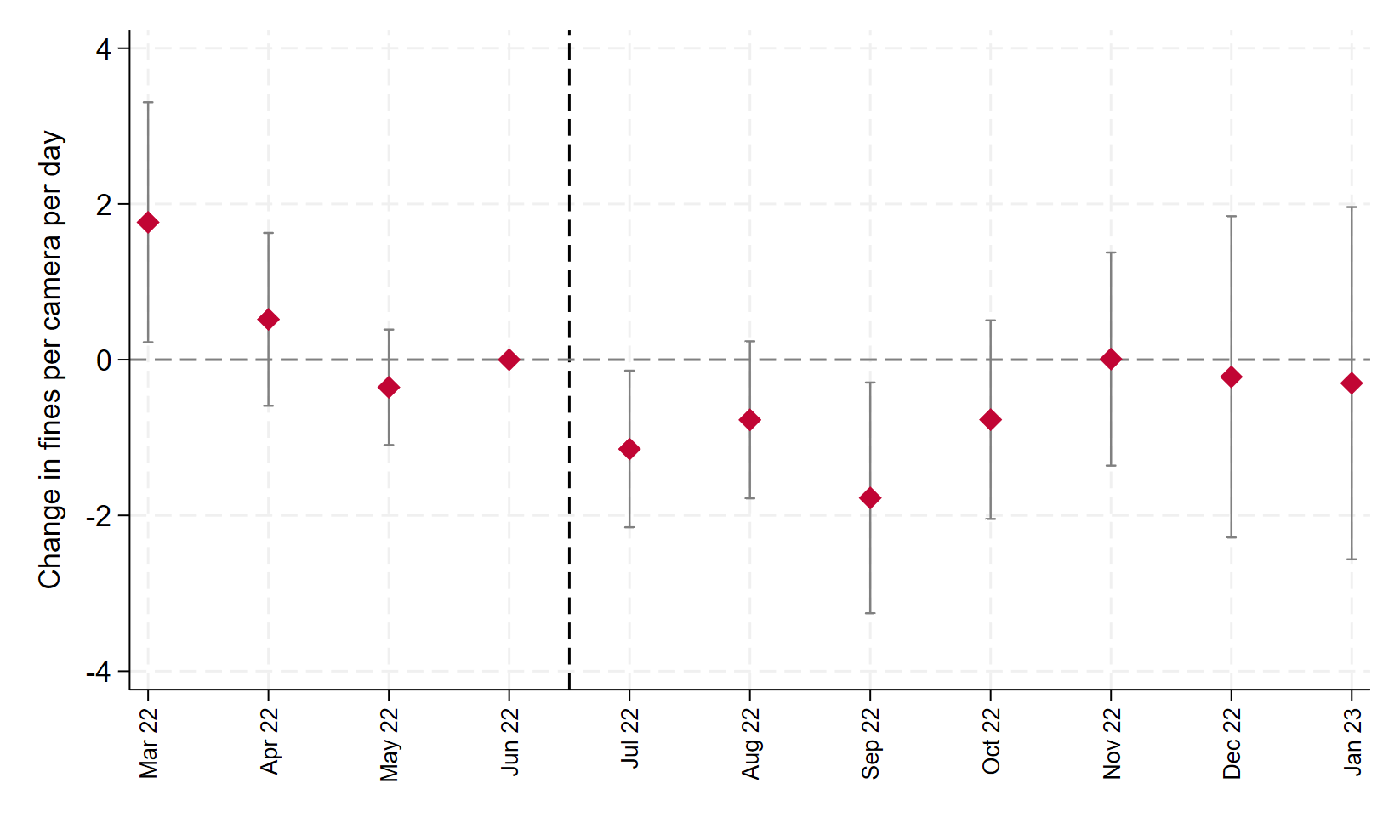}
  \end{subfigure}
  \caption{Fines per fixed camera per day. RDiT (a) and event study (b).}
  \label{fig:fines_fixed}
\end{figure}

Second, we examine the number of fines issued per active fixed camera per day around the reform. This measure reflects both camera operation and driver behavior, since changes in fines per camera may result from changes in enforcement or from drivers becoming more or less likely to speed at fixed-camera locations. Figure \ref{fig:fines_fixed} presents the corresponding RDiT and event-study estimates for fixed cameras. The number of fines issued per active camera remained stable around the reform. Together with the absence of any change in the number of active cameras, this provides no evidence of a discontinuous change in fixed-camera enforcement or speeding behavior around the reform date.

\FloatBarrier
\subsubsection{Mobile Cameras\label{sec:gendetterenceandenforcement_mobile}}

\begin{figure}[ht]
    \centering
    \includegraphics[width=0.6\linewidth]{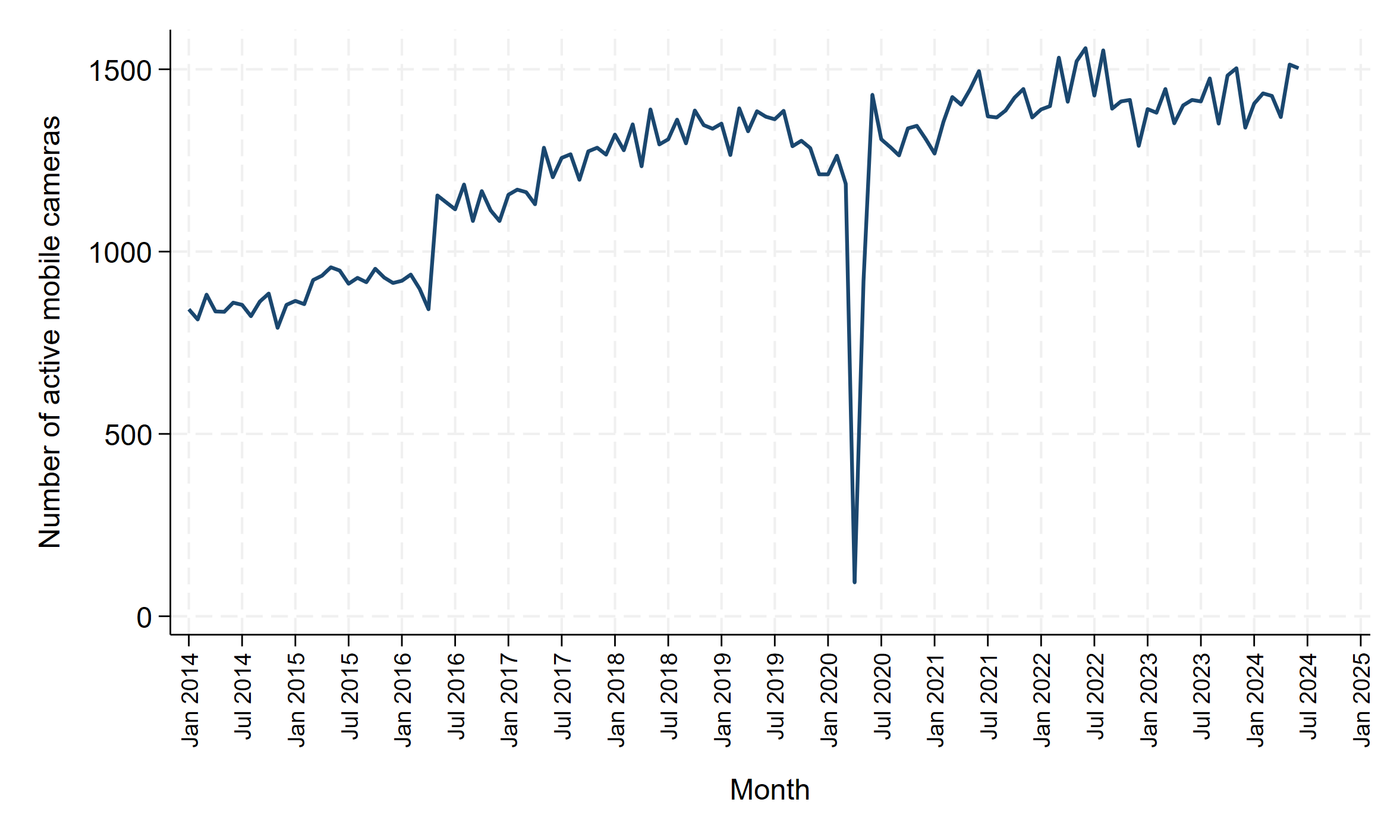}
    \caption{Number of active mobile cameras.\label{fig:active_mobilecameras}}
\end{figure}

As with fixed cameras, we first examine the evolution of the number of active mobile cameras over time. Figure \ref{fig:active_mobilecameras} shows the number of mobile cameras present each day. The mobile camera network grew from around 800 active units in 2014 to approximately 1,400-1,500 by 2019, before dropping sharply during the COVID-19 lockdowns in 2020 and recovering thereafter. Crucially, there is no visible discontinuity around July 2022, suggesting that the scale of the enforcement network did not change at the time of the reform.

\begin{figure}[ht]
  \centering
  \begin{subfigure}{0.49\linewidth}
    \centering
    \caption{RDiT: Fines per camera per day}
    \label{fig:RDiT_event_study_fines_mobile}
    \includegraphics[width=\textwidth]{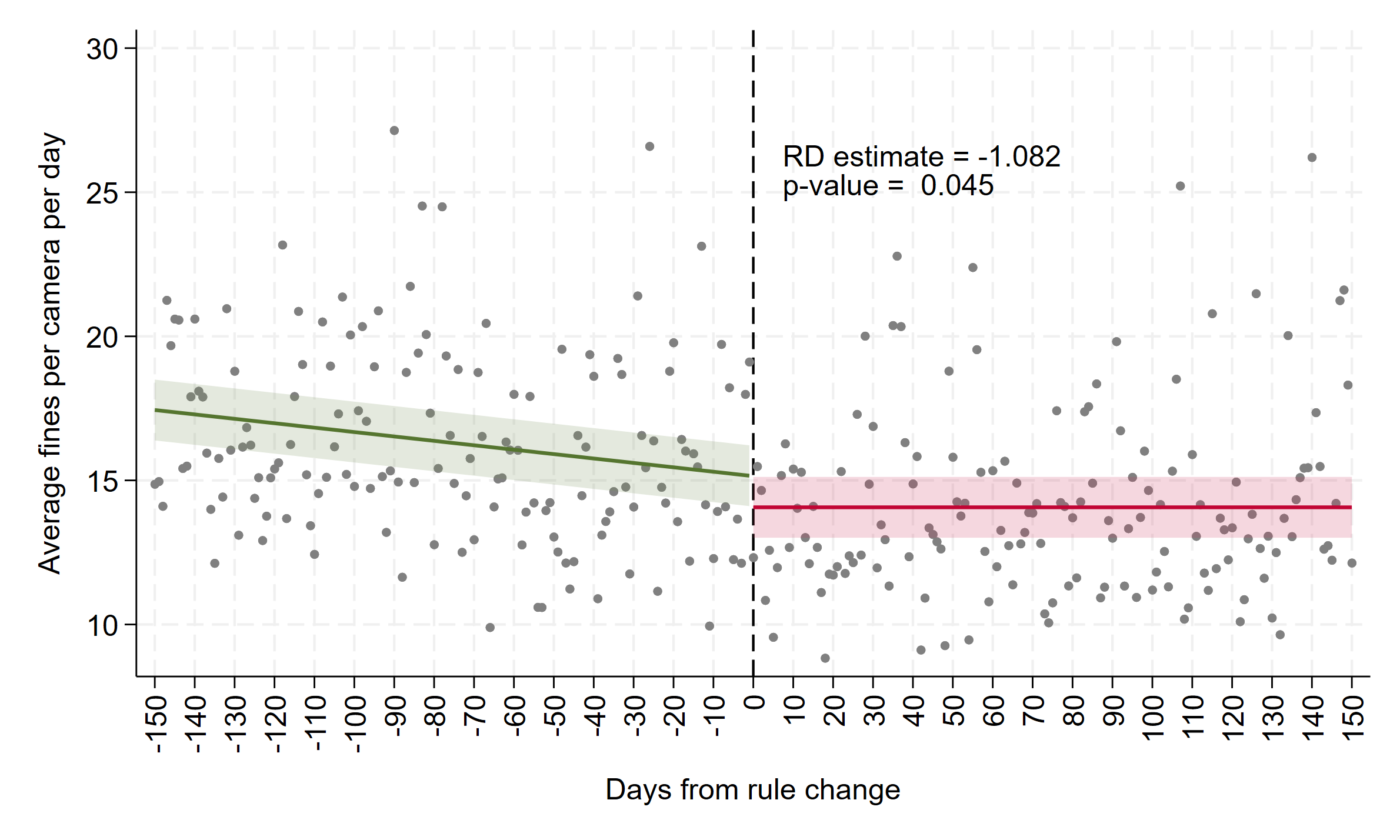}
  \end{subfigure}
  \hfill
  \begin{subfigure}{0.49\linewidth}
    \centering
    \caption{Event study: fines per camera}
    \label{fig:event_study_fines_mobile}
    \includegraphics[width=\textwidth]{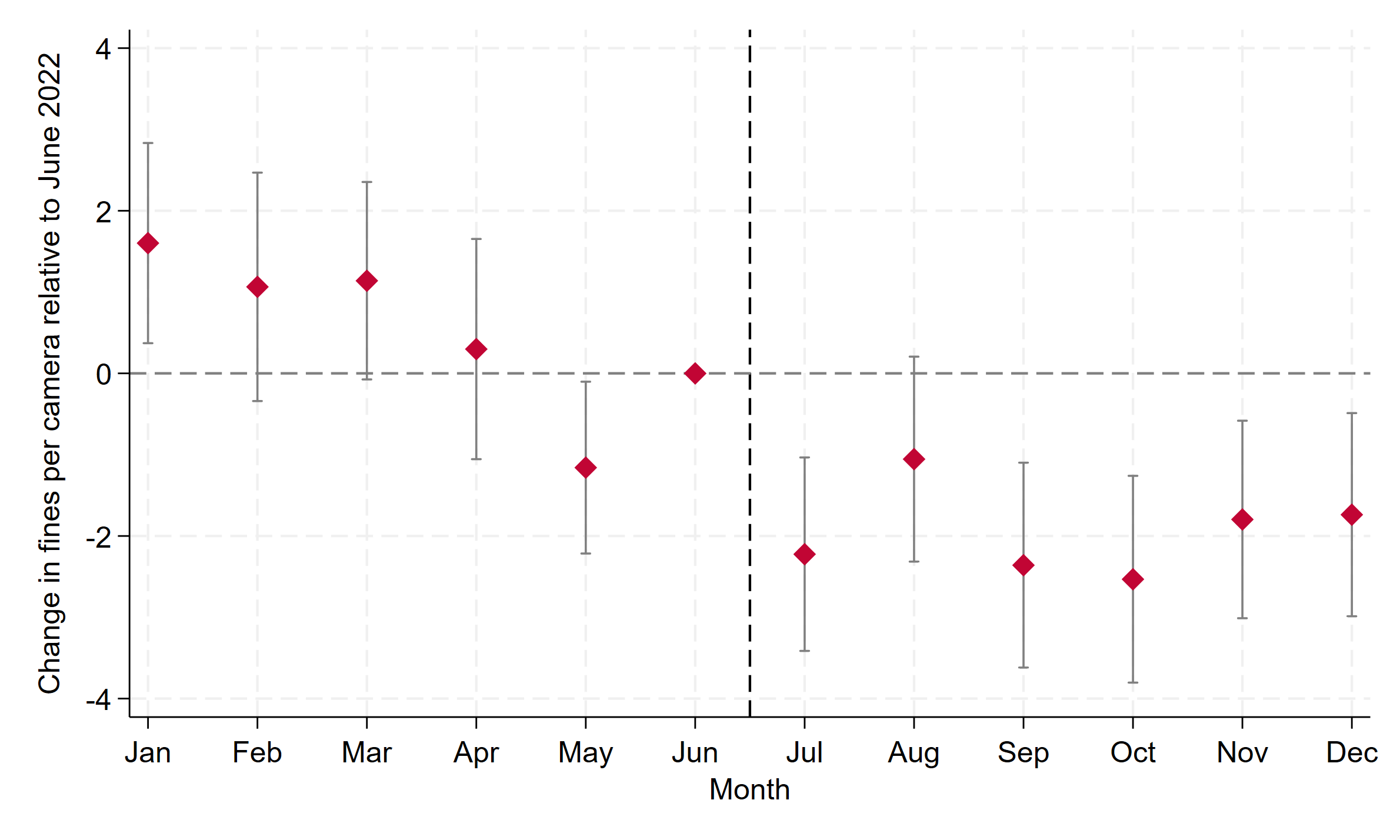}
    
  \end{subfigure}
  \caption{Fines per mobile camera per day. RDiT (a) and event study (b).}
  \label{fig:fines_mobile}
\end{figure}

Figure \ref{fig:fines_mobile}, however, shows a decline in the number of fines issued by mobile cameras around the cutoff. There are two potential explanations for this decline. First, it may reflect a general deterrence effect: following the increase in penalties, drivers may have become less likely to speed in locations monitored by mobile cameras. This channel is more plausible for mobile than fixed cameras. Fixed cameras are accompanied by advance warning signs, so drivers approaching them are already aware of their presence. Fines issued by fixed cameras therefore come from a selected group of drivers who continue to speed despite the warning. Mobile cameras, by contrast, operate at varying locations without the same advance warning. Drivers can, however, use real-time traffic applications, such as Google Maps, to identify reported mobile-camera locations and adjust their speed accordingly. Higher penalties may therefore increase drivers' incentives to use such information, leading to fewer mobile-camera detections even without a change in enforcement.

Second, the decline could reflect a change in enforcement intensity that is not captured by the number of active cameras, for example if active mobile cameras operated for fewer hours or were deployed at different time or location after the reform. The available data do not allow us to distinguish cleanly between these two mechanisms.\footnote{The location and duration of mobile-camera deployment can be inferred from the time and location of fines issued. However, this measure is itself affected by driver behavior: cameras that issue no fines are unobserved, and at locations or times when fewer drivers speed, a camera may appear to have been deployed for a shorter period.} We therefore remain agnostic as to whether the decline in mobile-camera fines reflects general deterrence, changes in enforcement intensity, or a combination of the two.

The decline in mobile-camera fines at the cutoff could threaten identification of effects on recidivism if it changed the composition of drivers fined before and after the reform. Our recidivism sample is insulated from this concern at the treatment-assignment stage because the initial offenses used in the RDiT estimation are restricted to fixed-camera detections. The baseline recidivism outcome does, however, include subsequent offenses detected by both fixed and mobile cameras. We therefore also reconstruct the outcome using only subsequent fixed-camera detections. Appendix \ref{app:reoffending_fixed} reports the results. The resulting estimates are quantitatively similar, indicating that the decline in mobile-camera detections does not drive our recidivism results.

We have shown that, for mobile cameras, fine transfers increase following the reform. Our interpretation is that drivers became more likely to transfer liability in response to the additional demerit points. An alternative explanation is that general deterrence or changes in enforcement altered the composition of drivers detected by mobile cameras around the cutoff. We cannot rule out such compositional changes entirely. However, for them to explain our transfer result, they would have to generate an increase in transfers specifically among drivers detected 11--12 km/h above the speed limit, the only category in which demerit points increased.

\begin{figure}[ht]
  \centering
  \begin{subfigure}{0.49\linewidth}
    \centering
    \caption{Fines per camera per day (11--12km/h)}
    \label{fig:fine_per_camera_mobile_11_12}
    \includegraphics[width=\linewidth]{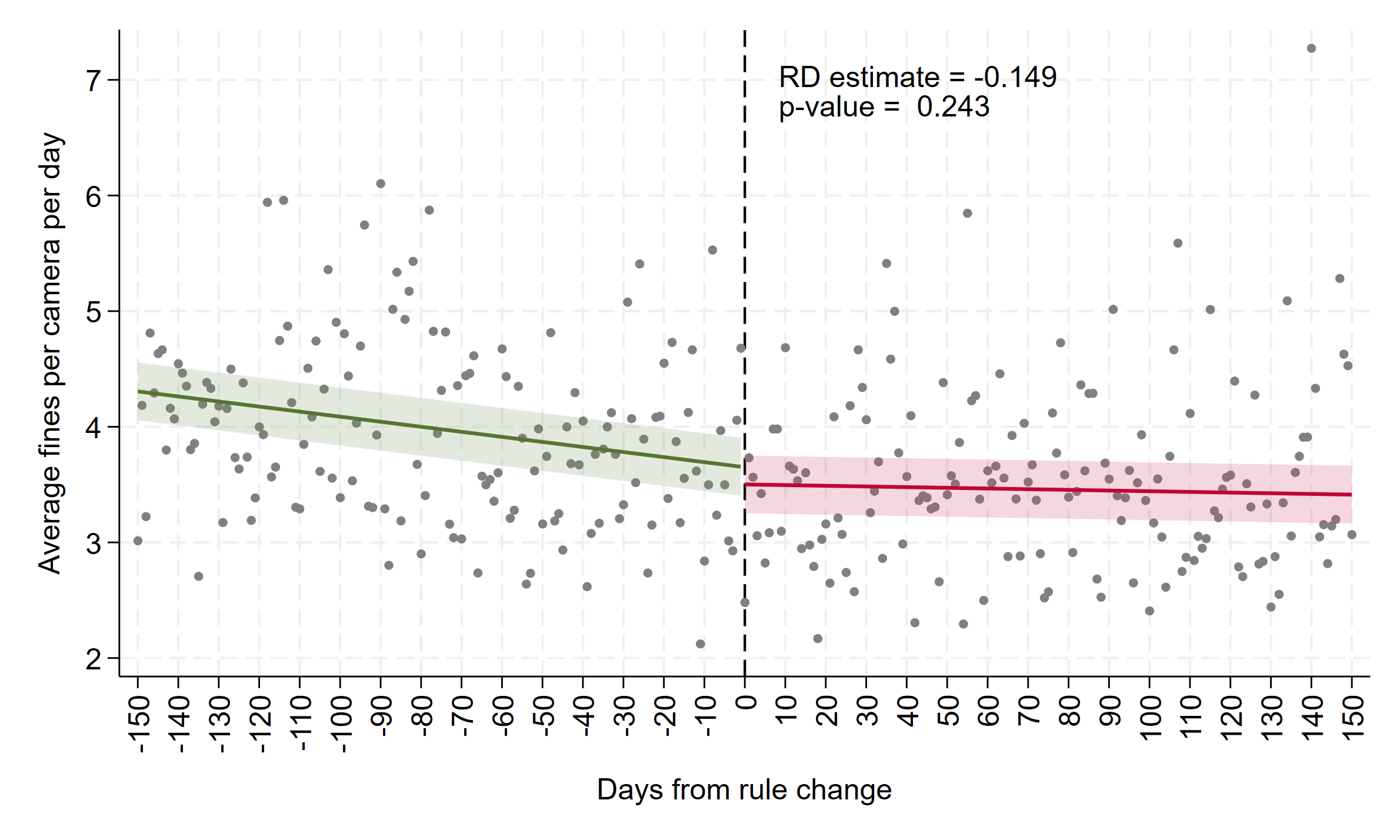}
  \end{subfigure}
  \hfill
  \begin{subfigure}{0.49\linewidth}
    \centering
    \caption{Fines per camera per day (other speeds)}
    \label{fig:fine_per_camera_mobile_not_11_12}
    \includegraphics[width=\linewidth]{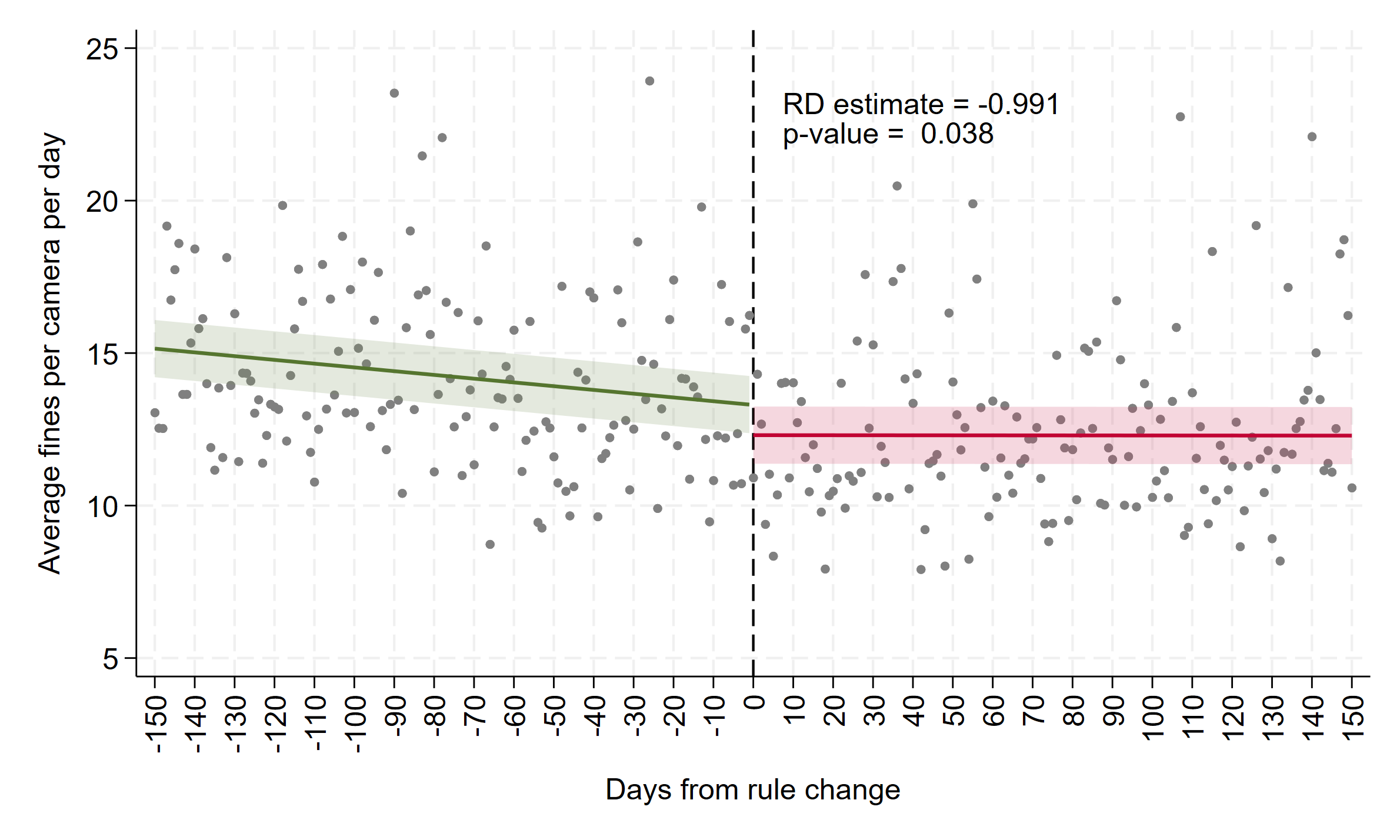}
  \end{subfigure}
  \caption{Fines per mobile camera per day, 11--12km/h (a) and other speeds (b).}
  \label{fig:fines_per_camera_mobile}
\end{figure}

Figure \ref{fig:fines_per_camera_mobile} examines the decline in mobile-camera fines separately for the 11--12 km/h category and all other speed bands. The estimated discontinuity is negative in both groups, although it is statistically significant only for other speed excesses. Importantly, however, the decline in mobile-camera fines is not concentrated in the 11--12 km/h category, whereas the increase in transfers is. This pattern is difficult to reconcile with a general change in enforcement intensity or deterrence alone.

\FloatBarrier
\section{Robustness and validity tests}

Following \citet{hausman2018} we conduct a set of robustness and validity checks: testing different bandwidths (Section \ref{sec:bandwidth}) and placebo checks (Section \ref{sec:placebo}). We complement these tests with an alternative identification strategy exploiting differential changes in sanctions across adjacent speed bands (Section \ref{sec:rd_did}), and a robustness check addressing differences in exposure to the post-reform penalty regime (Section \ref{sec:main_results_fixed_donut}).

\FloatBarrier
\subsection{Bandwidth Sensitivity\label{sec:bandwidth}}

Our main results use a bandwidth of 150 days on either side of the penalty change, chosen to balance the competing demands of sample size and local validity of the linear trend approximation. Unlike a standard cross-sectional RD, where data-driven bandwidth selectors can be derived, the RDiT design faces an additional challenge: because the running variable is calendar time, wider bandwidths increasingly risk capturing seasonal patterns, day-of-week effects, and longer-run trends that are unrelated to the policy change \citep{hausman2018}. The choice of bandwidth therefore involves a degree of judgement that cannot be fully resolved by standard selection procedures. We show that our results are not sensitive to this choice by estimating the main specification across a range of bandwidths from 10 to 180 days.

\begin{figure}[h!]
\centering
\begin{subfigure}{0.49\linewidth}
    \centering
    \caption{Transfer (11--12 km/h)}
    \includegraphics[width=1\linewidth]{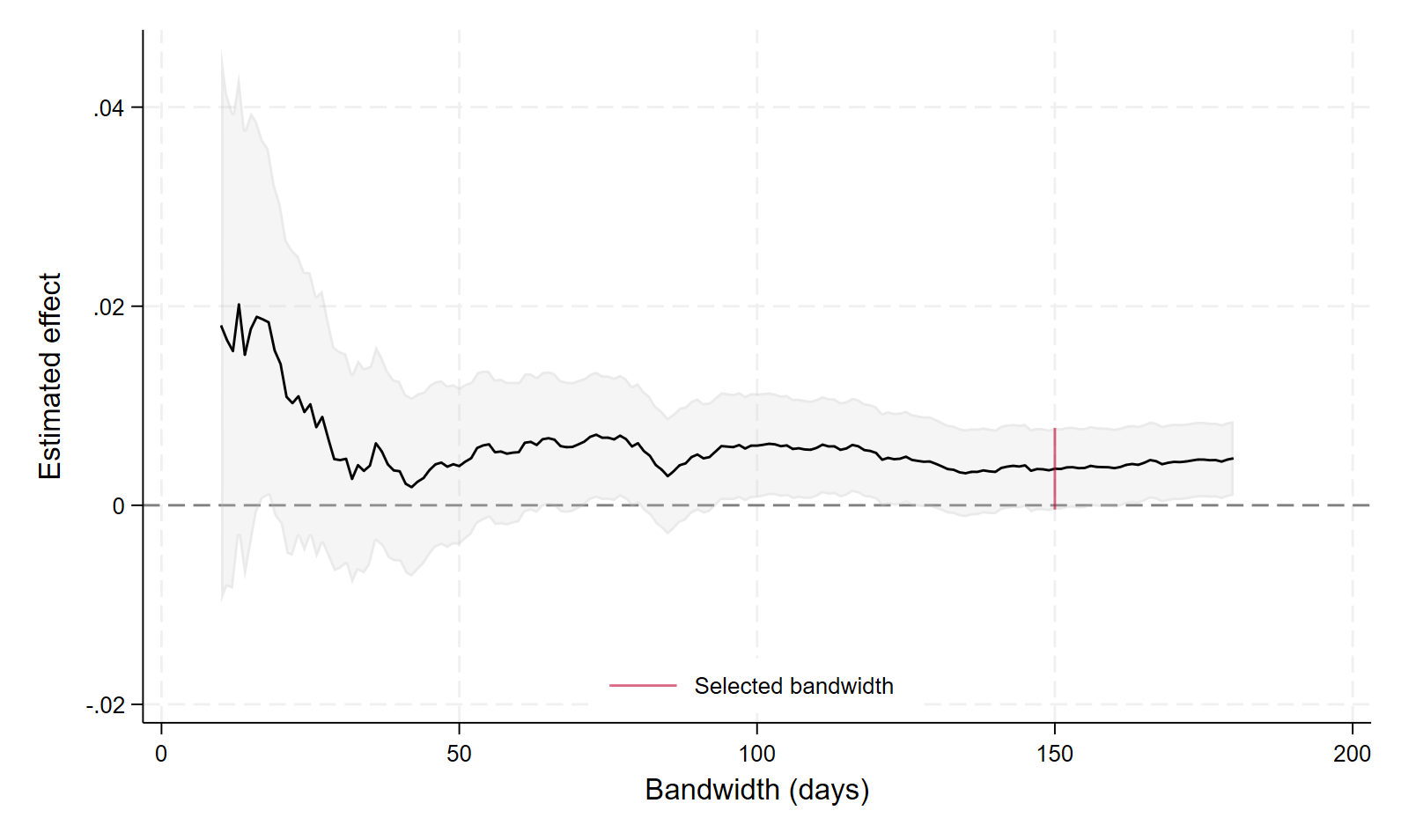}
\end{subfigure}
\hfill
\begin{subfigure}{0.49\linewidth}
    \centering
    \caption{Transfer (other speeds)}
    \includegraphics[width=1\linewidth]{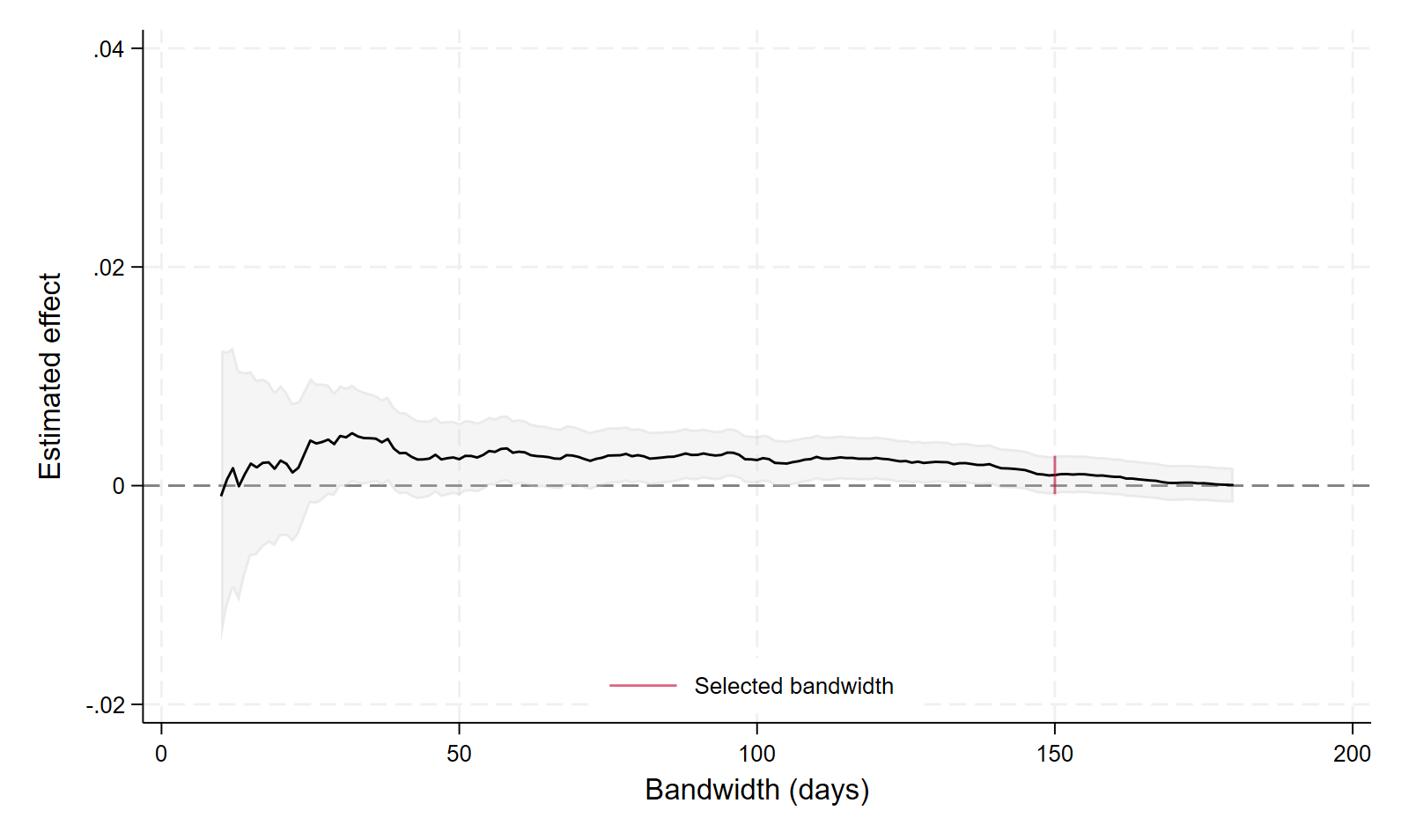}
\end{subfigure}
\begin{subfigure}{0.49\linewidth}
    \centering
    \caption{Reoffending (11--12 km/h)}
    \includegraphics[width=1\linewidth]{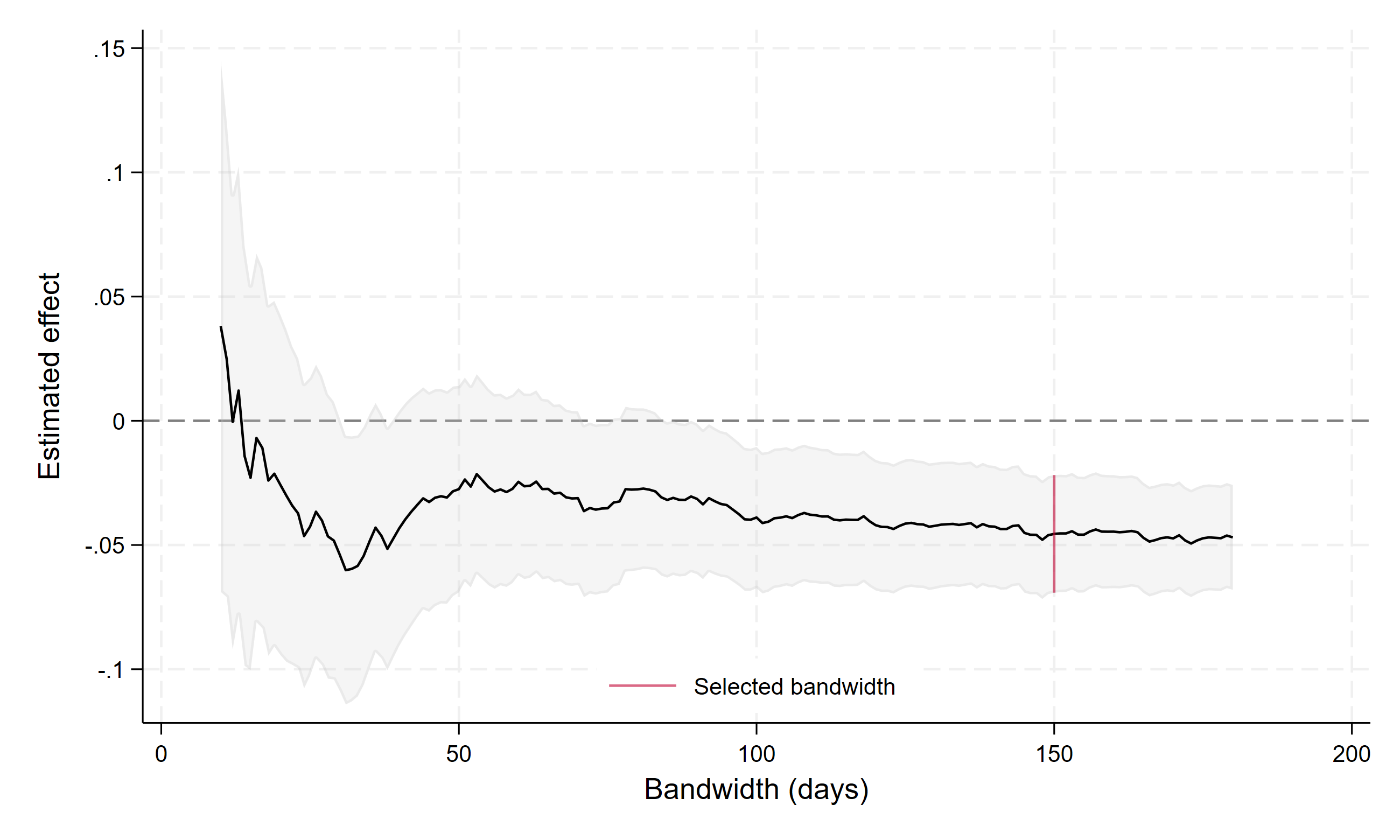}
\end{subfigure}
\begin{subfigure}{0.49\linewidth}
    \centering
    \caption{Reoffending (other speeds)}
    \includegraphics[width=1\linewidth]{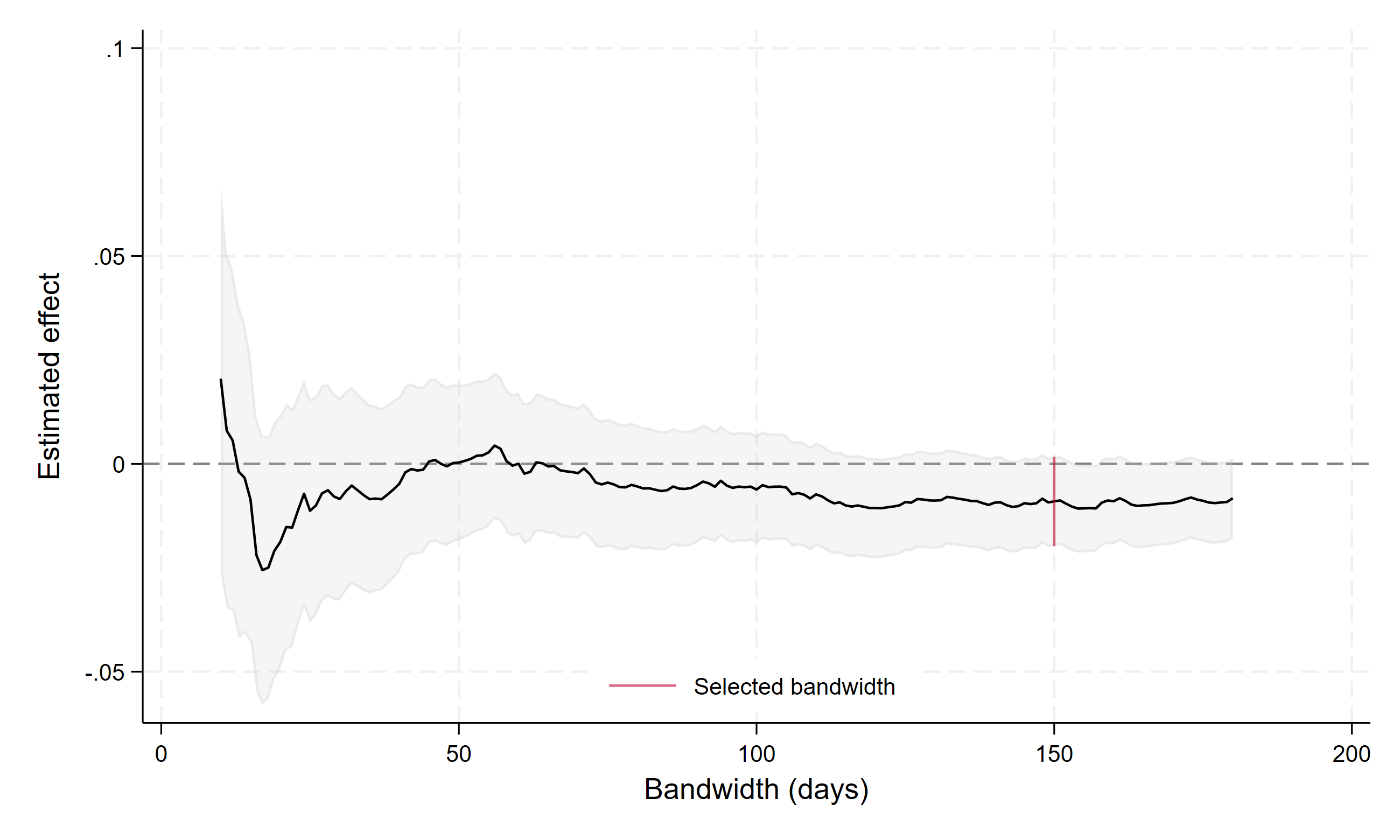}
\end{subfigure}
\caption{Effects of the increase in penalties on transfer rates (top panels) and reoffending (bottom panels), for offenses involving speed excesses of 11--12 km/h (left panels) and all other speed excesses (right panels), across bandwidths from 10 to 180 days.}
\label{fig:bandwidth_check_speeds}
\end{figure}

Figure \ref{fig:bandwidth_check_speeds} presents RD estimates for bandwidths ranging from 10 to 180 days, estimated using equation (\ref{eq:rdit}), separately for offenses at 11--12 km/h speed excess (left panels) and other speed excess tiers (right panels), for both transfer rates (top panels) and reoffending (bottom panels). The upper bound of 180 days is chosen to avoid overlap with the tail end of COVID-19 restrictions in late 2021, which could contaminate the pre-reform comparison window.

For transfer rates at 11--12 km/h (panel a), estimates are noisy at narrow bandwidths below 50 days, reflecting insufficient sample size. Above 50, the estimated effect is stabilised at a similar level with statistical significance changing over different bandwidth ranges. For other speed excess tiers (panel b), the estimated effects are smaller and show a disappearing effect after 150-day bandwidth.

For reoffending at 11--12 km/h (panel c), estimates are similarly noisy below 50 days. From around 100 days onwards, a negative and statistically significant effect emerges and remains stable throughout the tested range, with narrowing confidence intervals reflecting the gain in precision from larger samples. For other speed excess tiers (panel d), the estimated effects are small and predominantly insignificant across all bandwidths, confirming that the deterrence response is concentrated at the tier where demerit points increased.

The corresponding results pooled across all speed excess tiers are presented in Appendix \ref{app:bandwidth_check}.

\FloatBarrier

\subsection{Placebo tests: effects in previous years \label{sec:placebo}}

\subsubsection{Reoffending}

\begin{figure}[h!]
  \centering
  \includegraphics[width=0.7\linewidth]{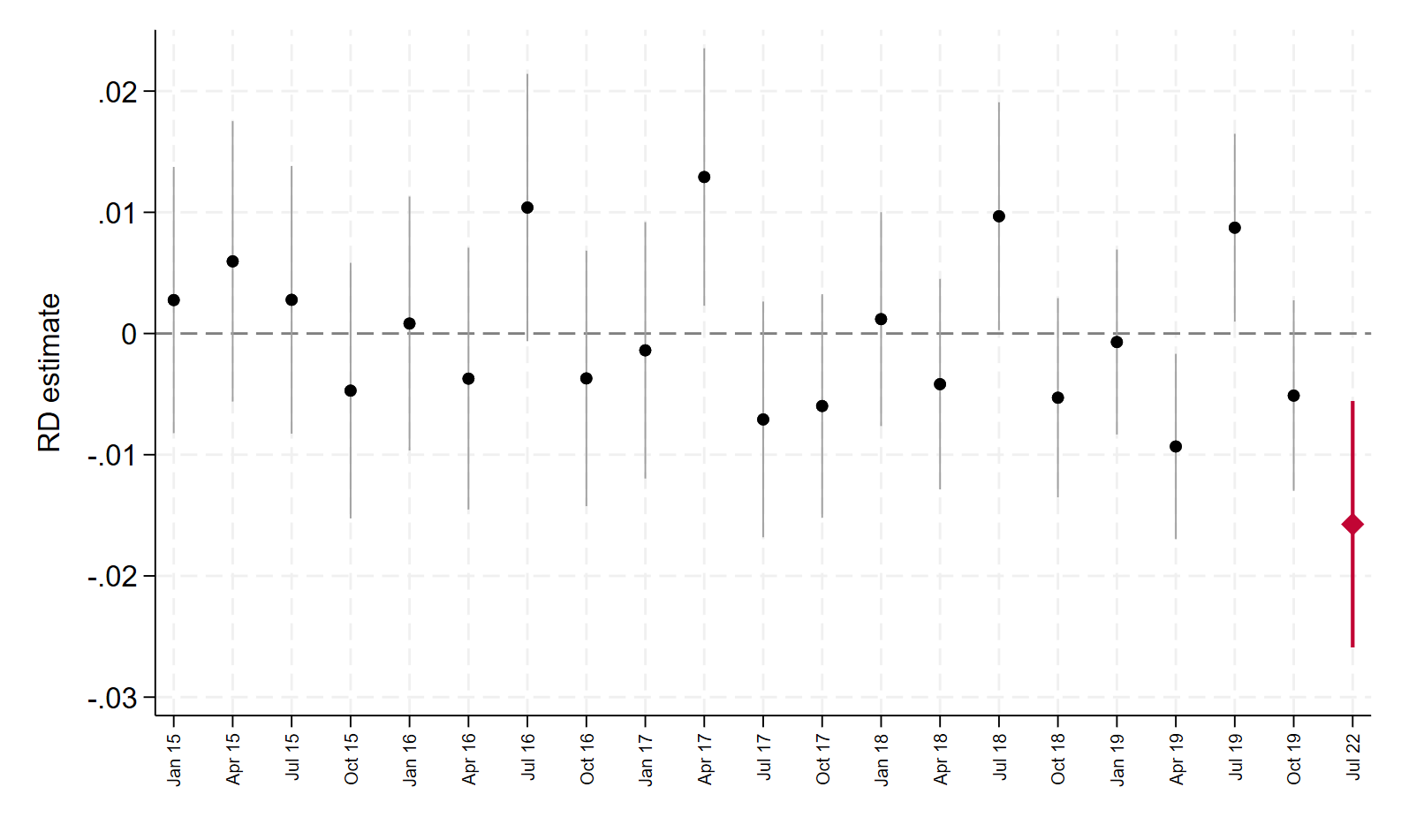}
  \caption{Reoffending effects at placebo periods for RDiT estimates.  July 2022 shown in red for comparison.}
  \label{fig:placebo_quarters_controls}
\end{figure}

Figure \ref{fig:placebo_quarters_controls} and Table \ref{tab:qvalues} report RDiT placebo estimates for all quarters across 2015-2019, with controls. We omit 2020 and 2021 due to the potential contamination from COVID-19 lockdowns, which disrupted normal traffic patterns and enforcement activity in ways unrelated to the penalty regime. For comparison, Figure \ref{fig:placebo_quarters_controls} also displays the July 2022 estimate in red. The July 2022 estimate is the only negative and statistically significant coefficient across all quarters examined, consistent with the penalty change driving the main result rather than a mechanical seasonal pattern around 1 July.

Across the 20 placebo quarters spanning 2015–2019, three estimates are statistically significant, all positive (showing the opposite sign to our main result). The July 2022 estimate is the only negative and statistically significant estimate. To account for the possibility that some significant results arise purely by chance when testing multiple hypotheses simultaneously, we compute q-values following \citet{anderson2008}, which implement the Benjamini-Hochberg false discovery rate correction \citep{benjamini1995}. Unlike p-values, which assess each hypothesis in isolation, q-values ask what proportion of the rejected hypotheses are likely to be false positives given the full set of tests conducted. No placebo estimate has a q-value below 0.18, confirming that none of the nominally significant placebo results survive correction for multiple testing. Taken together, the placebo evidence strongly supports the interpretation that the negative July 2022 estimate reflects a genuine causal response to the penalty reform rather than a statistical artefact.\footnote{When the p-value for the July 2022 estimate is included in the multiple testing correction alongside the 20 placebo p-values, the July 2022 result remains statistically significant at the 5\% level, while all placebo estimates remain insignificant.}

\begin{table}[htbp]\centering
\begin{tabular}{>{\raggedright\arraybackslash}p{1.5cm} >{\centering\arraybackslash}p{2.8cm} >{\centering\arraybackslash}p{2.8cm} >{\centering\arraybackslash}p{2.8cm} >{\centering\arraybackslash}p{2.8cm}}
\toprule
        & \multicolumn{1}{c}{1 Jan} & \multicolumn{1}{c}{1 Apr} & \multicolumn{1}{c}{1 Jul} & \multicolumn{1}{c}{1 Oct} \\
\midrule
2015    & 0.797 / 1.000 & 0.577 / 0.762 & 0.797 / 1.000 & 0.577 / 0.762 \\
2016    & 0.886 / 1.000 & 0.723 / 1.000 & 0.385 / 0.444 & 0.653 / 0.958 \\
2017    & 0.886 / 1.000 & 0.181 / 0.220 & 0.509 / 0.624 & 0.509 / 0.624 \\
2018    & 0.886 / 1.000 & 0.577 / 0.762 & 0.226 / 0.246 & 0.509 / 0.624 \\
2019    & 0.886 / 1.000 & 0.193 / 0.224 & 0.193 / 0.224 & 0.509 / 0.624 \\
\bottomrule
\multicolumn{5}{l}{\footnotesize BH q-value / sharpened q-value from \citet{anderson2008}.}\\
\multicolumn{5}{l}{\footnotesize No q-value falls below 0.10, indicating no placebo result survives multiple testing correction.}\\
\end{tabular}
\caption{Placebo test: BH and sharpened q-values.}
\label{tab:qvalues}
\end{table}

\subsubsection{Transfers}

\begin{figure}[h!]
  \centering
  \includegraphics[width=0.7\linewidth]{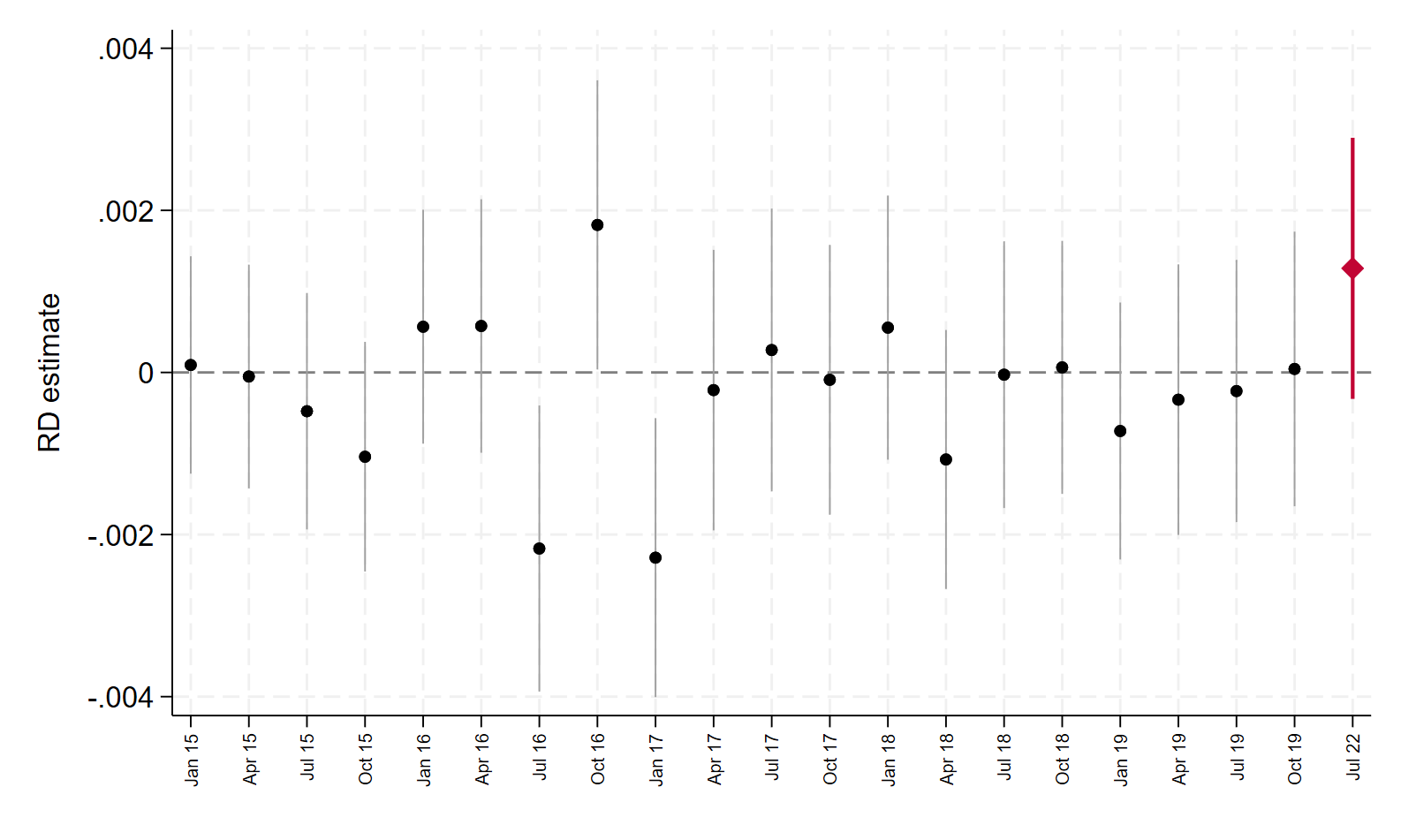}
  \caption{Transfer effects at placebo periods for RDiT estimates.  July 2022 shown in red for comparison.}
  \label{fig:placebo_quarters_transfers}
\end{figure}

Figure \ref{fig:placebo_quarters_transfers} presents RDiT placebo estimates for transfer rates across all quarters spanning 2015-2019, with July 2022 shown in red for comparison. The placebo coefficients are primarily small and statistically indistinguishable from zero throughout. 

Table \ref{tab:qvalues_transfers} reports BH and sharpened q-values for the 20 placebo quarter-year combinations. No q-value falls below 0.15, confirming that none of the placebo estimates survive correction for multiple testing across the full set of hypotheses. While the July 2022 estimate is positive, it does not reach conventional levels of statistical significance once included in the multiple testing correction.

\begin{table}[htbp]\centering
\begin{tabular}{>{\raggedright\arraybackslash}p{1.5cm} >{\centering\arraybackslash}p{2.8cm} >{\centering\arraybackslash}p{2.8cm} >{\centering\arraybackslash}p{2.8cm} >{\centering\arraybackslash}p{2.8cm}}
\toprule
        & \multicolumn{1}{c}{1 Jan} & \multicolumn{1}{c}{1 Apr} & \multicolumn{1}{c}{1 Jul} & \multicolumn{1}{c}{1 Oct} \\
\midrule
2015    & 0.975 / 1.000 & 0.975 / 1.000 & 0.975 / 1.000 & 0.752 / 1.000 \\
2016    & 0.975 / 1.000 & 0.975 / 1.000 & 0.159 / 0.189 & 0.304 / 0.376 \\
2017    & 0.159 / 0.189 & 0.975 / 1.000 & 0.975 / 1.000 & 0.975 / 1.000 \\
2018    & 0.975 / 1.000 & 0.752 / 1.000 & 0.975 / 1.000 & 0.975 / 1.000 \\
2019    & 0.975 / 1.000 & 0.975 / 1.000 & 0.975 / 1.000 & 0.975 / 1.000 \\
\bottomrule
\multicolumn{5}{l}{\footnotesize BH q-value / sharpened q-value. Both control the false discovery rate \citep{benjamini1995}.}\\
\multicolumn{5}{l}{\footnotesize No q-value falls below 0.10, indicating no placebo result survives multiple testing correction.}\\
\end{tabular}
\caption{Placebo test: BH and sharpened q-values for transfer rates}
\label{tab:qvalues_transfers}
\end{table}

\subsection{Alternative Identification Strategy: DiD\label{sec:rd_did}}

As an alternative identification strategy, we exploit the fact that the demerit point increase applied exclusively to the 11--12~km/h speed excess tier, while drivers caught at 9-10~km/h faced a fine increase but no change in demerit point allocation. Drivers do not precisely control whether they are detected at 9-10 or 11--12~km/h above the speed limit.\footnote{This assumption is supported by survey evidence showing that 66.7\% of Queensland drivers could not correctly identify the fine associated with the lowest speeding bracket, and by the fact that vehicle speedometers are deliberately calibrated to over-read speed, making precise speed management around the threshold implausible.} The 9-10~km/h group therefore serves as a control for the 11--12~km/h treatment group. We estimate a difference-in-differences specification interacting a treatment indicator (speed excess 11--12~km/h) with monthly time dummies, using June 2022 as the baseline month, which allows us to trace out the differential evolution of outcomes between the two groups before and after the reform.

\begin{figure}[h!]
\centering
\begin{subfigure}{0.49\linewidth}
    \centering
    \caption{Transfer rates}
    \includegraphics[width=1\linewidth]{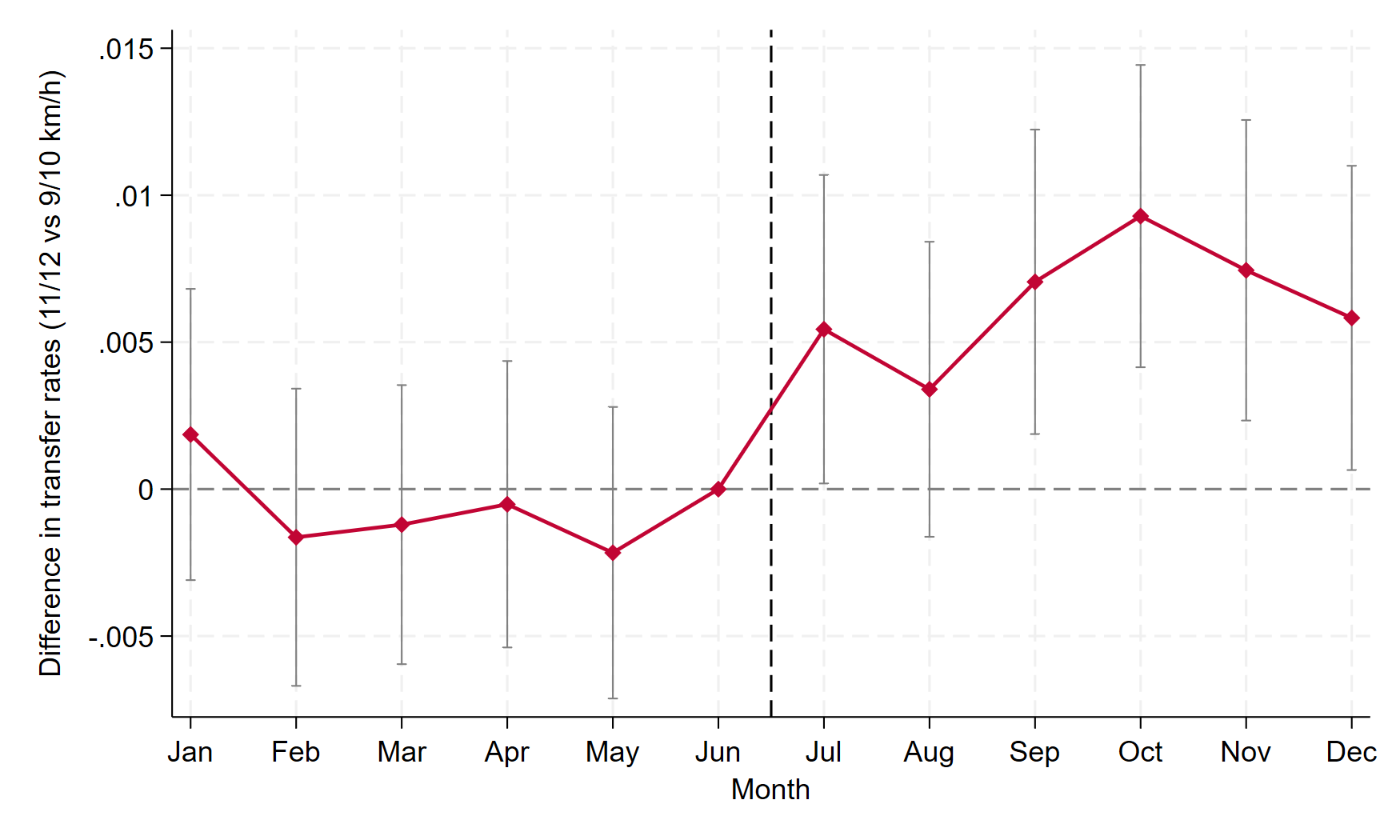}
\end{subfigure}
\hfill
\begin{subfigure}{0.49\linewidth}
    \centering
    \caption{Reoffending rates}
    \includegraphics[width=1\linewidth]{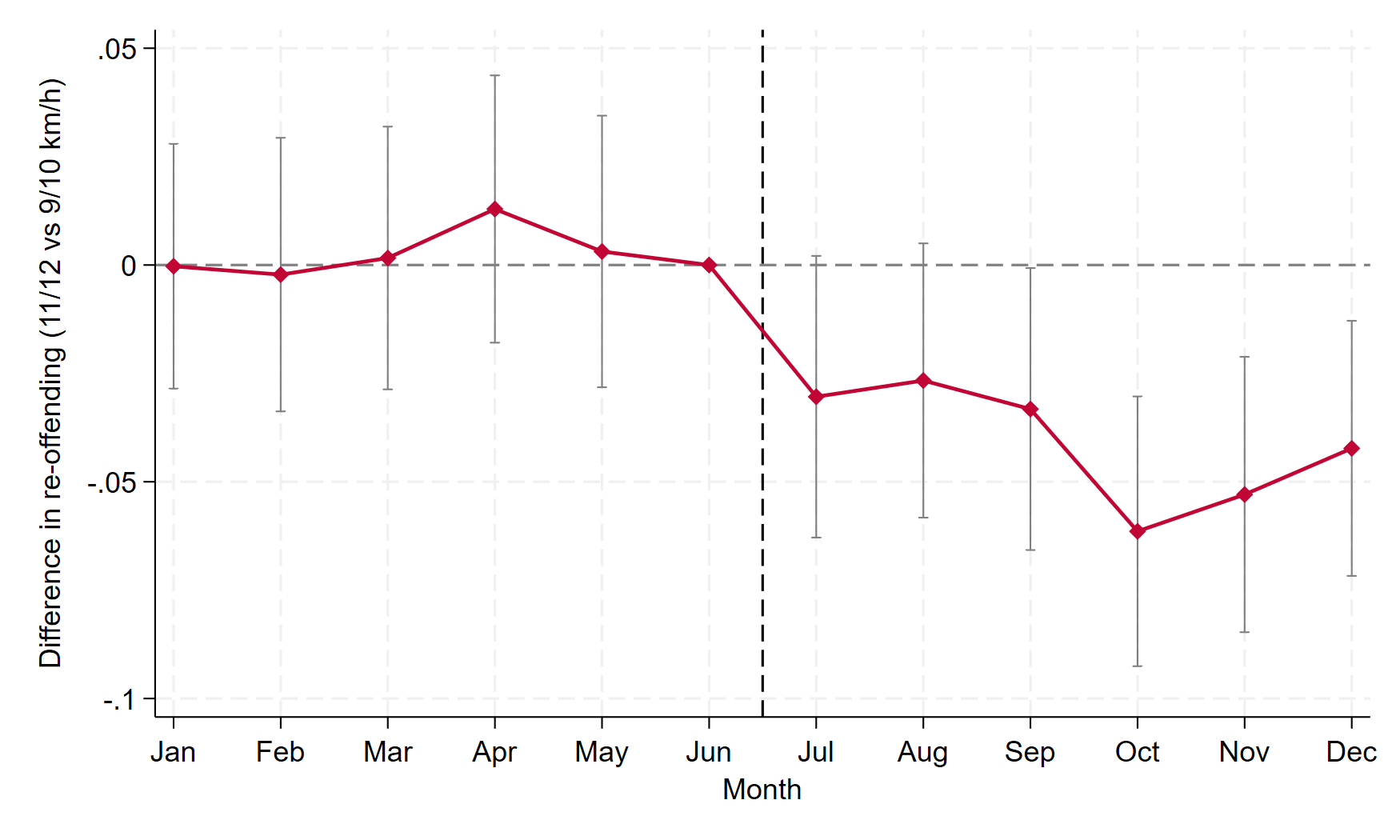}
\end{subfigure}
\caption{Estimated effects of increase in demerit points using a difference-in-differences framework.}
\label{fig:did_by_month}
\end{figure}

Figure \ref{fig:did_by_month} presents the results. Panel (a) shows the differential effect on transfer rates. The pre-reform coefficients are small and close to zero, consistent with the absence of pre-existing differential trends between the two groups. Following the reform in July 2022, the differential transfer rate increases persistently and becomes statistically significant from September onwards, indicating that drivers caught at 11--12~km/h became increasingly more likely to nominate another driver relative to those caught at 9--10~km/h. Panel (b) shows the corresponding results for reoffending rates. The pre-reform coefficients again show no systematic pattern, supporting the parallel trends assumption. After July 2022, the differential reoffending rate falls persistently, with statistically significant negative effects from July onwards, suggesting that the demerit point increase generated a stronger specific deterrence effect among 11--12~km/h offenders relative to the control group. Together, these results corroborate the main RDiT findings and provide additional evidence that the demerit point increase, rather than the fine increase alone, was the primary driver of both the transfer and deterrence responses.

\subsection{Exposure to the Post-Reform Penalty Regime \label{sec:main_results_fixed_donut}}

Table~\ref{tab:main_results_fixed_donut} in the Appendix addresses a potential concern about the timing of reoffending relative to the reform date. Since our reoffending outcome is measured over the 12 months following the initial offense, drivers caught before the reform may face the old, lower penalties during part of their follow-up period, whereas those caught after the reform face the new penalties throughout. This difference in exposure could affect the interpretation of our estimates. To address this concern, we exclude reoffending within the first 150 days following the initial offense, ensuring that all subsequent offenses counted in the outcome occur under the new penalty regime. The results remain qualitatively unchanged.

\FloatBarrier
\section{Conclusion}
\enlargethispage{2\baselineskip}

This paper examines how drivers adjust their behavior in response to a sharp increase in speeding sanctions, exploiting a major policy change introduced in Queensland on 1 July 2022. The policy substantially increased monetary fines across all speed-excess categories and raised demerit point allocations for selected offenses. Using comprehensive administrative data on speeding infringements and a regression discontinuity in time (RDiT) design, we estimate the causal effect of these changes on both subsequent reoffending and strategic liability transfer behavior.

Overall, the policy reduced the probability of reoffending by 1.6 percentage points, corresponding to a 5 per cent decline relative to the pre-intervention mean. Importantly, these effects are highly heterogeneous across offense types. The deterrent response is concentrated among drivers exceeding the speed limit by 11–12 km/h---precisely the group exposed to the largest increase in sanctions, both in monetary fines and demerit points. For these offenders, reoffending falls by 4.6 percentage points, or 13 per cent relative to baseline. By contrast, we find no evidence of behavioral responses among drivers facing higher fines alone with unchanged demerit points.

Alongside this deterrent effect, we document a substantial increase in strategic liability transfer following the policy change. Higher sanctions, and in particular the increase in demerit point allocations in the 11–12 km/h tier, induced registered vehicle owners to nominate another driver at significantly higher rates, with transfer rates rising by approximately 30–60 per cent relative to baseline among affected offenders. 

From a policy perspective, our results carry two main implications. First, they caution against relying solely on higher fines as a tool for improving road safety, especially in settings where baseline penalties are already substantial. Cumulative sanctions, such as demerit points, may provide an important additional margin of deterrence. In the United States, some states, such as Georgia and New York, operate demerit point systems, while others, such as Kansas, do not. Belgium similarly lacks such a system. Jurisdictions without demerit point systems may be forgoing an important source of deterrence.

Second, policymakers should account for the transferability of liability when designing penalty systems. When sanctions can be reassigned to another party, higher penalties may also induce strategic evasion, potentially undermining their intended deterrent effects and raising concerns about the integrity of the enforcement system.

\appendix
\newpage

\section*{Appendix}

\section{Increase in penalty units\label{app:pu_change}}

Table \ref{tab:pu_change} reports the change in penalty unit (PU) allocations by speed excess category introduced on 1 July 2022. PU amounts increased across all speed excess categories, with the percentage increase ranging from 23\% for the highest tier (above 40~km/h) to 125\% for the 11--12~km/h range. The two lowest categories (1-10~km/h and 13-20~km/h) both saw a 50\% increase, while the 21-30~km/h and 31-40~km/h ranges increased by 35\% and 61\% respectively. Since monetary fines are determined by multiplying the PU amount by the prevailing penalty unit value, these changes directly translate into the fine increases reported in Table \ref{tab:fee_change}.

\begin{table}[h!]
\centering
\begin{tabular}{|c|c|c|c|}
\hline
Speed excess & PU amount & PU amount & \% difference \\
&  (before 1 July 2022) & (from 1 July 2022) &  \\
\hline
1-10 & 1 1/3 & 2 & 50\% \\
\hline
11--12 & 1 1/3 & 3 & 125\% \\
\hline
13-20 & 2 & 3 & 50\% \\
\hline
21-30 & 3 1/3 & 4 1/2 & 35\% \\
\hline
31-40 & 4 2/3 & 7 1/2 & 61\% \\
\hline
$>$40 & 9 1/3 & 11 1/2 & 23\% \\
\hline
\end{tabular}
\caption{PU change for speeding offenses in Queensland on 1 July 2022.}
\label{tab:pu_change}
\end{table}

\section{Reoffending on fixed cameras\label{app:reoffending_fixed}}

Table \ref{tab:main_results_fixed} presents a robustness check restricting the reoffending analysis to fixed speed camera detections only. Since mobile camera offenses are more susceptible to strategic liability transfer, this restriction ensures that the outcome is not contaminated by nomination behavior. The estimated effects are somewhat smaller in magnitude than the main results, as expected given the different composition of offenders detected by fixed cameras, but remain negative and statistically significant for the 11--12~km/h tier, confirming the robustness of the main findings.

\begin{table}[htbp]
\centering
\caption{Effect of policy change on reoffending on fixed speed cameras.}
\label{tab:main_results_fixed}
\resizebox{\textwidth}{!}{%
\begin{tabular}{>{\raggedright\arraybackslash}p{3cm} *{5}{>{\centering\arraybackslash}p{1.8cm}}}
\toprule
 & \multicolumn{3}{c}{Overall} 
 & \multicolumn{2}{c}{Speed excess: 11--12 km/h} \\
\cmidrule(lr){2-4}\cmidrule(lr){5-6}
 &  &  & Exclude transfers & Only & Excluded \\
\midrule
Reoffending  & \mbox{-0.006} & \mbox{-0.009**} & \mbox{-0.010**} & \mbox{-0.028***}  & -0.005  \\
after & (0.004) & (0.004) & (0.004) & (0.010) & (0.004) \\
\midrule
N & 149,755 & 149,755 & 146,752 & 27,077 & 122,678 \\
\midrule 
Controls &  & \checkmark & \checkmark & \checkmark & \checkmark \\
\bottomrule
\multicolumn{6}{l}{\footnotesize Standard errors clustered at driver level in parentheses.}\\
\multicolumn{6}{l}{\footnotesize $^\dagger$ p$<$0.10, * p$<$0.05, ** p$<$0.01, *** p$<$0.001}\\
\end{tabular}%
}
\end{table}

As an additional robustness check, Table \ref{tab:main_results_fixed_donut} excludes reoffending occurring within the first 150 days after the initial offense. The estimates remain quantitatively similar, indicating that the main results are not driven by short-run repeat offending.

\begin{table}[htbp]
\centering
\caption{Effect of policy change on reoffending between 150-365 days after the initial offense.}
\label{tab:main_results_fixed_donut}
\resizebox{\textwidth}{!}{%
\begin{tabular}{>{\raggedright\arraybackslash}p{3cm} *{5}{>{\centering\arraybackslash}p{1.8cm}}}
\toprule
 & \multicolumn{3}{c}{Overall} 
 & \multicolumn{2}{c}{Speed excess: 11--12 km/h} \\
\cmidrule(lr){2-4}\cmidrule(lr){5-6}
 &  &  & Exclude transfers & Only & Excluded \\
\midrule
Reoffending  & \mbox{-0.010**} & \mbox{-0.012***} & \mbox{-0.013***} & \mbox{-0.046***}  & -0.005  \\
after & (0.004) & (0.004) & (0.004) & (0.010) & (0.005) \\
\midrule
N & 149,755 & 149,755 & 146,752 & 27,077 & 122,678 \\
\midrule 
Controls &  & \checkmark & \checkmark & \checkmark & \checkmark \\
\bottomrule
\multicolumn{6}{l}{\footnotesize Standard errors clustered at driver level in parentheses.}\\
\multicolumn{6}{l}{\footnotesize $^\dagger$ p$<$0.10, * p$<$0.05, ** p$<$0.01, *** p$<$0.001}\\
\end{tabular}%
}
\end{table}

\section{Bandwidth selection\label{app:bandwidth_check}}

\begin{figure}[h!]
\centering
\begin{subfigure}{0.49\linewidth}
    \centering
    \caption{Transfer Behavior}
    \includegraphics[width=1\linewidth]{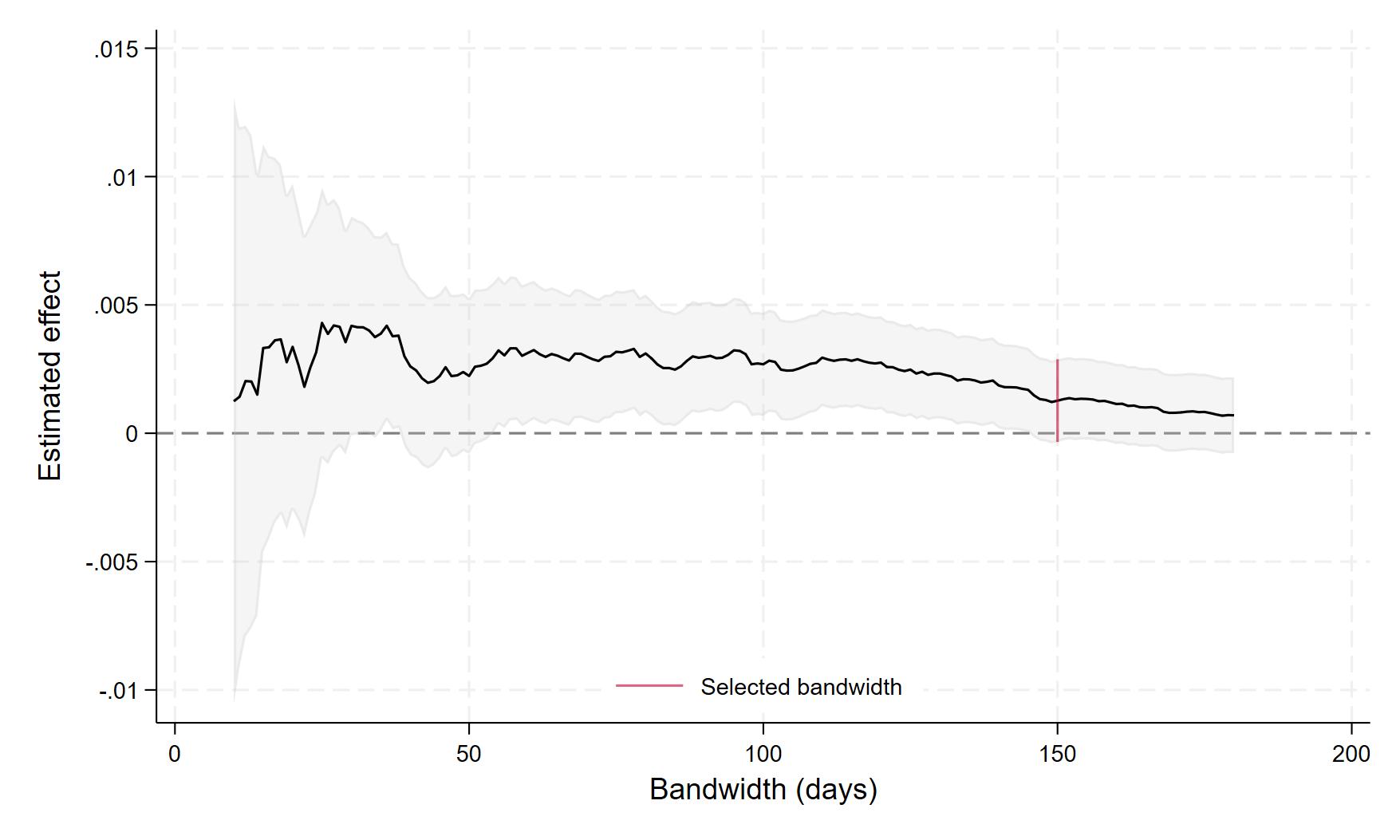}
\end{subfigure}
\hfill
\begin{subfigure}{0.49\linewidth}
    \centering
    \caption{Reoffending Behavior}
    \includegraphics[width=1\linewidth]{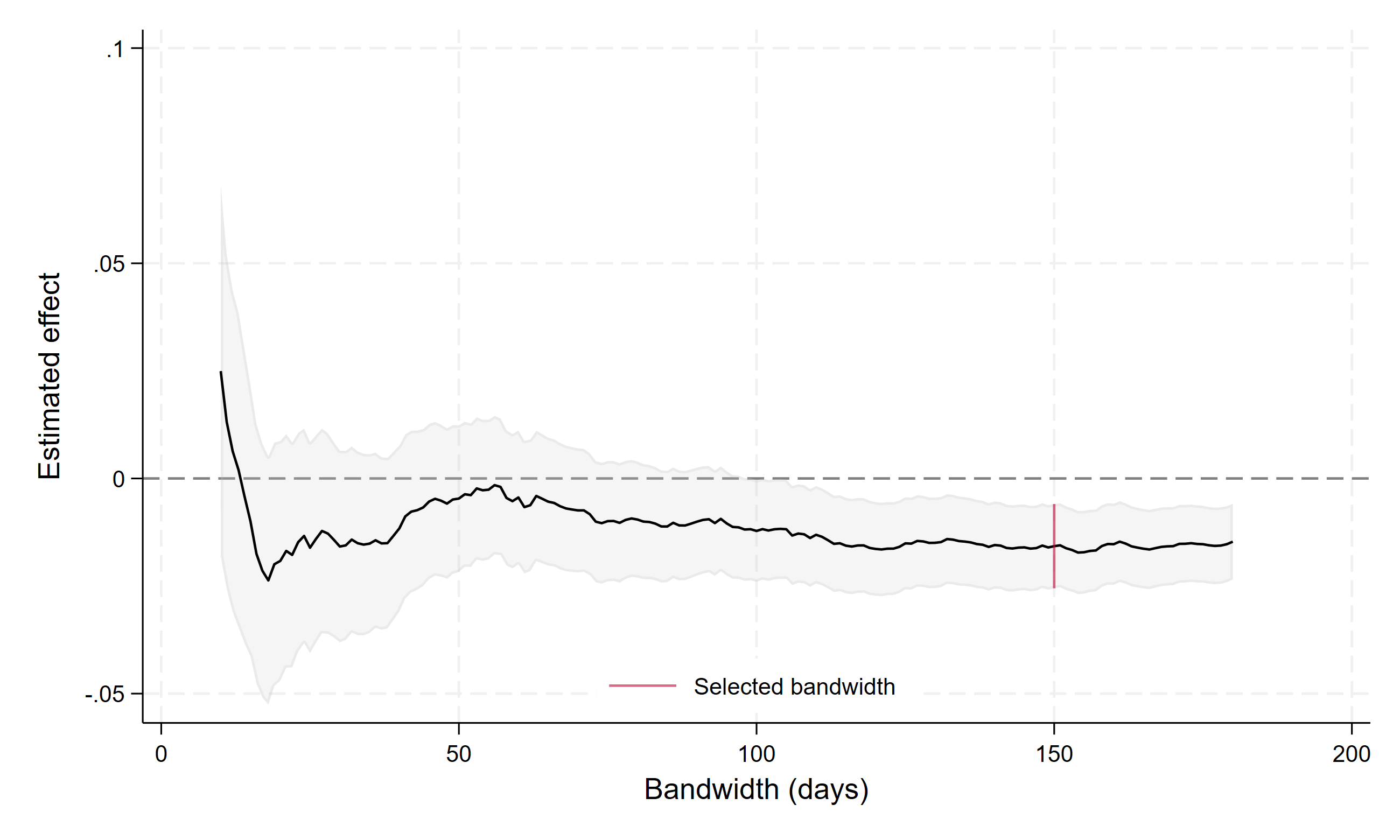}
\end{subfigure}
\caption{Effects of the increase in penalty on transfer rate (a) and reoffending (b) by bandwidths between 10 and 180 days.}
\label{fig:bandwidth_check}
\end{figure}

\clearpage
\bibliographystyle{apalike} 
\bibliography{bibliography}

\newpage


\clearpage
\setcounter{section}{0}
\setcounter{table}{0}
\setcounter{figure}{0}
\setcounter{page}{1}

\renewcommand{\thesection}{OA.\Alph{section}}
\renewcommand{\thetable}{OA.\arabic{table}}
\renewcommand{\thefigure}{OA.\arabic{figure}}

\begin{center}
    {\LARGE\textbf{Online Appendix}}\\[0.5em]
    {\large Sanctions, Evasion, and Deterrence: Evidence from Speeding Enforcement}
\end{center}

\vspace{1em}

\section{Effect heterogeneity\label{app:hetereogeneity}}

\begin{table}[htbp]\centering
\def\sym#1{\ifmmode^{#1}\else\(^{#1}\)\fi}
\caption{Heterogeneity analysis}
\begin{tabular}{>{\raggedright\arraybackslash}p{1cm} *{8}{>{\centering\arraybackslash}p{1.55cm}}}
\toprule
       & \multicolumn{2}{c}{Age} & \multicolumn{2}{c}{Demerit pts} & \multicolumn{2}{c}{Income} & \multicolumn{2}{c}{Aus. born} \\
\cmidrule(lr){2-3}\cmidrule(lr){4-5}\cmidrule(lr){6-7}\cmidrule(lr){8-9}
        &\multicolumn{1}{c}{$\geq$ 25}&\multicolumn{1}{c}{$<$ 25}&\multicolumn{1}{c}{$<$ 2}&\multicolumn{1}{c}{$\geq$ 2}&\multicolumn{1}{c}{$\geq$ \$794}&\multicolumn{1}{c}{$<$ \$794}&\multicolumn{1}{c}{$\geq$ 68\%}&\multicolumn{1}{c}{$<$ 68\%}\\
\midrule
Effect &      -0.015\sym{**}&      -0.017         &      -0.023\sym{***}&      -0.002         &      -0.029\sym{***}&      -0.002         &      -0.004         &      -0.024\sym{***}\\
                    &     (0.005)         &     (0.017)         &     (0.006)         &     (0.008)         &     (0.007)         &     (0.007)         &     (0.008)         &     (0.007)        \\
\midrule
Obs.        &     134,000         &      15,237         &      81,831         &      67,419         &      72,362         &      76,888         &      72,023         &      77,227        \\
Pre-treatment mean & 0.324 & 0.341 & 0.255 & 0.407 & 0.340 & 0.310 & 0.317 & 0.334 \\
Relative effect (\%) & -4.6 & -5.0 & -9.1 & -0.4 & -8.4 & -0.7 & -1.2 & -7.1 \\
\bottomrule
\multicolumn{9}{l}{\footnotesize Standard errors clustered at driver level in parentheses.}\\
\multicolumn{9}{l}{\footnotesize $^\dagger$ p$<$0.10, * p$<$0.05, ** p$<$0.01, *** p$<$0.001}\\
\end{tabular}
\label{tab:heterogeneity_reoffending}
\end{table}

\begin{table}[htbp]\centering
\def\sym#1{\ifmmode^{#1}\else\(^{#1}\)\fi}
\caption{Heterogeneity analysis: transfers, mobile speed cameras}
\begin{tabular}{>{\raggedright\arraybackslash}p{1.5cm} *{8}{>{\centering\arraybackslash}p{1.4cm}}}
\toprule
                    & \multicolumn{2}{c}{Age} & \multicolumn{2}{c}{Demerit pts} & \multicolumn{2}{c}{Income} & \multicolumn{2}{c}{Aus. born} \\
\cmidrule(lr){2-3}\cmidrule(lr){4-5}\cmidrule(lr){6-7}\cmidrule(lr){8-9}
                    & \multicolumn{1}{c}{$\geq$ 25} & \multicolumn{1}{c}{$<$ 25} & \multicolumn{1}{c}{$<$ 2} & \multicolumn{1}{c}{$\geq$ 2} & \multicolumn{1}{c}{$\geq$ \$794} & \multicolumn{1}{c}{$<$ \$794} & \multicolumn{1}{c}{$\geq$ 68\%} & \multicolumn{1}{c}{$<$ 68\%} \\
\midrule
Effect &       0.001         &       0.004  &       0.001         &       0.002  &       0.001         &       0.001         &       0.001         &       0.001         \\
                    &     (0.001)         &     (0.002)         &     (0.001)         &     (0.001)         &     (0.001)         &     (0.001)         &     (0.001)         &     (0.001)         \\
\midrule
Obs.        &     385,354         &      51,370         &     258,661         &     178,162         &     200,353         &     236,582         &     275,795         &     161,140         \\
Pre-treatment mean & 0.016 & 0.015 & 0.017 & 0.015 & 0.017 & 0.015 & 0.015 & 0.018 \\
Relative effect (\%) & 6.1 & 26.4 & 4.5 & 13.7 & 8.1 & 8.2 & 9.0 & 6.2 \\
\bottomrule
\multicolumn{9}{l}{\footnotesize Standard errors clustered at driver level in parentheses.}\\
\multicolumn{9}{l}{\footnotesize $^\dagger$ p$<$0.10, * p$<$0.05, ** p$<$0.01, *** p$<$0.001}\\
\end{tabular}
\label{fig:heterogeneity_transfers}
\end{table}

\begin{landscape}
\section{Literature Review\label{app:littreview}}


\begin{longtable}{>{\raggedright\arraybackslash}p{2.5cm} >{\centering\arraybackslash}p{1.5cm} >{\centering\arraybackslash}p{2.5cm} >{\centering\arraybackslash}p{1cm} >{\raggedright\arraybackslash}p{3cm} >{\raggedright\arraybackslash}p{8cm}}

\caption{Literature Review} \label{tab:littreview} \\ 

\toprule
Paper & Margin & Country/State & Dem. Pt. & Identification Strategy & Result \\
\midrule
\endfirsthead

\multicolumn{6}{l}{\footnotesize \textit{Continued from previous page}} \\
\toprule
Paper & Margin & Country/State & Dem. Pt. & Identification Strategy & Result \\
\midrule
\endhead

\midrule
\multicolumn{6}{r}{\footnotesize \textit{Continued on next page}} \\
\endfoot

\bottomrule
\multicolumn{6}{l}{\footnotesize Notes: Int = specific deterrence (intensive margin); Ext = general deterrence (extensive margin); Dem. Pt. = demerit points.} \\
\endlastfoot

\cite{dusek2021}
& Int/Ext
& Czech Republic
& No
& RDD penalty threshold \& event study
& Intensive margin: higher fines produce a small but imprecisely estimated additional effect. Extensive margin: the speeding rate drops by a third and re-offense rates fall by 70\%. \\
\addlinespace

\cite{Kaila2024}
& Int
& Finland
& No
& RDD penalty threshold
& Fines that are 200 euros larger decrease reoffending by 15\% in the following six months. After 12 months, the effect disappears. \\
\addlinespace

\cite{gehrsitz2017}
& Int
& Germany
& Yes
& RDD
& One-month license suspensions lower the probability of recidivating within a year by 20\%. \\
\addlinespace

\cite{gonccalves2023police}
& Int
& Florida
& No
& Discretionary vs.\ non-discretionary officers
& A \$125 increase in fine amount reduces the likelihood of any new traffic offense in the following year by 1.6 percentage points. \\
\addlinespace

\cite{abay2018}
& Int/Ext
& Denmark
& Yes
& DID
& Drivers reduced their frequency of traffic offenses by 9-34 percent depending on accumulated demerit points. \\
\addlinespace

\cite{studdert2017}
& Ext
& Queensland
& Yes
& RDD date of notification
& Sharp decrease in reoffending of 25\% for the cohort as a whole and up to 40\% for some subgroups of drivers. \\
\addlinespace

\cite{Lawpoolsri01042007}
& Ext
& Maryland
& No
& Non-causal
& Drivers who received a speeding citation in May 2002 had almost twice the risk of receiving a speeding citation during follow-up, compared with those not cited. \\

\cite{killias2016} 
&  Int
&  Switzerland
& No
&  Non-causal
&  Speed-limit violations remained unaffected in the longer run by higher fines. When cities without relevant enforcement activity were excluded, a reduction by 17 \% was observed.
\\

\cite{ELVIK2007689} 
&  Int
&  Norway
& No
&  Non-causal
&  For speeding in general, no effect of increasing fixed penalties can be found.
\\

\cite{Bhalla53} 
&  Int
&  Russia
& No
&  Non-causal
&  Eliminating penalties - overall speeding increased significantly by 13
percentage points.

\end{longtable}
\end{landscape}

\end{document}